\documentclass[useAMS,usenatbib]{mn2e}

\usepackage{amssymb}
\usepackage{amsmath}
\usepackage{mathtools}
\usepackage{graphicx}
\usepackage{longtable}
\usepackage{hyperref}

\newcommand{\apj}{ApJ}
\newcommand{\apjs}{ApJS}

\newcommand{\aap}{A{\&}A}

\newcommand{\mnras}{MNRAS}
\newcommand{\aj}{AJ}
\newcommand{\araa}{ARAA}

\newcommand{\pasj}{PASJ}

\title[Kinematic Ram Pressure Measurements]{A Kinematic Measurement of Ram Pressure in the Outer Disk of Regular Galaxies}

\author[Haan \& Braun]{S. Haan$^{1}$ \& R. Braun$^{1,2}$\\
$^{1}$CSIRO Astronomy and Space Science, ATNF, PO Box 76, Epping 1710, Australia\\
$^{2}$SKA Organisation, Jodrell Bank Observatory, Lower Withington, Macclesfield, Cheshire, SK11 9DL, UK}

\begin{document}

%\onecolumn

\pagerange{\pageref{firstpage}--\pageref{lastpage}} \pubyear{2013}

\maketitle

\label{firstpage}

\begin{abstract}
While most ram pressure studies have focused on ram pressure stripping in galaxy clusters, we devise a novel approach based on a kinematic measurement of ram pressure perturbations in HI velocity fields for intergalactic material (IGM) densities and relative velocities that are one to two orders of magnitude lower than in galaxies showing ram pressure stripping. Our model evaluates ram pressure induced kinematic terms in gas disks with constant inclination as well as those with a warped geometry. Ram pressure perturbations are characterized by kinematic modes of even order, m=0 and m=2, corresponding to a ram wind perpendicular and parallel to the gas disk, respectively. Long-term consequences of ram pressure, such as warped disks as well as uncertainties in the disk geometry typically generate uneven modes (m=1 and m=3), that are clearly distinguishable from the kinematic ram pressure terms. We have applied our models to three nearby isolated galaxies, utilizing Markov Chain Monte Carlo fitting routines to determine ram pressure perturbations in the velocity fields of NGC~6946 and NGC~3621 of $\sim$30~km~s$^{-1}$ (effective line-of-sight velocity change) at HI column densities below (4--10)$\times10^{20}$~cm$^{-2}$ (at radial scales greater than $\sim$15~kpc). In contrast, NGC~628 is dominated by a strongly warped disk. Our model fits reveal the three-dimensional vector of the galaxies' movement with respect to the IGM rest-frame and provide constraints on the product of speed with IGM density, opening a new window for extragalactic velocity measurements and studies of the intergalactic medium.

\end{abstract}


\section{Introduction}
The boundary between the interstellar material (ISM) and the surrounding intergalactic material (IGM) is often characterized by ram pressure interaction which plays an important role in galaxy evolution via the stripping of gas and regulating the gas reservoir for star-formation and fueling of active galactic nuclei (AGN). The strength of ram pressure that is exerted on the ISM depends primarily on two physical properties, the density of the ISM and the relative velocity of the galaxy, as first suggested by \cite{Gun72}. While the ISM mass at small radii is dominated by molecular gas, the outer disks of galaxies are primarily atomic. Neutral atomic hydrogen with its 21 cm line emission (HI) is the most sensitive tracer of the dynamics in the outskirts of galaxies and provides an ideal tool to identify interaction features since it is often the most spatially extended component of a galaxy's disk and is very sensitive to interactions because of its dissipative nature. \par 

In particular, observations of the nearest galaxy cluster, the Virgo cluster, have brought tremendous insight about the stripping of ISM from spiral galaxies \citep[e.g.][]{Cay90, Vei99, Vol99, Vol03, Ken04, Vol04, Cro05, Cor06, Cro08, Roe08, Vol09, Abr11, Arr12, Vol13} as well as elliptical galaxies  \citep[e.g.][]{Ran95, Luc05, Mac06}. The consequence of ram pressure stripping on galaxy evolution can be seen, for example, as a deficiency of atomic neutral hydrogen in galaxies and a declining star formation activity towards the center of a galaxy cluster \citep[see e.g.][]{Hel81, Hay86, Hof88, Nak06, Vol12}. Moreover, ram pressure winds can possibly lead to asymmetric dense molecular arms \citep{Hid02} as well as metal enrichment of the IGM which is suggested by simulations \citep{Sch05, Dom06} and observations of galaxy clusters \citep[see. e.g][]{Cui10}.\par 

Ram pressure stripping has currently been studied in circumstances of dense inter-group/cluster gas and high velocity galaxy orbits. Much less is known about the impact of inter-galactic gas on the ISM in more isolated galaxies, where forces are insufficient to overcome the gravitational potential or significantly disturb the ISM distribution. Here we propose a novel approach utilizing kinematic features caused by ram wind interaction with diffuse, outer HI disks at IGM densities that are an order of magnitude lower than in galaxies that show ram pressure stripping. In principle this provides a unique tracer of the properties of the intergalactic medium and a three-dimensional measurement of the direction of a galaxies' movement with respect to the rest-frame of the IGM. In particular the amplitude of the velocity change in the ISM due to ram pressure can provide important information about the density of the IGM and the galaxy's relative speed with respect to the IGM.\par

The distribution and kinematics of the outer disk of galaxies often show warped disks which can be described by a systematic change in inclination and position angle of the disk as function of radius \citep[see e.g.][]{Bri90, Kam92, Joz07, Kru07, Kam13}. However, not all anomalies in the observed velocity field can be attributed to warped disks. One important question is how to distinguish kinematic features of warped disks from ram interactions. In this study we develop a kinematic model of ram pressure interaction in the outer HI disk of galaxies and test these models on nearby galaxies. Our kinematic disk model of ram pressure interaction is described in \S~\ref{sec:mathmodel} which includes the derived mathematical description (\S~\ref{subsec:math}) and the implication on the HI velocity field (\S~\ref{subsec:velfield}), models of velocity perturbations due to ram pressure and warped disks (\S~\ref{subsec:model}), and the velocity decomposition using Monte Carlo fitting routines (\S~\ref{subsec:mcmc}). In \S~\ref{sec:comp} we compare our ram pressure models to the observed HI kinematics of nearby disk galaxies and discuss the implications for deriving the properties of the IGM in \S~\ref{sec:dis}. The main conclusions of this study are summarized in \S~\ref{sec:sum}.

\section{A Kinematic Model of Ram Pressure Interaction}
\label{sec:mathmodel}

\subsection{Mathematical Description of the Impact of Ram Pressure on the ISM}
\label{subsec:math}
Ram pressure is exerted on the interstellar medium (ISM) of a galaxy due to its motion relative to an intergalactic medium (IGM). Since the relative velocities are typically 100's of km~s$^{-1}$, while the sound speed in the neutral ISM \citep[with temperature less than about $10^4$~K required for a significant neutral fraction][]{Wol03} is less than about 10~km~s$^{-1}$, this is a highly supersonic interaction. The kinematic outcome will depend on whether the interaction is adiabatic or radiative. If the gas is collisionally heated to a high enough temperature that radiative losses are insignificant, such as may occur in cluster - cluster mergers for example, then the velocity difference can be calculated on the basis of kinetic energy conservation. If, on the other hand, the heat of the collision is efficiently radiated, as must apply to any residual atomic ISM, then only the condition of momentum conservation applies to the calculation of the perturbed velocity. The interaction causes a drag force to be exerted on the outer gas of a galaxy and can, in extreme cases, strip much of its interstellar gas. The strength of interaction between the IGM and the gas bound by the gravitational potential of a galaxy depends primarily on the following three parameters: 1) the relative velocity, $v_{Ram}$, of the galaxy with respect to the IGM, 2) the density of the IGM, $\rho_{IGM}$, and 3) the exposed surface area, $S$, of the ISM towards the ram wind. The ram pressure, $P$, is usually defined as \citep{Gun72},
\begin{equation}
P= \rho_{IGM}\;v_{Ram}^2
\label{eq:vram1}
\end{equation}
and the component of the ram force parallel to the galaxy's disk on a gas element is given by
\begin{equation}
F_{Ram\parallel} \propto\rho_{IGM} \; v_{Ram}^2\;S \cos\gamma_{Ram}
\label{eq:vram2}
\end{equation}
and perpendicular by
\begin{equation}
F_{Ram\perp} \propto \rho_{IGM}\; v_{Ram}^2\;S \sin\gamma_{Ram}
\label{eq:vram3}
\end{equation}
where $\gamma_{Ram}$ is the angle between the wind direction and the galaxy's plane. 

\begin{figure}
\begin{center}
\includegraphics[scale=0.45]{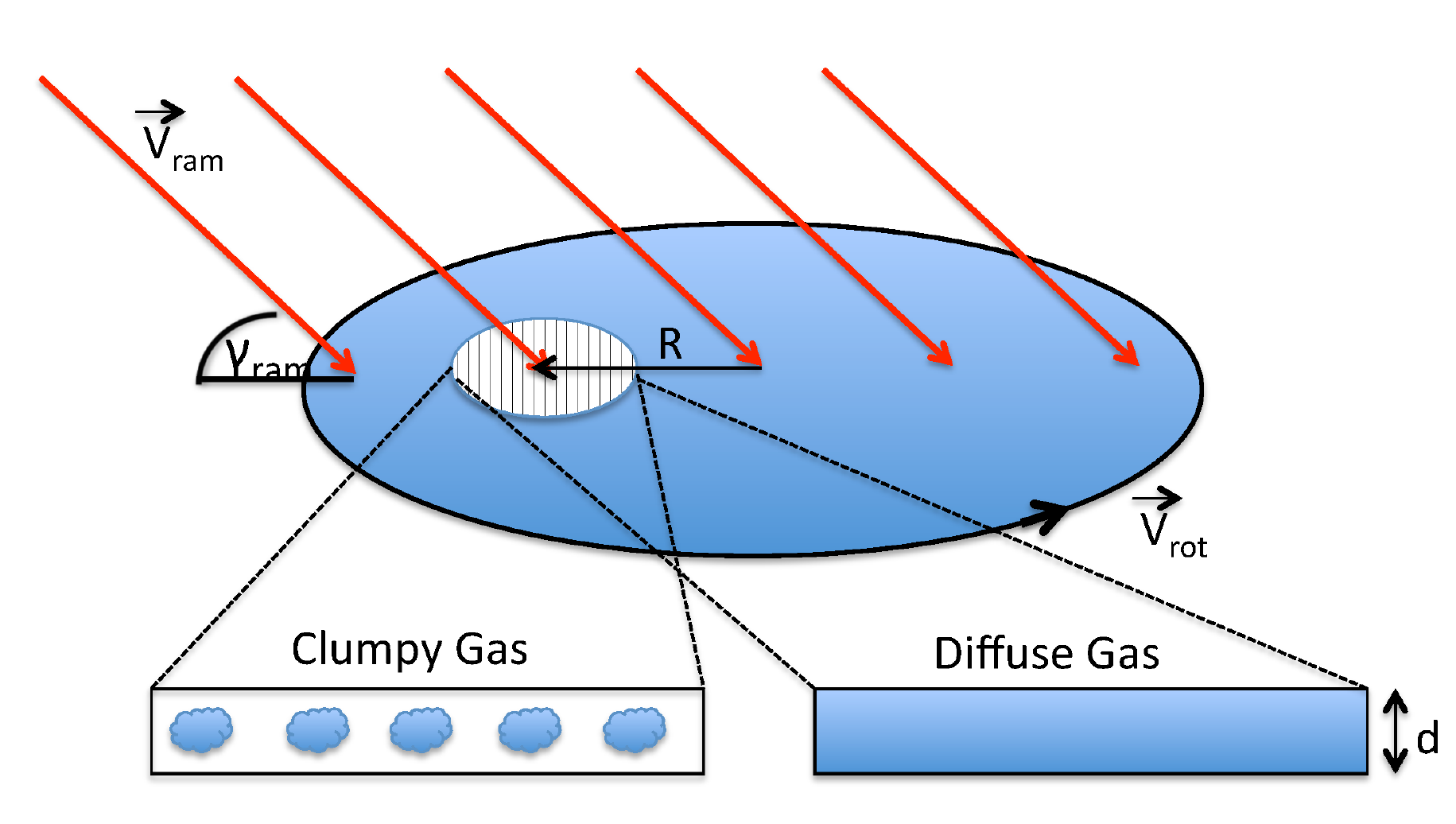}
\caption{Illustration of a disk galaxy moving through the intergalactic medium. The ram wind has a different impact on the state of gas, depending on whether it is diffuse or distributed as clumps.}
\label{illustration}
\end{center}
\end{figure}

The gas within a volume element with a surface area $A$ and a thickness $d$ of the gas disk will be assumed to be either diffuse gas (homogeneously distributed over a volume element) or clumpy (see Fig.~\ref{illustration} for illustration). In the diffuse case, the gas has a constant density $\rho_{Diff}$ within a volume element and the surface area, $S_{Diff}=A$. In the clumpy case, the gas is distributed over a number of individual gas clouds within the volume element, with volume filling factor, $f_{3D}$. For a population of isotropic objects, these will have a surface covering factor, $f_{2D} = f_{3D}^{2/3}$, so that $S_{Clmp}=A f_{2D}$. In the context of two neutral atomic phases in approximate pressure equilibrium \citep{Wol03}, these two cases might be identified with the warm and the cool neutral medium (WNM and CNM) respectively with a volume density contrast of about 100:1. For the WNM, $f_{3D} \sim f_{2D} \sim 1$, while for the CNM, $f_{3D} = 0.01$ and $f_{2D} = 0.05$. 

We will assume that the galactic gas is distributed as a thin plane that is not exactly edge-on relative to the IGM motion. A relevant question to consider is the depth in the disk to which the ram pressure wind will propagate. For a diffuse disk of thickness, $d = 200$ pc and sound speed, $c = 10$ km~s$^{-1}$, the disk penetration time, $\tau_P = d/(c\sin\gamma_{Ram})  = 20/\sin\gamma_{Ram}$ Myr, is much shorter than a disk rotation period for $\gamma_{Ram}\ge20\deg$. In the case of a clumpy ISM with a small volume filling factor like the CNM noted above, the disk penetration timescale is unlikely to be different, since most of the disk volume is filled with a low density medium with a high shock propagation speed. The probability condition for not having multiple clumps along the same line-of-sight is given approximately by $P = f_{2D}/\sin\gamma_{Ram} < 0.5$ implying $\gamma_{Ram}\ge6\deg$ for the CNM. Even in cases of approximate alignment of clumps, the inter-clump shock would tend to bend around obstructions and act on the entire disk volume within this characteristic penetration time. Therefore, it is unlikely that gas cloud shadowing plays a significant role in influencing the outcome of a ram pressure interaction with a galactic disk. 

Our analysis will also be confined exclusively to the low-ram pressure regime, i.e. at ram pressure accelerations that do not lead to significant stripping of gas. In the event that the relative galaxy-IGM velocity greatly exceeds the galactic rotation speed, as is the case in galaxy clusters, where the ram pressure is one to two orders of magnitude larger than in galaxy groups, significant asymmetries of the stripped material will occur, such as leading side-trailing asymmetries as seen for some galaxies in the Virgo Cluster \citep[see e.g.][]{Ken04, Vol12}.\par

The force opposing the ram pressure is the gravitational binding force of the galaxy, which is roughly equal to the centrifugal force,
\begin{equation}
F_{Grav}\sim m_{ISM} v_{Rot}^2/r
\end{equation}  
where $v_{Rot}$ is the rotation velocity of the ISM at a radius $r$ from the galactic center and $m_{ISM}$ the mass of the gas in the volume element with $m_{ISM}=\rho_{Diff}\;S\;d$ is independent of whether the gas is diffuse or clumpy.
The ratio $\mu=F_{Ram}$/$F_{Grav}$ describes the relative impact of the ram pressure on the ISM in the galactic disk and is basically a measure of the strength to disturb the motion of the ISM. 
For diffuse gas, this effective ram parameter is given by 
\begin{equation}
 \mu=\frac{\rho_{IGM} \; v_{Ram}^2\;S \cos(\gamma_{Ram})}{m_{ISM} v_{Rot}^2/r}= \frac{\rho_{IGM}}{\rho_{Diff}}\frac{v^2_{Ram}}{v^2_{Rot}} \frac{r}{d}\cos(\gamma_{Ram})
\end{equation}
and for clumpy gas by,
\begin{equation}
 \mu=\frac{\rho_{IGM} \; v_{Ram}^2\;S \cos(\gamma_{Ram})}{m_{ISM} v_{Rot}^2/r}= \frac{\rho_{IGM}}{\rho_{Diff}}\frac{v^2_{Ram}}{v^2_{Rot}} \frac{r f_{2D}}{d}\cos(\gamma_{Ram})
\end{equation}
that differs only by the clump surface covering factor, $f_{2D} = 0.05$, that applies to the CNM.
Since diffuse gas has a much higher surface covering factor than a clumpy gas, it requires much smaller ram forces to disturb the motion of the diffuse ISM.
In extreme cases, i.e. $\mu>>1$, this can lead to ram pressure stripping as seen for high $\rho_{IGM}$ as e.g. demonstrated in galaxy clusters.\par

The effective velocity change $\Delta v$ of the ISM due to ram pressure can be described via momentum conservation in the approximation of an fully dissipative collision between the ISM and IGM,
\begin{equation}
\vec{p}_{IGM} +\vec{p}_{ISM} = \vec{p}_{Tot}.
\end{equation}
In the rest-frame of the galaxy ($\vec{p}_{ISM}=0$), this equation reduces to
\begin{equation}
m_{IGM}\;\mid v_{Ram} \mid=(m_{IGM} + m_{ISM})\mid \Delta v \mid,
\end{equation}
and using $m_{ISM}=m_H\;n_H\;S\;d = m_H\;N_{ISM}$ (diffuse gas) and $m_{IGM}=m_H\;N_{IGM}$,  the effective velocity change is given as
\begin{equation}
\mid \Delta v \mid = \frac{\mid v_{Ram} \mid}{1+\frac{N_{ISM}}{N_{IGM}}}.
\end{equation}
For $N_{ISM} >> N_{IGM}$, the effective velocity change of the ISM can be approximated by 
\begin{equation}
\mid \Delta v \mid \simeq \mid v_{Ram} \mid \frac{N_{IGM}}{N_{ISM}}.
\end{equation}
\par

The effective velocity perturbation induced by this momentum transfer in each portion of the disk can only be integrated over a fraction of the rotational period of the gas around the galaxy's center due to the continuously changing relative orientation of the ram wind. On longer timescales, the orbital paths themselves will be modified by the accumulated ram pressure acceleration \citep[see][]{Haa13} together with gravitational restoring forces of the galaxy potential. We will discuss the long-term consequences of the ram pressure interaction and the implications for the interpretation of a galaxy's velocity field in detail in \S~\ref{sec:dis}. For the following we assume that at all times there is a basic orbital configuration of essentially circular gas orbits on which is superposed a kinematic perturbation that is constant on timescales that are a small fraction of a rotational period.

\subsection{The modified HI Velocity Field}
\label{subsec:velfield}
Neutral Hydrogen (HI) is often the most spatially extended component of a galaxy's disk and is very sensitive to interactions because of its dissipative nature. In particular the kinematics of diffuse gas may respond in a measurable way to ram forces long before significant displacement or removal of gas from the disk has occurred. Thus one might expect to be able to measure the ram pressure and hence the properties of the IGM  at much smaller densities than those found in galaxy clusters ($n_{IGM}<<n_{ICM}$). This might be accomplished via detailed study of the HI kinematics in the outer part of galaxies, where we find diffuse atomic gas at low column densities (and hence large $\mu$).\par

In the following we develop a model to quantify the effect of ram pressure on the HI velocity field of a galaxy. To describe the transition from a clumpy to a diffuse ISM we use a smooth step function as given by,
\begin{equation}
\eta(\rho_{ISM})=0.5+0.5\tanh((\rho_{Crit}-\rho_{ISM})/f_{Smooth}).
\label{eq:sfunct}
\end{equation}
This function returns values between 0 and 1 where the parameter $\rho_{Crit}$ and $f_{Smooth}$ define the critical density threshold and the smoothing range, respectively. This models the rapid decline of the HI column density and surface covering factor of clumps in the outer disk, which scales the effective ram pressure to 0\% for very dense regions and 100\% for low density diffuse gas. However, the critical density $\rho_{Crit}$ where such a change occurs is not known beforehand and can be only determined by fitting simultaneously the columns density distribution and amplitude of velocity perturbation in observed galaxies.
We have tested other scale functions in order to describe the impact of the ram force on the gas kinematics as function of HI gas density (e.g various power laws), but came to  the conclusion that a smooth step function provides a reasonable match to the observed kinematics (see \S~\ref{sec:comp}). However, the overall 2-dimensional pattern as seen in the velocity field is roughly the same, independent of the HI density scale function, since it parametrizes only the dependence on the clumpiness of the HI gas. The line-of-sight velocity, which is the only velocity component that is directly observable
\begin{equation}
\label{eq:circular}
v_{LoS}=v_{Sys}+v_{Rot}(r)\cos(\theta)\sin(i) + v_{Exp}(r)\sin(\theta)\sin(i),
\end{equation}
with 
\begin{eqnarray}
x&=&-(X-X_0)\sin(\phi) + (Y-Y_0)\cos(\phi),  \nonumber \\
y&=&\frac{-(X-X_0)\cos(\phi) - (Y-Y_0)\sin(\phi)}{\cos(i)},  \nonumber \\
r&=&\sqrt{x^2+y^2},  \nonumber \\
\cos(\theta)&=&\frac{x}{r},  \nonumber\\   
\sin(\theta)&=&\frac{y}{r}.  \nonumber  
\label{eq:sintheta}
\end{eqnarray}  
Here $X_0$ and $Y_0$ are the position of the rotation center in the plane of the sky (X,Y), $r$ the radius in the plane of the disk (x,y), $v_{Sys}$ the systemic velocity, $v_{Rot}(r)$ the rotational velocity at a radius $r$, $v_{Exp}$ the rdial velocity, $i$ the inclination angle and $\phi$ the position angle of the receding major axis of the galaxy measured in anti-clockwise direction from north toward the east. In the following we will set $X_0=0$ and $Y_0=0$ and we will assume first a constant PA and inclination of the disk. The more general case where both PA and inclination can change as function of radius (e.g. due to a warped disk) is described in detail in App.~\ref{app:b} and is labelled as $\theta_W$ throughout the paper. \par

\begin{figure}
\begin{center}
\includegraphics[scale=1.0]{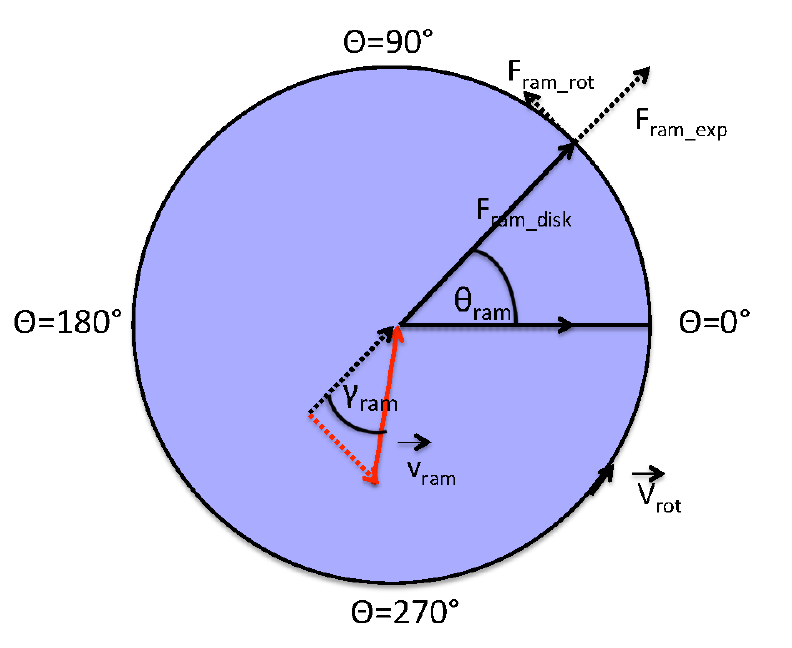}
\caption{Illustration of the contribution from ram wind parallel to the disk, $F_{ram disk}$, to the rotation ($F_{ram rot}$) and radial velocity component $F_{ram exp}$ of a galaxy. The ram wind vector $\mathbf{v}$ (red, out of the plane) is defined by the inclination angle $\gamma_{ram}$ between disk and ram wind and the azimuthal angle $\theta_{ram}$ (see text). }
\label{geometry}
\end{center}
\end{figure}

The velocity due to the ram wind $\Delta v$ can be described as an additional component to the HI kinematics (see Fig.~\ref{geometry}). Here we assume that the entire gas disk is exposed to a constant ram pressure field which contributes a velocity component to the velocity field of the galaxy. The magnitude of this ram velocity component is a function of the HI gas density as described by the scale function $\eta(\rho_{ISM}$). Here we decompose the ram velocity vector into three different components: 1) perpendicular to the gas disk as given by,
 \begin{equation}
v_{Ram\bot}=\mid \Delta v\mid \eta(\rho_{ISM})\sin(\gamma_{Ram}),
\end{equation}  
 2) an additional rotational velocity component,
\begin{equation}
v_{RamRot}=\mid \Delta v\mid \eta(\rho_{ISM})\sin(\theta-\theta_{Ram})\cos(\gamma_{Ram}),
\end{equation}  
and 3) an additional radial velocity component,
\begin{equation}
v_{RamExp}=\mid \Delta v\mid \eta(\rho_{ISM})\cos(\theta-\theta_{Ram})\cos(\gamma_{Ram}).
\end{equation}  
Note that the ram wind vector field is fixed with respect to the sky plane but that the angles $\theta_{Ram}$ and $\gamma_{Ram}$ depend on the alignment of the plane of the galaxy with respect to the sky plane as described in App.~\ref{app:c}.
The entire velocity field, including the ram component, is then given as
\begin{multline}
\label{eq:ramvelocity}
%\begin{split}
v_{LoS} =v_{Sys}+[v_{Rot}(r)+v_{RamRot}]\cos(\theta)\sin(i) \\
 + [v_{Exp}(r)+v_{RamExp}]\sin(\theta)\sin(i) + v_{Ram\bot} \cos(i).
%\end{split}
\end{multline}

If we subtract the circular rotation $v_{Rot}$ and the systemic $v_{Sys}$ and set the non-ram wind radial velocity $v_{Exp}$ term to zero, we derive a line-of-sight residual velocity field $v_{Res}$ that includes only ram pressure terms which can be written as,  
\begin{eqnarray}
v_{Res}&=&\bigl[ v_{RamRot}\cos(\theta) + v_{RamExp}\sin(\theta)\bigr] \sin(i) \nonumber \\
&&+ v_{Ram\bot} \cos(i)  \nonumber \\
&=&\mid\Delta v\mid \eta(\rho_{ISM}) \nonumber \\
&& \times  \Bigg( \cos(\gamma_{Ram})\sin(i) \bigg( \sin(\theta-\theta_{Ram})\cos(\theta) \nonumber \\ &&+\cos(\theta-\theta_{Ram})\sin(\theta) \bigg) + \sin(\gamma_{Ram})\cos(i) \Bigg) \nonumber \\
&=&\mid \Delta v\mid \eta(\rho_{ISM}) \bigl[ \cos(\gamma_{Ram})\sin(i)\sin(2\theta-\theta_{Ram}) \nonumber \\
&& +\sin(\gamma_{Ram})\cos(i) \bigr].
\label{eq:ramresidual}
\end{eqnarray} 
 
%\begin{equation}
%\label{eq:ramresidual}
%\begin{split}
%v_{Res} & = & \bigl[ v_{RamRot}\cos(\theta) + v_{RamExp}\sin(\theta)\bigr] \sin(i) + v_{Ram\bot} \cos(i) \\
% & = & \mid\Delta v\mid \eta(\rho_{ISM})\lbrace\cos(\gamma_{Ram})\sin(i)\left[\sin(\theta-\theta_{Ram})\cos(\theta) +\\
%& & \cos(\theta-\theta_{Ram})\sin(\theta)\right] + \sin(\gamma_{Ram})\cos(i) \rbrace \\
% & = & \mid \Delta v\mid \eta(\rho_{ISM}) \bigl[ \cos(\gamma_{Ram})\sin(i)\sin(2\theta-\theta_{Ram})+\sin(\gamma_{Ram})\cos(i) \bigr].
%\end{split}
%\end{equation} 
The first term of $v_{Res}$, with its $2\theta$ dependence, is a second order Fourier component of azimuthal angle in the galaxy plane, while the second term is independent of azimuth corresponding to a zero order Fourier component.\par

A significant fraction of disk galaxies have warped disks that are characterized by a change in PA and inclination as function of radius. In the following we provide a mathematical description of the velocity field and residual velocity field which includes the change in PA and inclination as function of radius. A complete description is given in App.~\ref{app:c}. The line-of-sight velocity component for a combination of a warped disk and ram wind  (extending Eq.~\ref{eq:ramvelocity} to the case of a warped disk)
\begin{multline}
\label{eq:ramwarpvelocity}
%\begin{split}
v_{LoS}=v_{Sys}+(v_{Rot}(r)+v_{RamRot})\cos[\theta_W(r)]\sin[i(r)] \\
+ v_{RamExp}\sin[\theta_W(r)]\sin[i(r)] + v_{Ram\bot} \cos[i(r)].
%\end{split}
\end{multline}
where $\theta_W(r)$ is a function of the PA $\phi$(r) of the warped disk that changes as a function of radius $r$ (see equation \ref{eq:thetawarp}), while the inclination $i(r)$ is a function of radius as well. 
We find that the residual velocity field can be decomposed into three Fourier components of $\theta$
\begin{multline}
\label{eq:ramwarp_res}
%\begin{split}
v_{Res}=c_0(r) + c_1(r)\cos[\theta_{warp*}(r)] \\
+ c_2(r)\sin[2\theta_W(r) - \theta_{Ram}(r)]
%+\mid \Delta v\mid \eta(\rho_{ISM}) \bigl[ \cos(\alpha)sin(i(r))sin(2\theta(r)-\theta_{Ram}) +  \cos(i(r)\;sin(\alpha) \bigl]
%\end{split}
\end{multline}
with 
\begin{eqnarray}
c_0(r)&=&\mid \Delta v\mid \eta(\rho_{ISM}) \cos[i(r)]\sin[\gamma_{Ram}(r)]\\
c_1(r)&=&v_{Rot}(r)\sin[i(r)]\;2\sin[\phi_W(r)/2]\\
c_2(r)&=&\mid \Delta v\mid \eta(\rho_{ISM}) \cos[\gamma_{Ram}(r)]\sin[i(r)]
\end{eqnarray}
where the first order describes the main residual of a warped disk in comparison to a circular rotating disk with constant inclination and position angle, while the zero and second order characterize the ram interaction terms of a warped disk perpendicular and parallel to the disk, respectively. The inclination and position angle of the warp are a function of radius 
\begin{eqnarray}
i(r)&=&i_0 + i_W(r),\\
\phi(r)&=&\phi_0 + \phi_W(r).
\end{eqnarray}
and can be described, e.g., as transition function between an inner and outer disk with
\begin{eqnarray}
i_W(r)&=&\Delta i \{0.5+0.5\tanh[(r -r_W)/s_W]\},\\
\phi_W(r)&=&\Delta \phi \{0.5+0.5\tanh[(r -r_W)/s_W]\}.
\end{eqnarray}
where $\Delta\phi$ and $\Delta i$ characterize the amplitude of change in PA and inclination, respectively. The complete derivation is provided in App.~\ref{app:b}. We note that this is just a first order approximation of the geometry of a true warp, which can have, for instance, different transition scale lengths between the change in PA and inclination, depending also on the orientation of the disk. \par

%\clearpage

\subsection{Model of velocity field pattern including ram pressure}
\label{subsec:model}

\begin{figure}
\begin{center}
\includegraphics[scale=0.26]{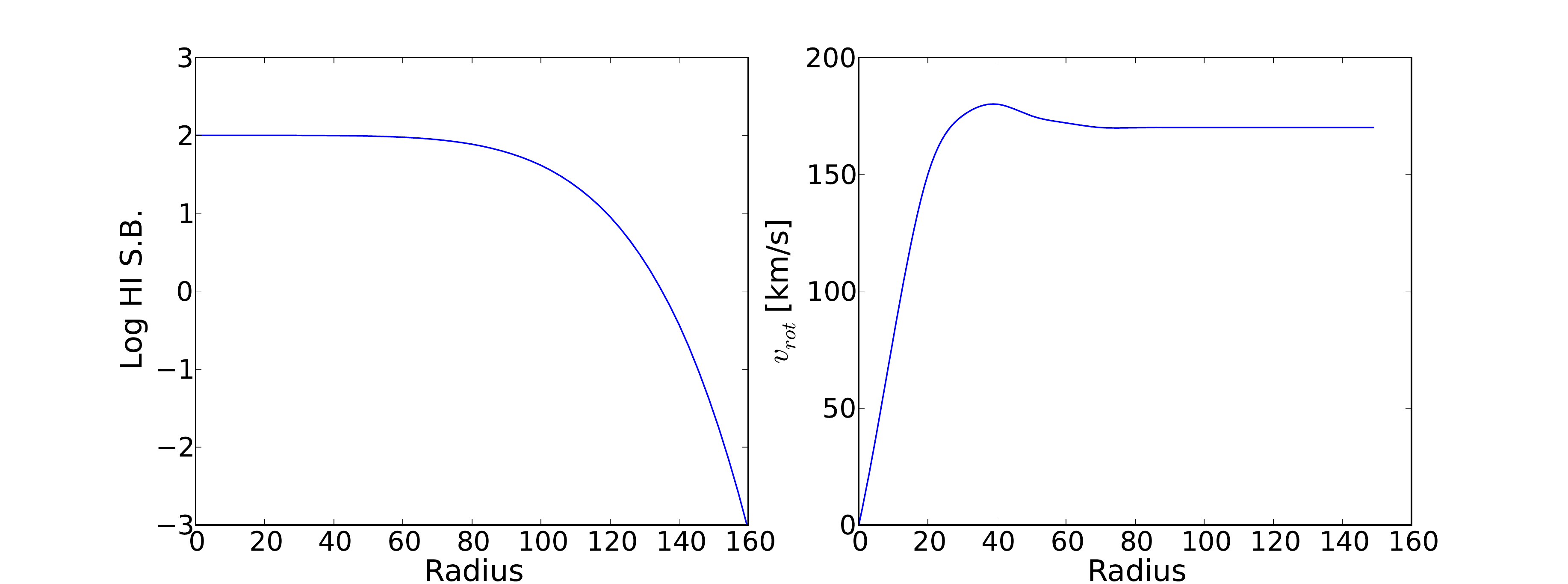}
\includegraphics[scale=0.25]{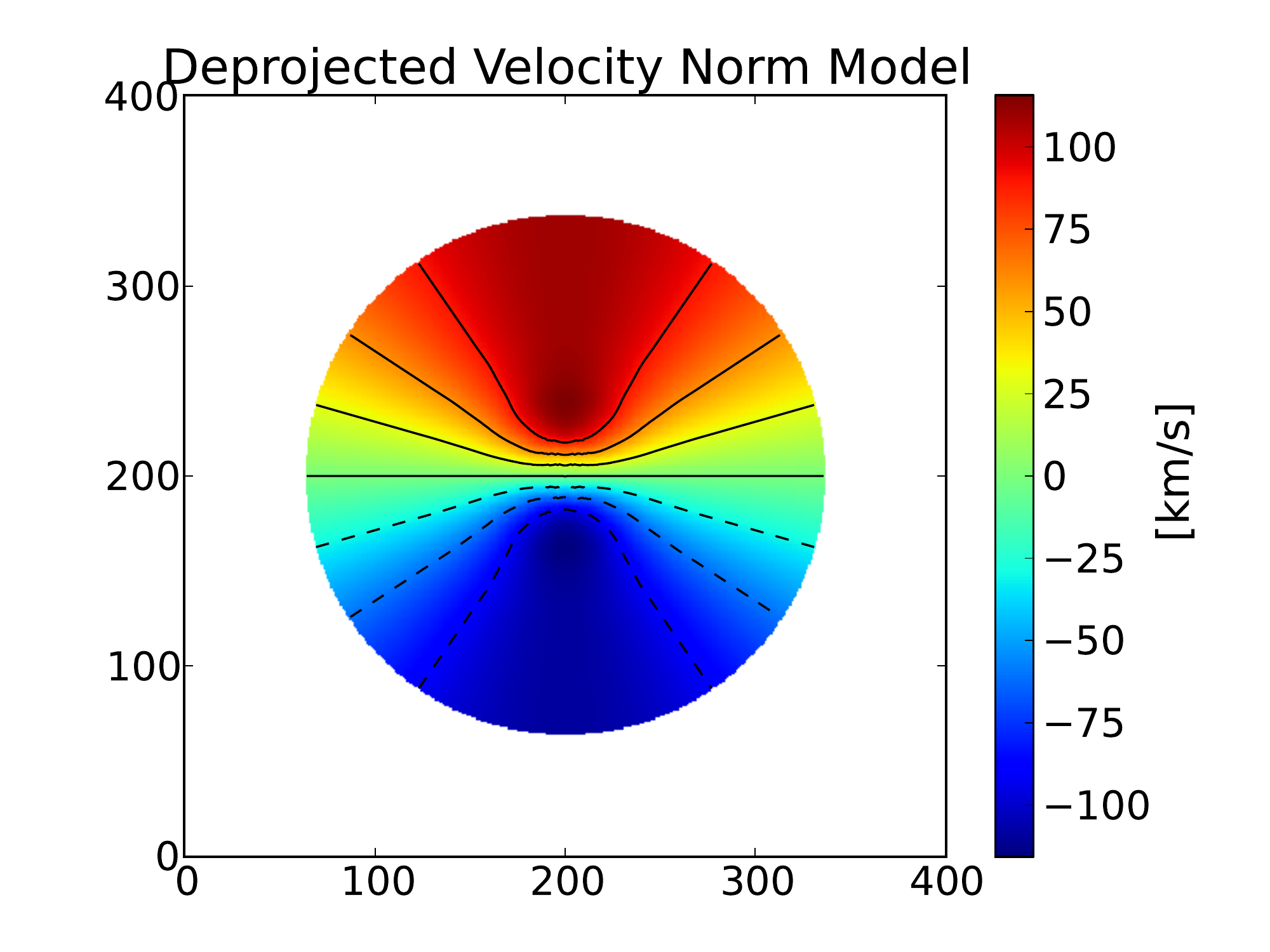}
\caption{Top: Radial profile of the HI column density  (left) and rotation velocity (right) of a typical disk galaxy as used in the model. Bottom: the model residual velocity field with no ram pressure interaction or warped disk.}
\label{radialmodel}
\end{center}
\end{figure}

\begin{figure}
\begin{center}
\includegraphics[scale=0.33]{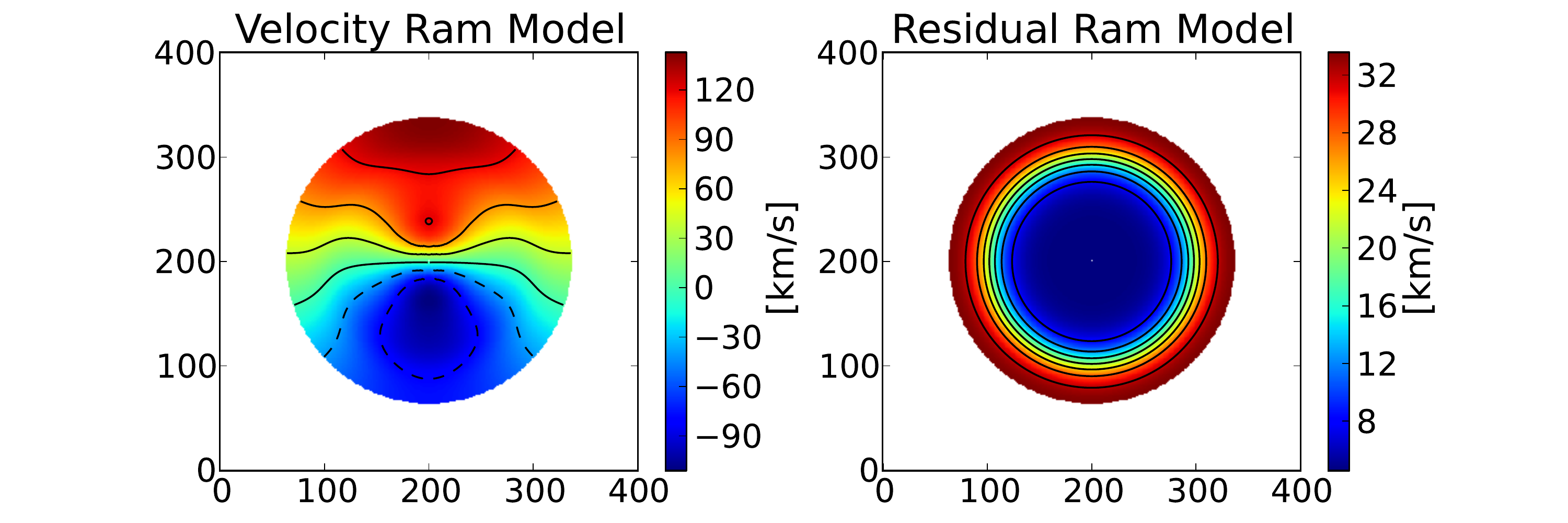}
\caption{Model velocity (left) and residual field (right) with ram wind perpendicular to disk.} 
\label{modelram}
\end{center}
\end{figure}

\begin{figure}
\begin{center}
\includegraphics[scale=0.33]{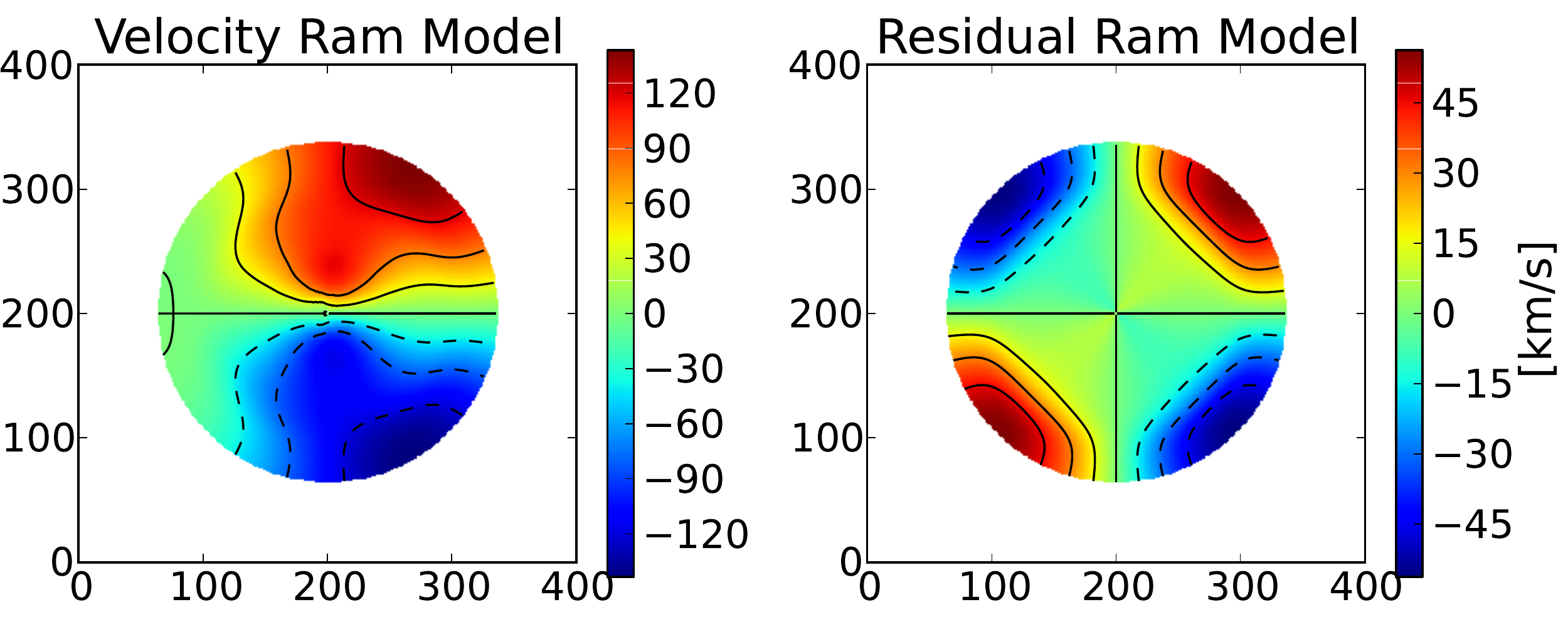}
\includegraphics[scale=0.33]{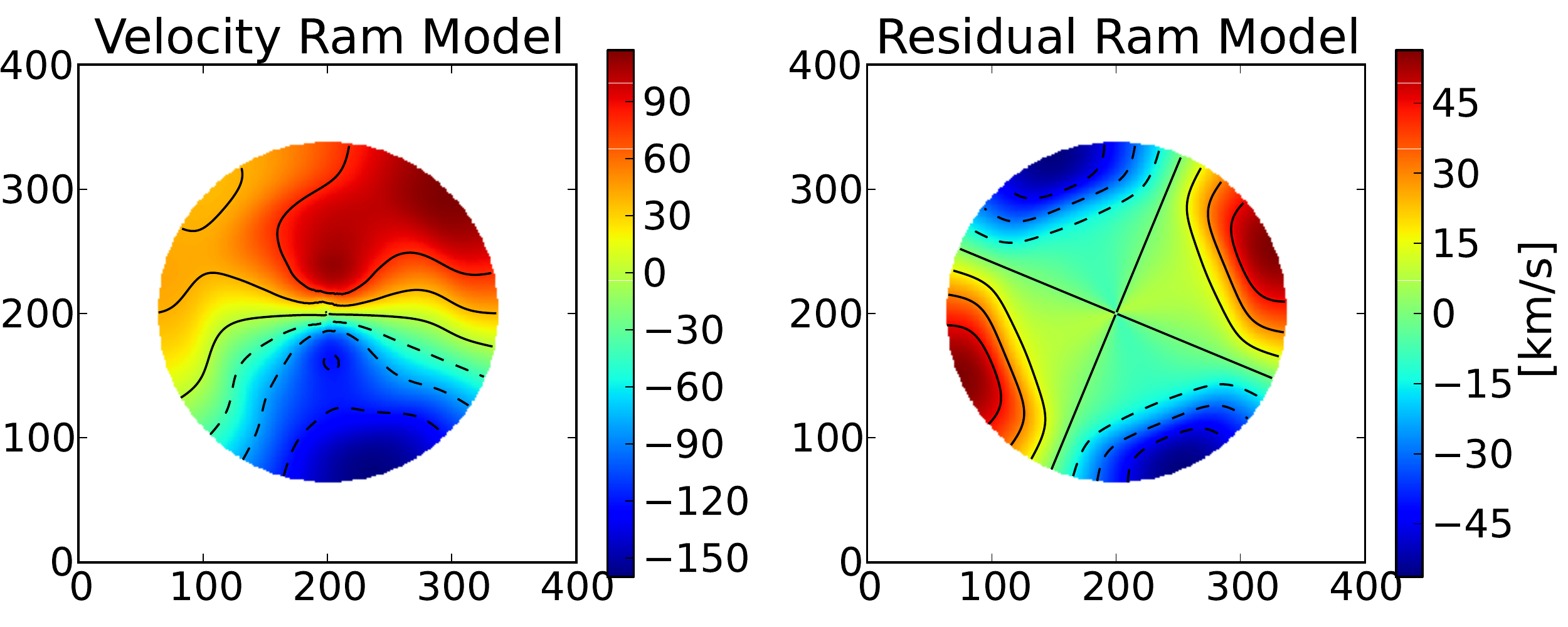}
\includegraphics[scale=0.33]{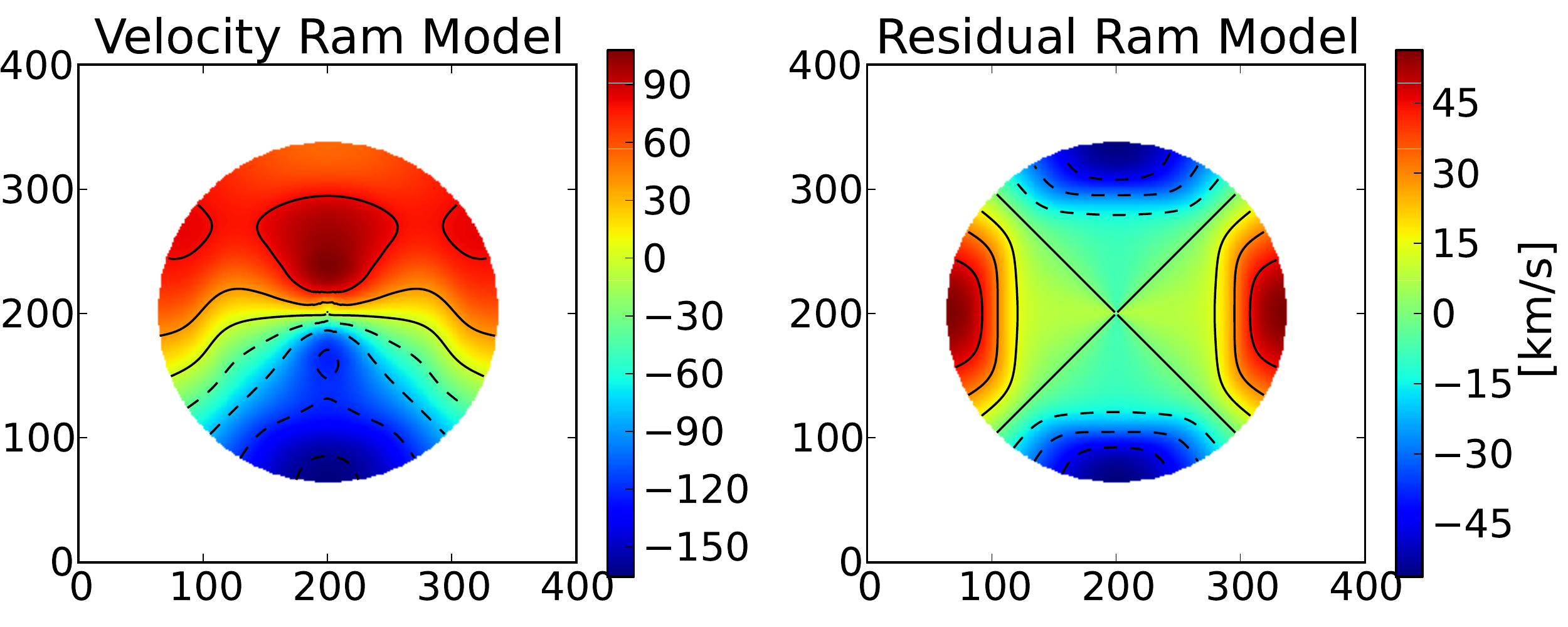}
\includegraphics[scale=0.33]{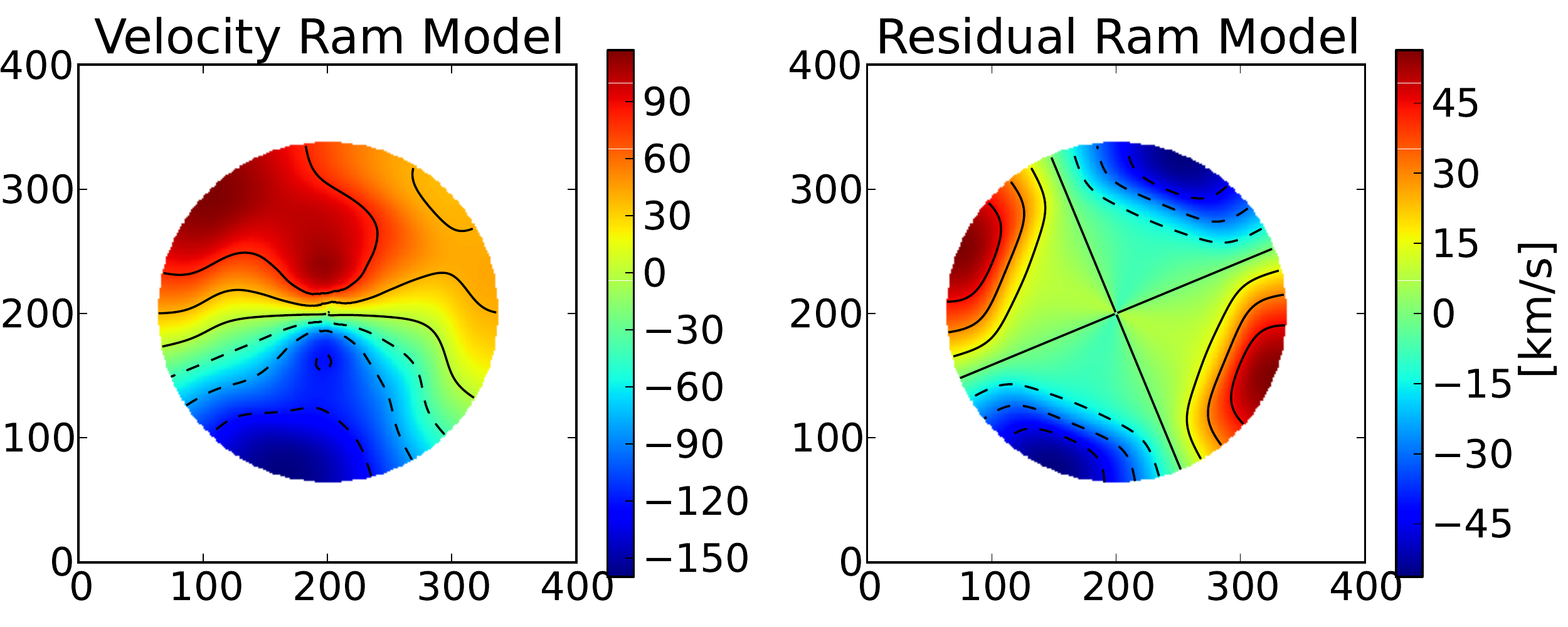}
\caption{Model velocity (left) and residual field (right) with ram wind parallel to disk. The ram wind is shown for 4 different angles. From top to bottom: $\theta_{ram}=0\deg$, $\theta_{ram}=45\deg$, $\theta_{ram}=90\deg$, and $\theta_{ram}=135\deg$. } 
\label{modelram2}
\end{center}
\end{figure}

\begin{figure}
\begin{center}
\includegraphics[scale=0.33]{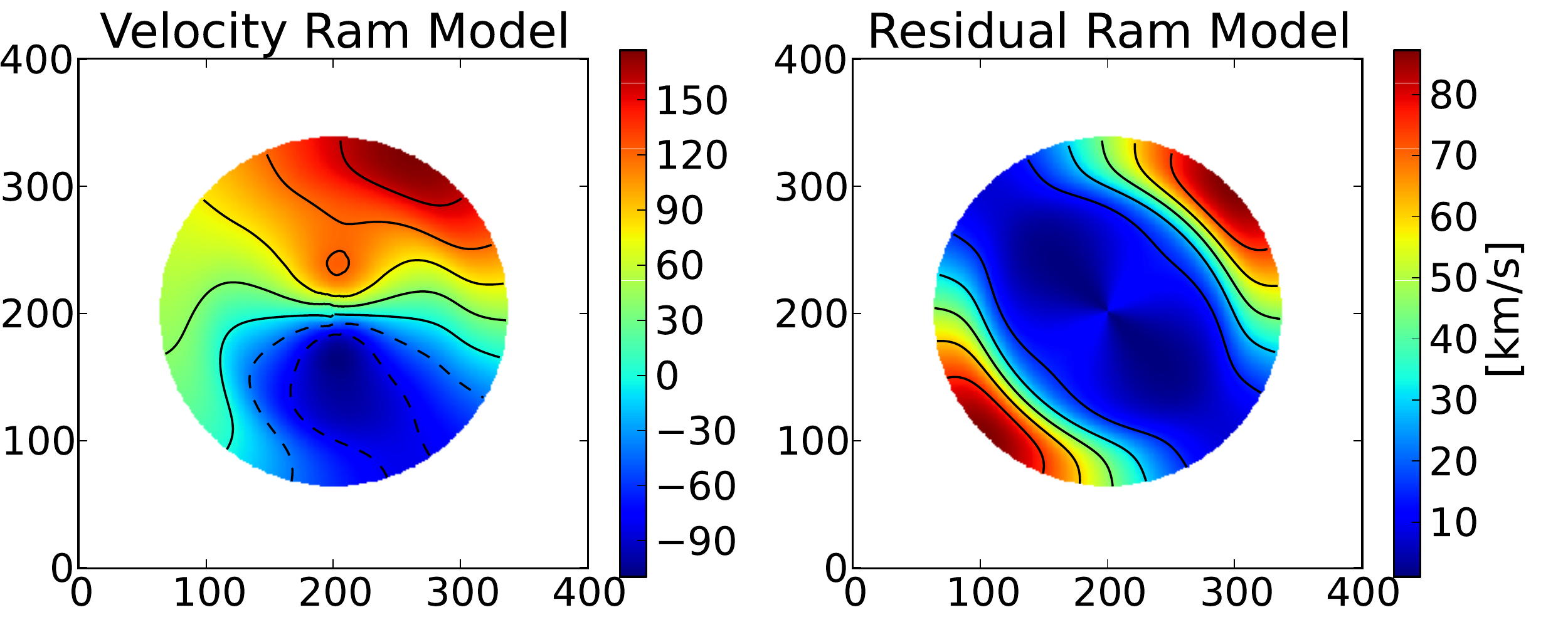}
\includegraphics[scale=0.33]{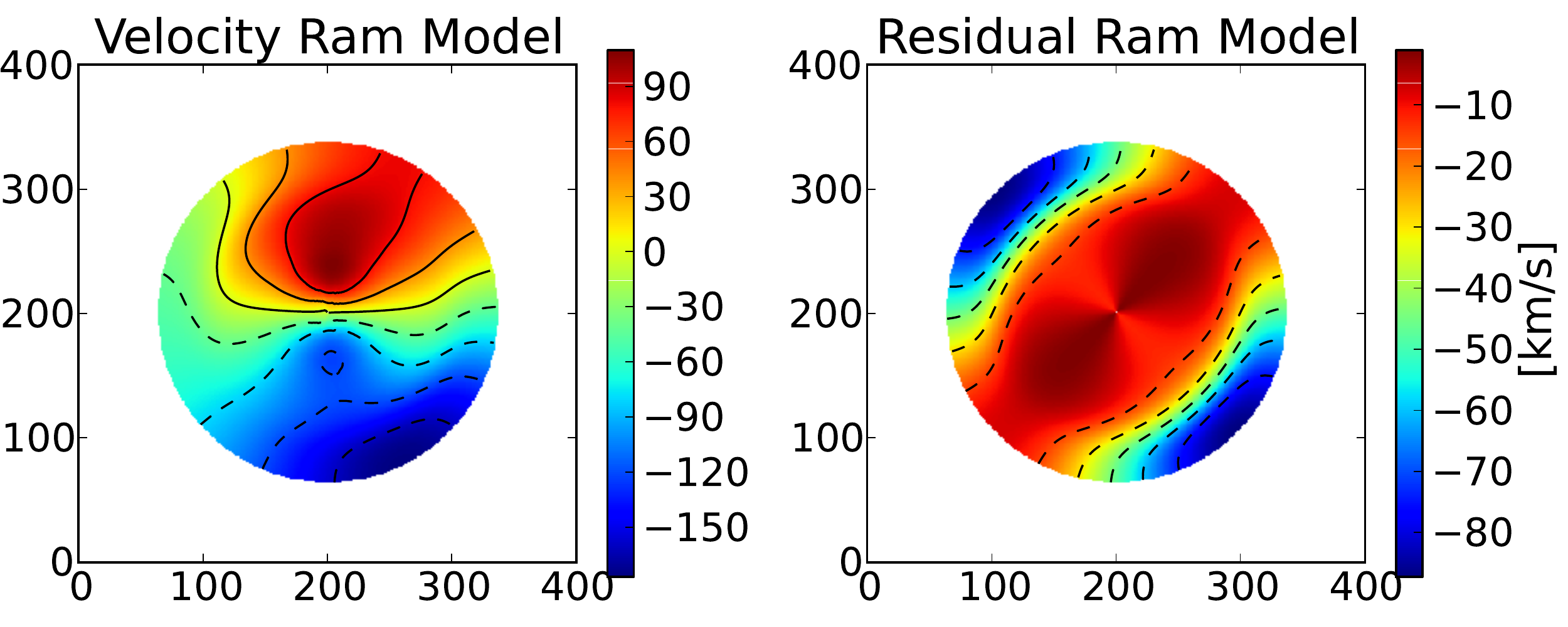}
\caption{Model velocity and residual field with ram wind from an angle of $\gamma_{ram}=45\deg$ (top panel) and $\gamma_{ram}=-45\deg$ (bottom panel) inclined to disk and an azimuthal angle of $\theta=0\deg$.} 
\label{modelram3}
\end{center}
\end{figure}

Here we develop a model HI velocity field modified by a ram pressure interaction. We start with a HI density profile and a typical rotation curve for disk galaxies as shown in Fig.~\ref{radialmodel}
The HI density field model follows a S\'ersic relation 
\begin{equation}
I(r)=I_0\;exp[-(1-n)(r/r_i)]^{1/n}
\label{eq:radmodel_HIdesn}
\end{equation}  
where $I_0$ is the central intensity, $r$ the radius, $n$ is the Sersic index, and $r_i$ the radius where the slope reaches maximum steepness before flattening off. We have chosen for our model a Sersic index of $n=0.18$ as measured by \cite{Por09} using the THINGS sample, however, any other Sersic index would reveal the same result.
The undisturbed velocity field is computed using Equation \ref{eq:circular} and is shown in Fig.~\ref{radialmodel}. In a second step we add to this velocity field a ram velocity vector as defined in equation \ref{eq:ramvelocity} and model the modified velocity field as function of the ram direction given by $\theta$ and $\gamma$. The effective ram force depends on the HI gas density as described by Equation \ref{eq:ramvelocity}. Fig~\ref{modelram}, Fig~\ref{modelram2}, and Fig~\ref{modelram3} show the line-of-sight velocity component of the models for different angles of the ram wind.
%Note that the figure depicts the kinematic projected velocities but keeps the disk geometrically unprojected.
We can identify distinct patterns in the outer HI velocity fields due to the ram interaction and its direction, which can be classified into distinct harmonic components of the azimuthal angle, $\theta$:
\begin{itemize}
\item A ram wind perpendicular to the disk results in an $m=0$ Fourier component that is independent of azimuth with amplitude proportional to gas column density (see Fig.~\ref{modelram}).
\item A ram wind parallel to the disk results in an $m=2$ Fourier component of azimuth with amplitude proportional to gas column density (see Fig.~\ref{modelram2})
\end{itemize}

\begin{figure}
\begin{center}
\includegraphics[scale=0.33]{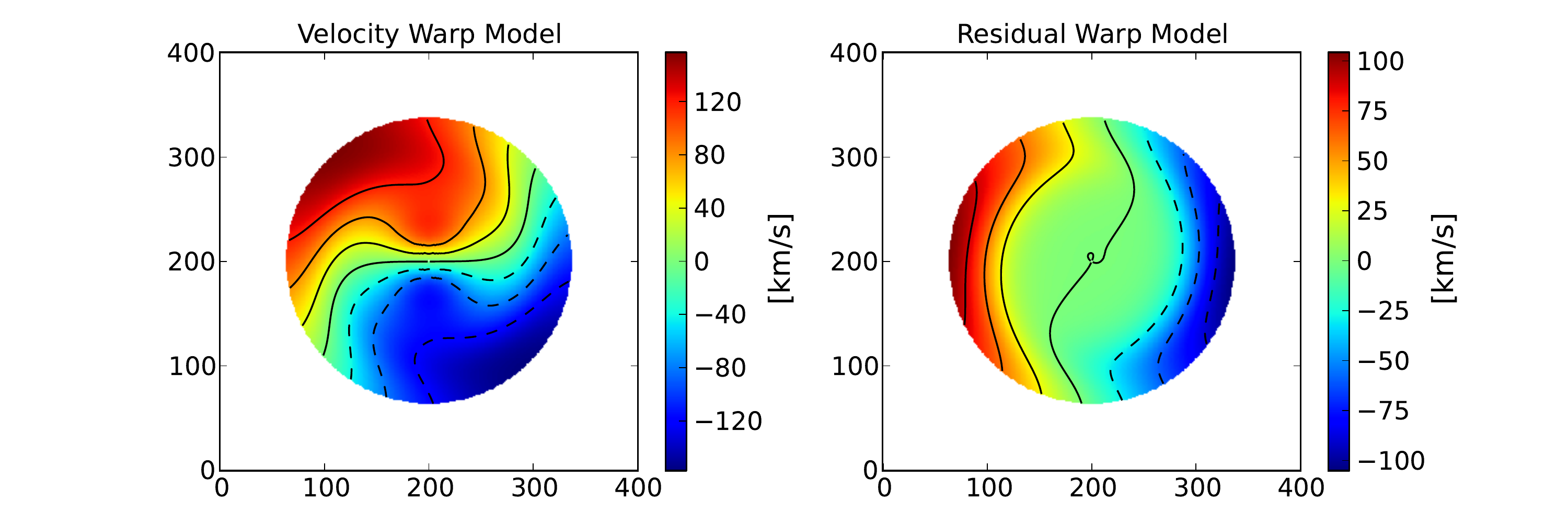}\\
\includegraphics[scale=0.33]{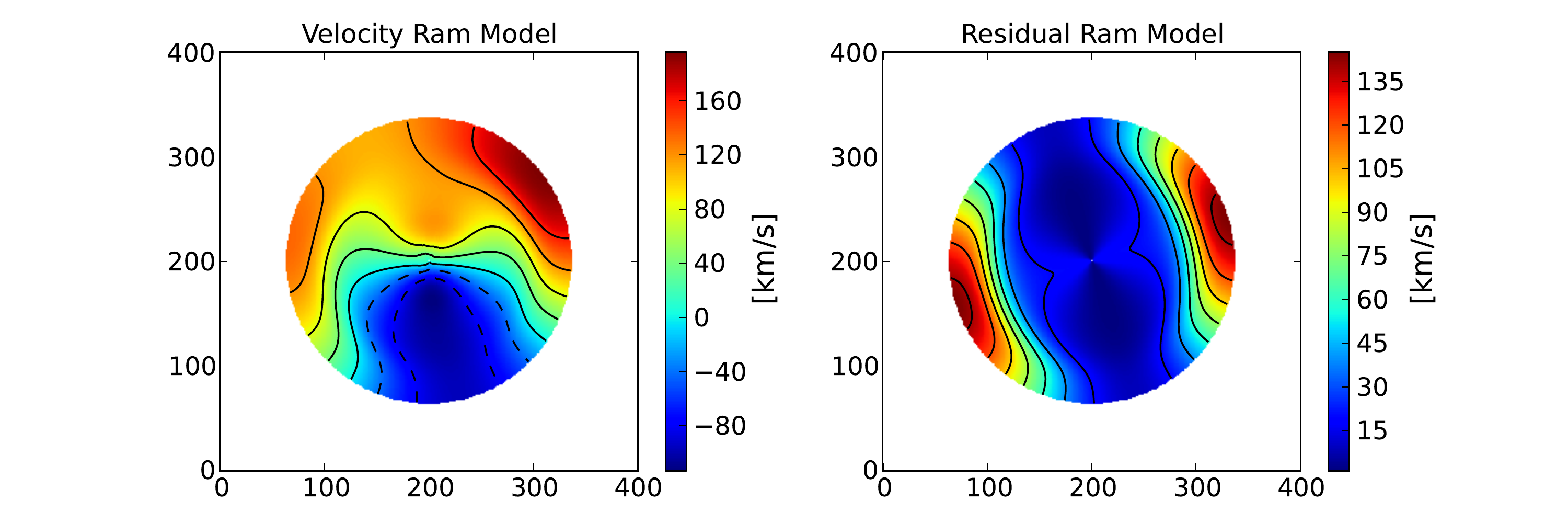}\\
\includegraphics[scale=0.33]{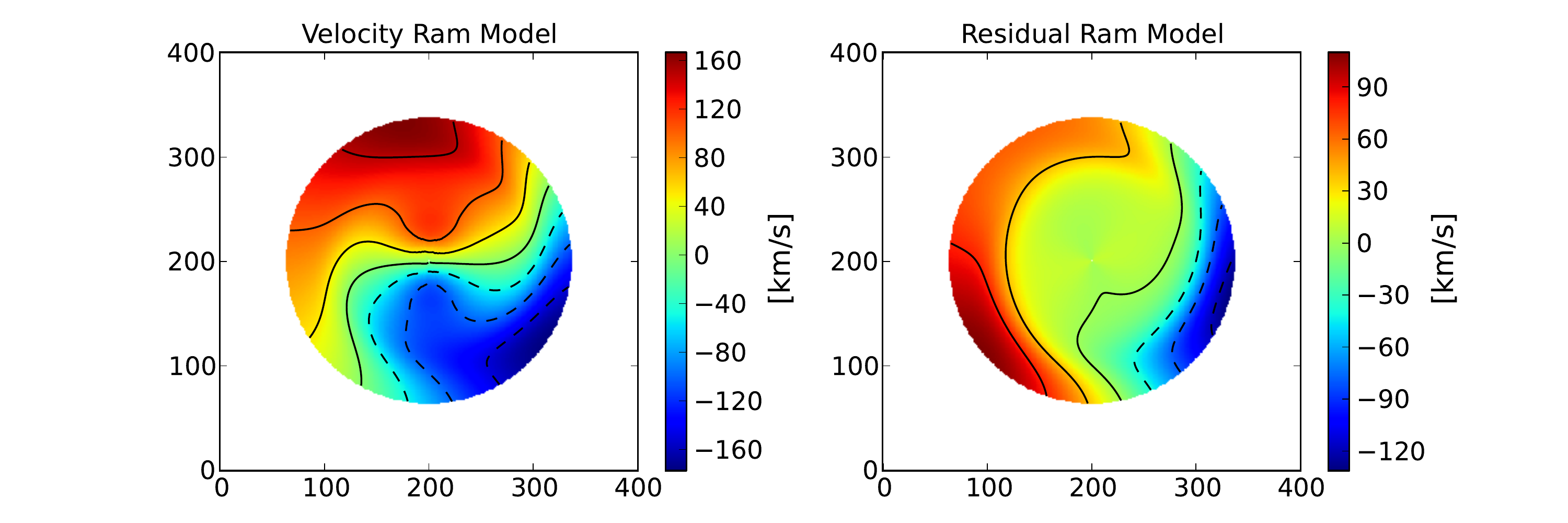}
\caption{Top: Model velocity (left) and residual field (right) for a simple warped disk model (no ram wind) with a change of  45$\deg$ in position angle and 30$\deg$ in inclination (top). Middle:  Model with ram wind ($\gamma_{Ram}=35\deg$ and $\theta_{ram}=55\deg$), no warp. Bottom: Combined warp+ram model. A detailed description of the model is given in App.\ref{app:a}} 
\label{modelwarp}
\end{center}
\end{figure}

One of the main questions is whether ram interaction can be distinguished from the kinematic signature of a warped disk due to systematic change in inclination and position angle of the disk as function of radius. To test this we have created a simple warped disk model with a transition in inclination and position angle between an inner and an outer disk, which is described in detail in App.~\ref{app:a}. The warp model velocity field shown in Fig.~\ref{modelwarp} shows a typical pattern as expected from observations of prominent warped disks \citep[see e.g.][]{Bra91, Kam92, Joz07}. The residual velocity field results in an $m=1$ component (bimodal) in galaxy azimuth, $\theta$, that twists with radial distance, but is clearly distinguished from the observed ram interaction pattern ($m=0$ and $m=2$ components). These model velocity fields and the computed residual velocity models are in agreement with the mathematical derivation of the residual velocity field as described in Eq.\ref{eq:ramwarp_res}, which provides a physical description of the Fourier components $c0$, $c1$, and $c2$. We note that a warped gas disk can be produced over longer timescales (several orbits) by ram pressure as well \citep[see][]{Haa13}, which can be independently measured in our kinematic analysis, and will be discussed in more detail in \S~\ref{sec:dis}.
\par 

We note that harmonic components in the residual field can also be due to errors in the geometrical disk parameters and rotation velocities. An overview of the different components due to geometrical uncertainties can be found in \cite{Kru78}. However, these harmonic components usually occur over the entire disk rather than increasing towards the outer disk where ram pressure and warped disks are expected to dominate.  For instance an error in $PA$ or $i$ results in $m=1$ and $m=3$ components, respectively, while an offset of the center (lopsided disk) generates an $m=2$ component, usually decreasing towards larger radius. A radial inflow/outflow motion generates an $m=1$ component perpendicular to the $PA$ while an error in rotation velocity results in a $m=1$ component parallel to the $PA$. However, these components are clearly distinguishable from ram wind components since the former occur either systematically over the entire disk or are $m=1$ or $m=3$ components (odd orders) geometrically aligned with the $PA$ of the galaxy. Spiral arms and bars produce streaming motions but are localized in the stellar disk and can be distinguished from large scale motion.
Kinematic features of the ram wind are a combination of the $m=0$ or $m=2$ components (even orders) in the residual velocity field and are expected to become significant only towards large radius in conjunction with lower gas column densities. \par

\clearpage 

\subsection{Velocity Decomposition and MCMC Fitting}
\label{subsec:mcmc}

To search for velocity patterns that may be indicative of a ram pressure interaction as outlined in the previous section, we considered the following approach.
First, a simple model of the galaxy's velocity field is constructed, based on an average rotation velocity as function of radius and constant inclination $i$ and major axis position angle $\phi$. The residual velocity field is then defined as the difference between the observed velocity field and this nominal kinematic model. Systematic kinematic residuals are very sensitive to secondary velocity components, i.e. ram-pressure, warped disks, or errors in the geometrical disk parameters (center, systemic velocity, $i$, $\phi$, streaming motions).
Although this first characterization is done by eye, it already provides a fairly good overview about the galaxy's dynamical state.\par 

In a second step we compute the harmonic terms of the residual velocity field up to the 3rd order. The decomposition of the residual velocity components is then given as 
\begin{equation}
v_{res}=\eta(\rho_{ISM})\;\sum_{n=0}^3[c_n\cos(n(\theta-\theta_{n}))]
\label{eq:harm}
\end{equation}
where $c_i$ are the harmonic coefficients and $\theta$ the azimuthal angle with respect to the line-of-nodes, $\theta_0$.  The scaling function $F_{Ram}$ describes the transition from clumpy, high density to diffuse, low density disk and is given in Eq.~\ref{eq:sfunct}. As shown in \S~\ref{subsec:model}, the $c_0$ and $c_2$ components can be associated with the two ram pressure wind velocity components, perpendicular and parallel to the plane of the gas disk, respectively. Other harmonic components, i.e. those with uneven order, might indicate a warped disk or circular rotation velocity offset (strong $c_1$ or $s_1$ component), or a possible  component due to an offset in the inclination ($c_3$ or $s_3$) etc. Our harmonic decomposition model includes 10 free parameters: $\rho_{Crit}$,$f_{Smooth}$, $\theta_{0}$, plus the seven harmonic coefficients.\par 

\begin{figure*}
\begin{center}
\includegraphics[scale=0.55]{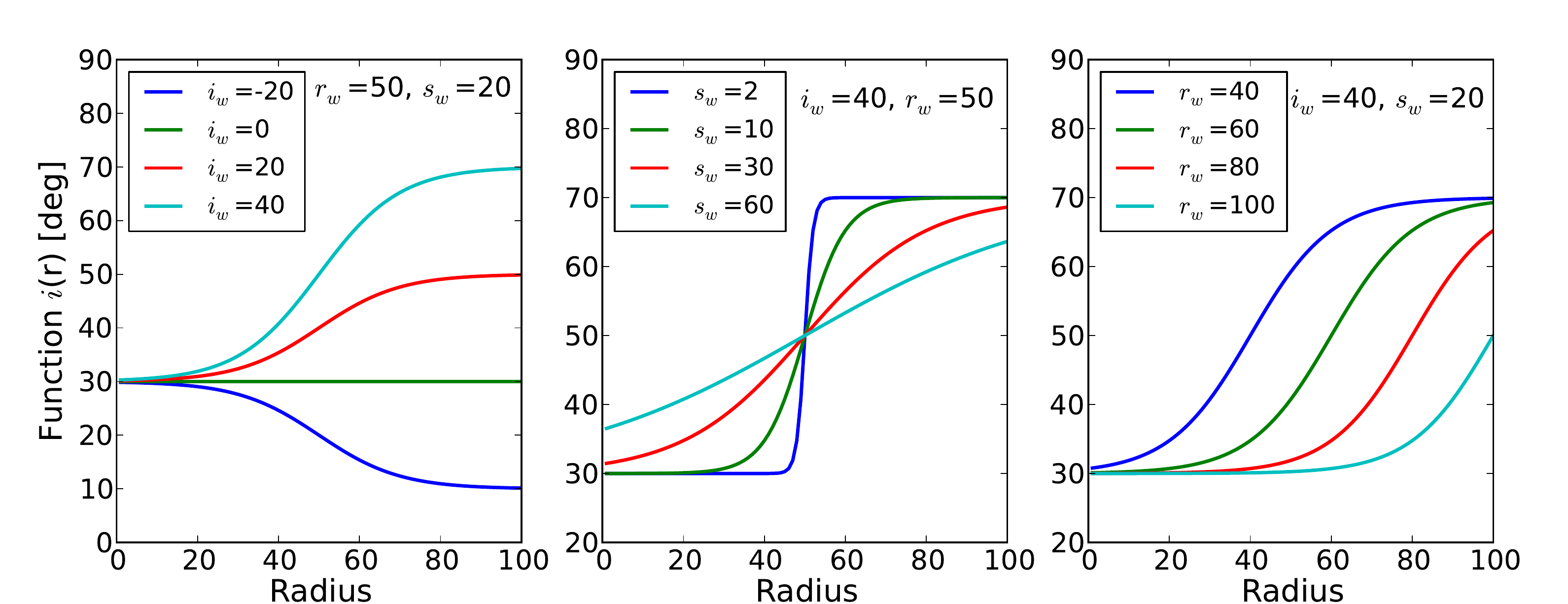}
\caption{Examples of the radial scaling function of the inclination $i(r)=30.+\Delta i(0.5+0.5\tanh[(r-r_W)/s_W])$ for different values of the amplitude $\Delta i$, the smoothness parameter $s_W$, and the transition radius $r_W$ of a warped disk.} 
\label{fig:sfunct_ex}
\end{center}
\end{figure*}

To quantify the relevant velocity components in more detail, i.e. in terms of the velocity offset due to ram pressure and/or warped disks we fit the residual velocity field with our derived model velocity field given in Eq.~\ref{eq:ramwarp_res}. This fit takes into account the ram interaction terms of a systematically warped disk as a function of radius. While the effective ram wind velocity can be described as function of the gas column density, a possible change in inclination and PA due to a warped disk are characterized as function of the radius $r$. Warped disks typically have an m=1 mode similar to an integral sign shape. The inclination $i(r)$ and position angle $\phi(r)$ of a warped disk as a function of radius can be described with a transition function as,
\begin{equation}
i(r)=i_0+\Delta i (0.5+0.5\tanh[(r-r_W)/s_W]),
\end{equation}
and
\begin{equation}
\phi(r)=\phi_0+\Delta\phi(0.5+0.5\tanh[(r-r_W)/s_W]),
\end{equation}
with the transition radius $r_W$ , the smoothness parameter $s_W$ (characterising the length of the transition), and $\phi_0$ and $i_0$ are the PA and inclination of the inner disk and $i_0+\Delta i$ and $\phi_0+\Delta\phi$ the PA and inclination of the outer warped disk, respectively.  These functions can span a large range of behaviours, from constant inclination and position angle to sharp transitions. Examples of the radial variation for different $r_W$ and $s_W$ are shown in Fig.~\ref{fig:sfunct_ex}. However, we note that these equations are just a rough approximation of a warped disk which can have more variations, e.g. different transition scales for position and inclination angle depending on the disk orientation on the sky. 

Altogether there are 9 free parameters ($\rho_{Crit}$,$f_{Smooth}$, $\theta_{Ram}$, $vc_{0}$, $vc_{2}$, $\Delta\phi$, $\Delta i$,  $r_W$,  $s_W$) and 2 independent input fields (column density $\Sigma_{ISM}$ and circular velocity $v_{Rot}(r,i_0,\phi_0)$) in this model. Given the ram wind velocity parameters $\Delta v\bot=\mid \Delta v\mid\sin(\gamma_{Ram})=c_0 / \cos(i)$ and $\Delta v\|=\mid \Delta v\mid\cos(\gamma_{Ram})=c_2 / \sin(i)$, the angle $\gamma_{Ram}$ can be derived from the ratio $\Delta v\bot/\Delta v\|=\tan(\gamma_{Ram})$. \par
 
The residual velocity field is fitted against the model by using a least square minimization of Markov Chain Monte Carlo (MCMC) samplers. We have applied this computation for both decompositions described above, the harmonic expansion up to the third order and the detailed harmonic fit with warped disk term. The MCMC sampling is performed using the open source Python implementation \textit{emcee} which is a very efficient and well tested algorithm for multi-dimensional fitting and global optimization problems \citep[for a detailed description see][]{For12}. We have tested this fit method with our ram pressure and warped disk models (see previous section) to find the best parametrization of the model and the critical number of independent walkers (typically a few hundred) and iterations of the MCMC samplings to obtain a good match between model and fit results.\par

In principle, it might be possible to include all additional parameters necessary to derive the rotation velocity in the fit: an ``all at once'' fit. However, this would require six additional fit parameters ($v_{sys}$, $x_{center}$, $y_{center}$,$i_0$, $\phi_0$, $v_{rot}$) to be fit simultaneously and two of these, inclination and rotation velocity, are degenerate. While this might be possible in the future, we follow a more conservative approach here, namely estimating the first order alignment of the disk and circular velocity carefully beforehand, and then studying possible second order effects, which include various uncertainties in the geometrical disk parameters as well as warped disk and ram pressure interaction terms. This allows us, at every stage of our study, to check carefully for possible second order velocity patterns and subsequently isolate independent kinematic features from each other. For instance, while a change in the inclination is usually indistinguishable from a change in rotation velocity, it is less likely that a change in inclination due to a warped disk is not accompanied by a change in position angle, unless the warped disk is exactly aligned with the major axis of the disk, which is statistically very unlikely. However, we want to emphasize that it is not a stringent requirement, nor the focus of our study, to break this degeneracy between inclination and circular rotation velocity, since both produce an m=1 mode in the residual velocity field which can be clearly distinguished from ram pressure patterns (m=0 and m=2 mode).

\section{Comparison with HI kinematics of Nearby Disk Galaxies}
\label{sec:comp}

\subsection{Sample Selection, data preparation and general fitting procedures}
To test whether we can find evidence for ram pressure interaction in regular galaxies, we have applied our kinematic analysis on several nearby disk galaxies for which moderately deep HI observations, reaching column densities of a few $10^{19}$~cm$^{-2}$, were available in the literature. In the following we will show three different cases: (a) NGC~6946 as an example for dominant kinematic terms due to ram pressure, (b) NGC~3621 as an example of a combination of kinematic features due to ram pressure and a warped disk, and (c) NGC~628 as an example of kinematic features due to a strong warped disk.\par

All three galaxies have been observed as part of the HI nearby galaxy survey \citep[THINGS, ][]{Wal08}, which combines HI observations using a combination of VLA B-,C-and D array configuration with a typical angular resolution of 7\arcsec and a spectral resolution of 5~km~s$^{-1}$. In a first step we have applied a constant inclination and position angle averaged over the entire HI disk scale and created geometrically deprojected circular rotation models and residual velocity maps for these galaxies. An overview of the properties of these galaxies is given in Tab~\ref{tab:obs_ov}.  Given the stellar distribution based on near-IR images (IRAC 3.6~$\mu$m) we have excluded in all subsequent analysis the central part of the galaxies (typically up to 7~kpc) which can be significantly disturbed by gravitational instabilities in the stellar distribution (e.g. due to stellar bars and spiral arms).  This is necessary since bars and spiral arms can produce strong gravitational torques and gas streaming motions \citep[see e.g.][]{Haa09}, complicating the decomposition of the HI kinematics, which is not the focus of this study. \par

While the high-spatial resolution HI images allow us to study the kinematics of the inner clumpy HI disk, they are not necessarily optimal for an accurate measurement of ram pressure interaction in the diffuse outskirts of the HI disk. To reach low HI column densities with large signal-to-noise ratios we have created new moment maps with a 60\arcsec resolution and obtain velocities at the peak HI intensity. 
We run our fit models on both, velocity and residual velocity fields, which reveal the same results as expected, since both fitting procedures require the circular rotation velocity as input. We have tested both fit models given Eq.~\ref{eq:ramwarp_res} and Eq.~\ref{eq:ramwarpvelocity}. While the errors given by the width of the normal distribution of the Monte Carlo samplers around the peak value are very small ($<$1\% of the mean value), the uncertainty of the observed velocity field can be much larger. To estimate the propagation of this error onto our fit results, we created a noise velocity field given by uniform random velocity distribution of 5~km~s$^{-1}$ (channel width) around the observed velocity field and then run the same fit on these alternate maps. The uncertainty of our fit results is then given by the difference in results of both fits, namely the observed velocity map and the alternate noise map.\par

The best fit models have been evaluated based on the normalized $\chi^2$ values of the least-square fit and the visual comparison between model and observed velocity field. However, in all cases our results converged to stable values with increasing number of independent MC walkers and iterations.  

\subsection{NGC~6946:  An example of kinematic ram pressure measurement in a regular spiral galaxy}
 
\begin{figure*}
\begin{center}
\includegraphics[scale=0.40]{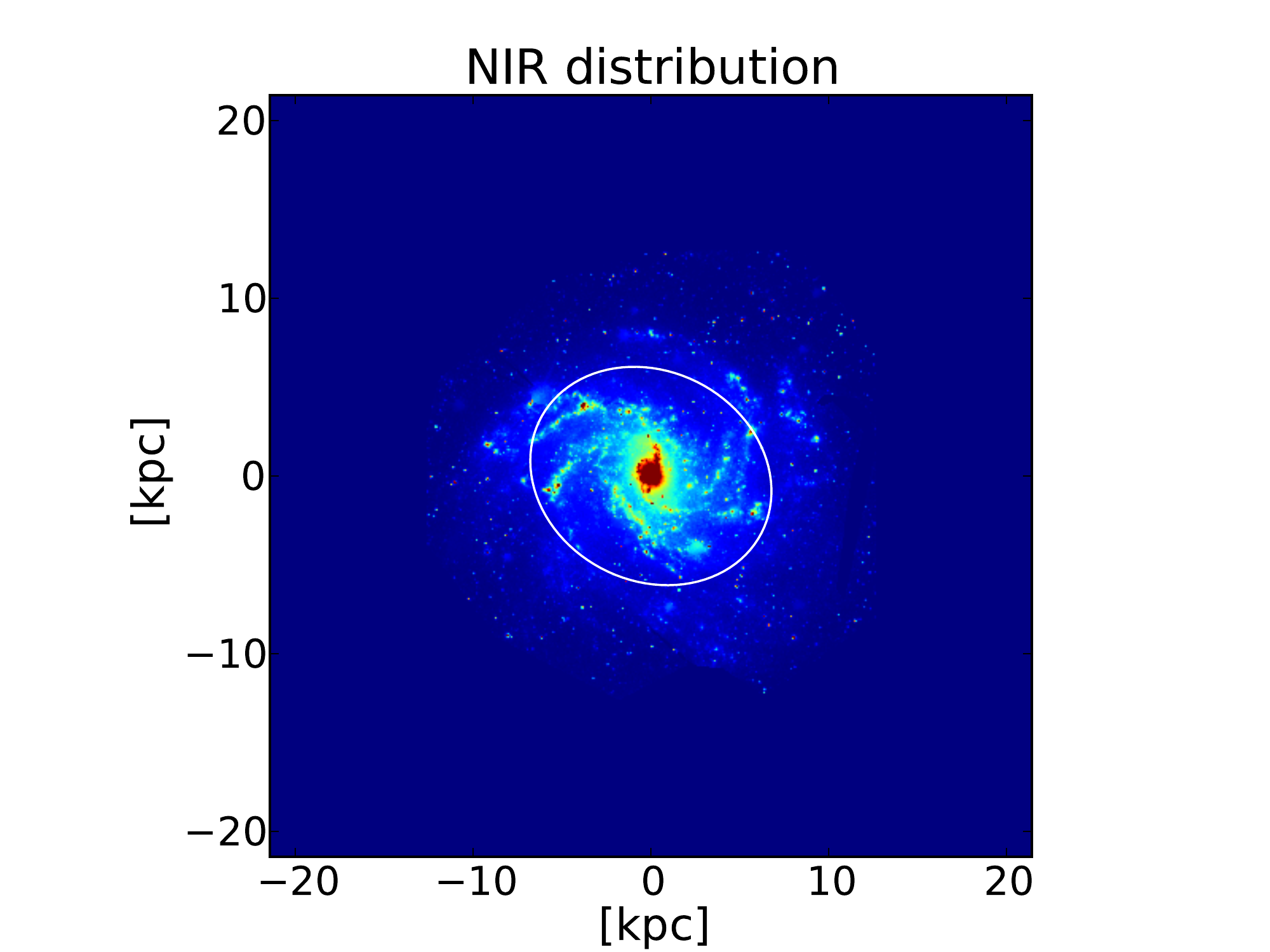}
\includegraphics[scale=0.40]{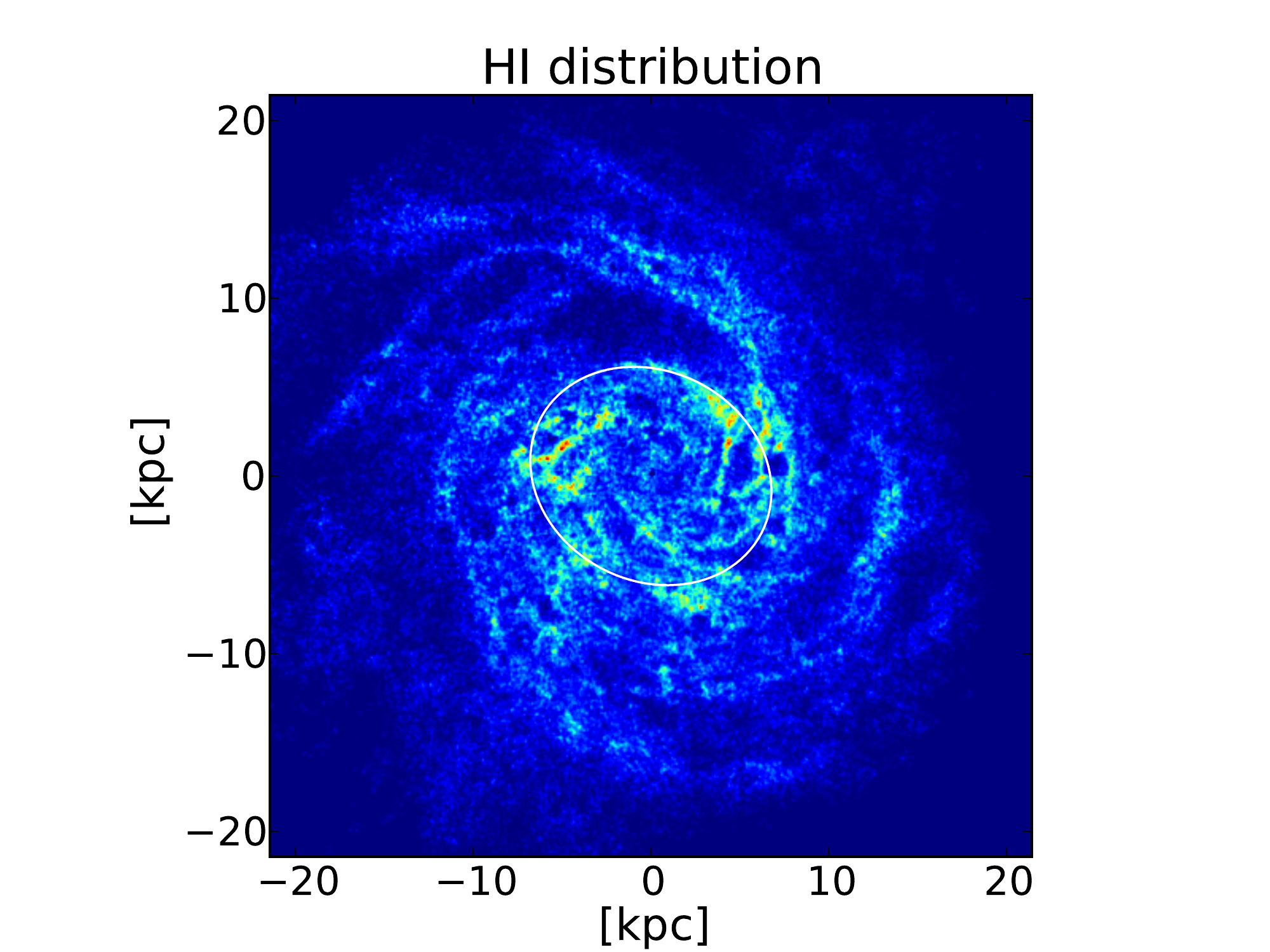}
\includegraphics[scale=0.40]{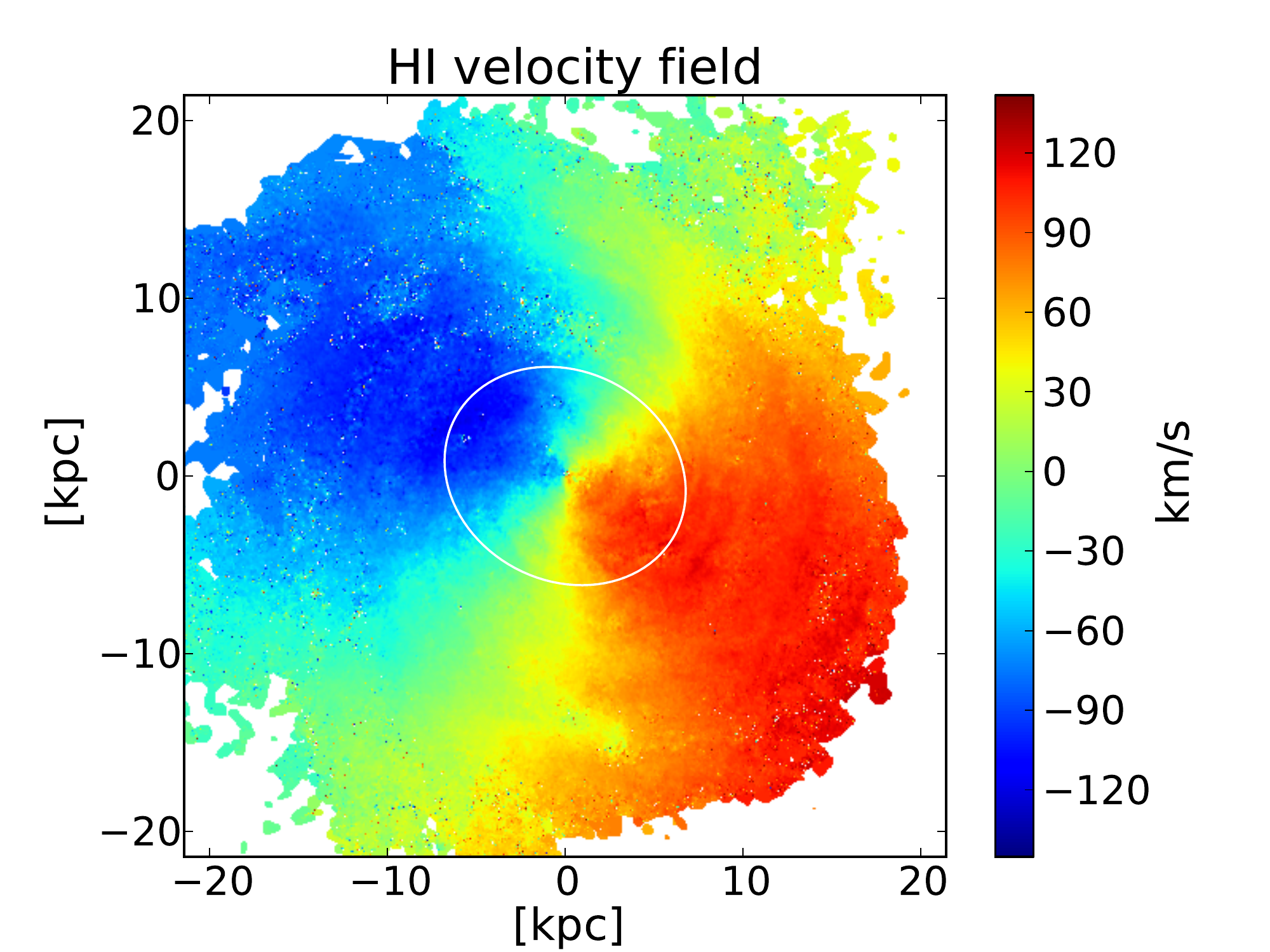}
\includegraphics[scale=0.40]{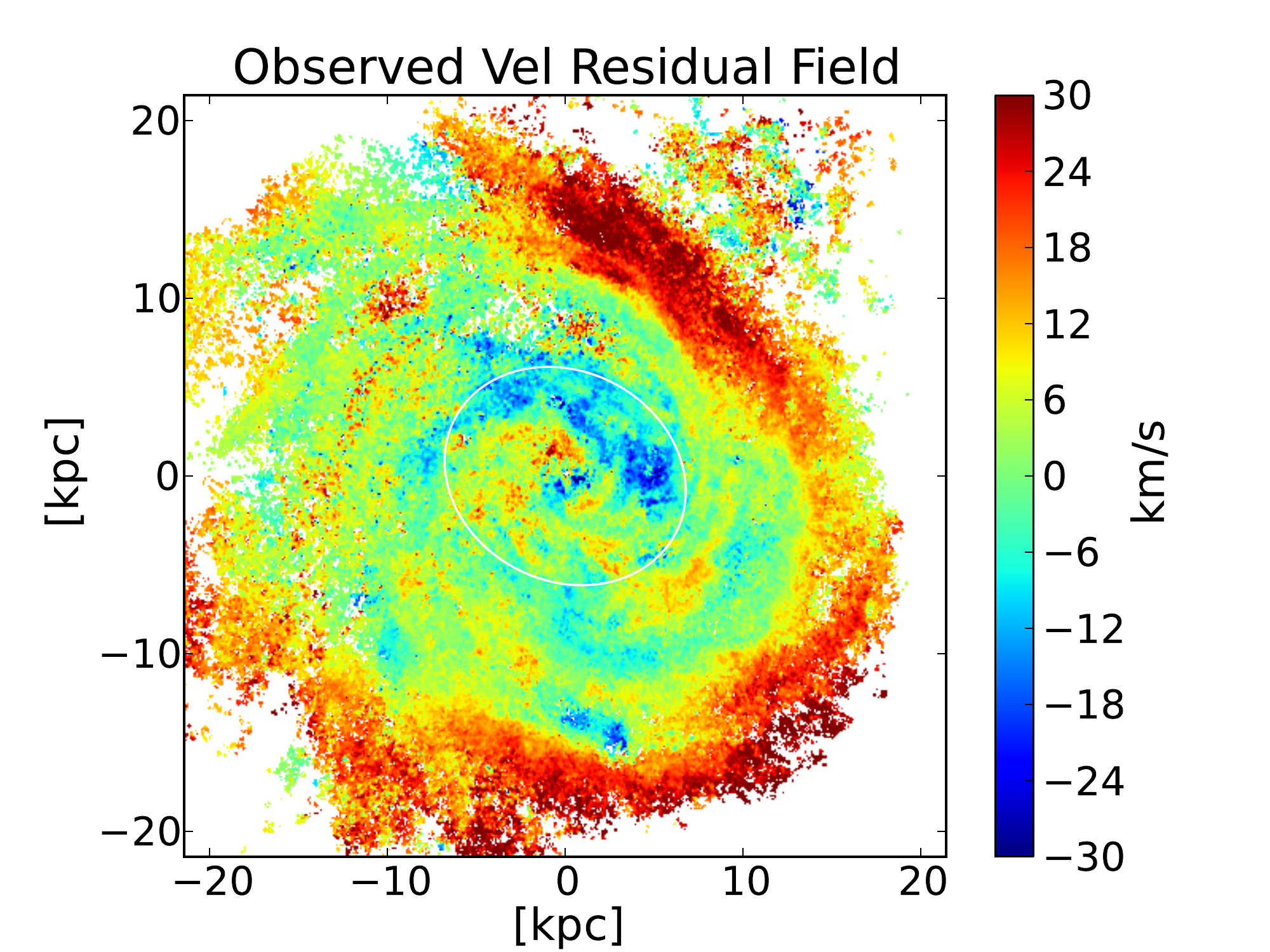}
\caption{NGC~6946: high spatial resolution maps. Top left: Stellar distribution as imaged in the near-IR by Spitzer IRAC at 3.6$\mu$m. Top right: atomic gas distribution (HI mom0 map). Bottom left: HI velocity field (mom1). Bottom right: HI velocity residual  field created from a circular rotation model with constant P.A. and inclination of 243$\deg$ and 33$\deg$, respectively. The HI data is based on the THINGS survey \citep{Wal08,deB08}. The white circle indicates the region within 7~kpc radius that has been excluded from the fit.} 
\label{fig:NGC6946dpj}
\end{center}
\end{figure*}

\begin{figure*}
\begin{center}
\includegraphics[scale=0.35]{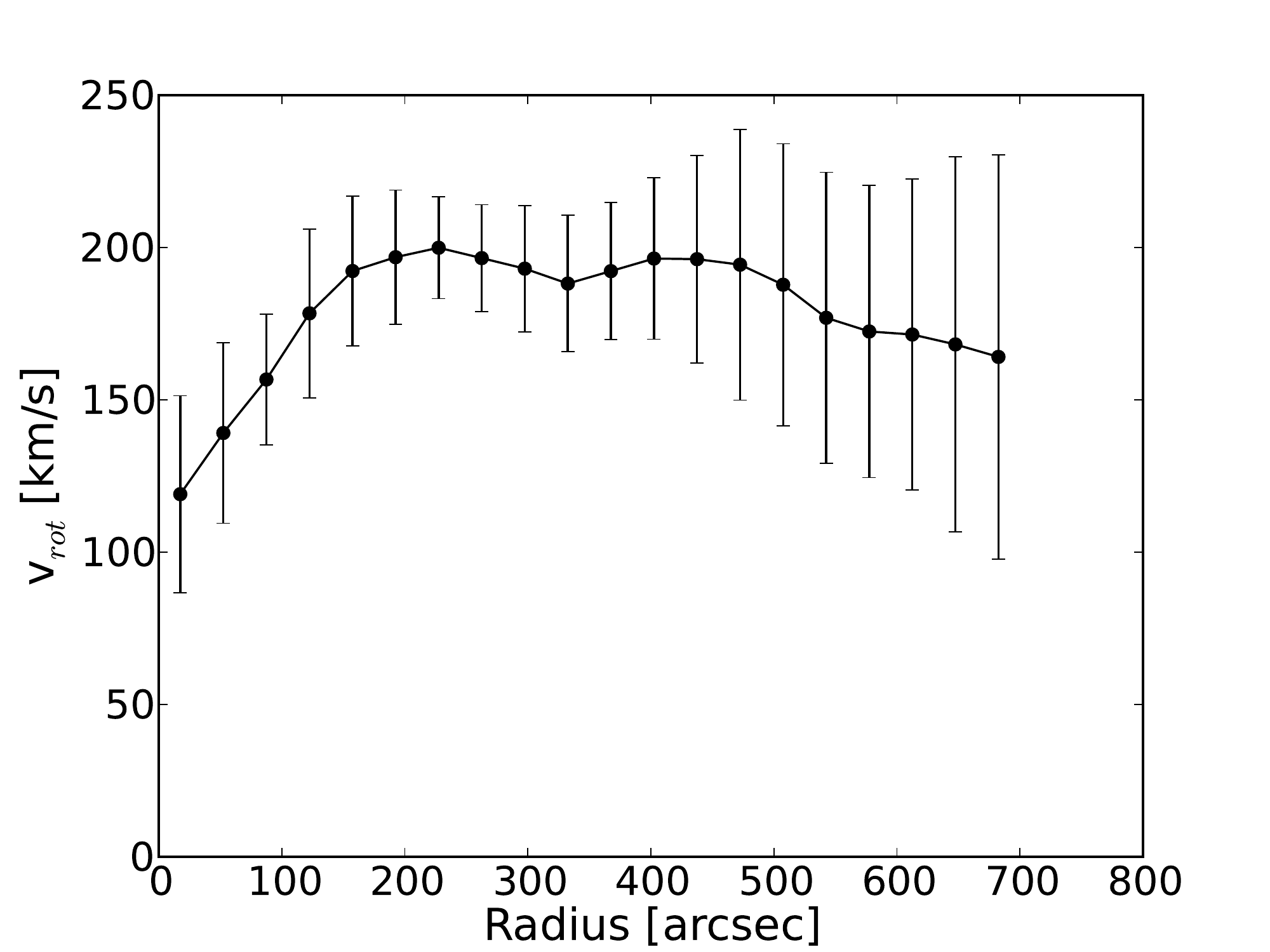}
\caption{NGC~6946. Rotation curve derived from velocity field with constant inclination and position angle. The errorbars represent the standard deviation of the velocity within each radial bin.}
% Right: The circular nominal velocity model in the observed projected orientation.} 
\label{fig:NGC6946rotc}
\end{center}
\end{figure*}

NGC~6946 is a nearby spiral galaxy at a distance of 5.9~Mpc, exhibiting a very regular stellar (as seen in the near-IR) and atomic gas distribution with no indications of interaction features or gas stripping.
%The rotation curve, $PA$ and inclination as function of radius is shown in de Blok et al. 2008.
An overview of the deprojected stellar distribution, atomic gas distribution and velocity field is shown in Fig.~\ref{fig:NGC6946dpj} which are derived from the high-spatial resolution HI moment-0 and moment-1 maps published in the THINGS survey.
%For convenience we present all maps and velocity fields geometrically deprojected. Note that this does not include a deprojection in the velocity space. 
We have derived a nominal circular rotation model based on a constant $PA$ and inclination of 243$\deg$ and 33$\deg$, respectively \citep[see also][]{deB08}. The rotation curve assumes a constant inclination and position angle and is shown in Fig.~\ref{fig:NGC6946rotc}. The circular rotation reaches a maximum of $\sim$200~km~s$^{-1}$ and remains approximately flat with a small decline of roughly 25~km~s$^{-1}$ towards the outskirts for a constant inclination and position angle. The residual velocity field is shown in Fig.~\ref{fig:NGC6946dpj}, which reveals an increasing residual velocity amplitude towards larger radii, suggesting the presence of a strong $m=0$ component. To obtain a first test of the dominant kinematic residual modes we have decomposed the residual velocity field into harmonic components. 
%as shown in Fig.~\ref{fig_NGC6946_harm}. 
We have excluded the HI kinematics within a radius of 7~kpc, which is dominated by the stellar distribution and the stellar gravitational potential as shown in Fig.~\ref{fig:NGC6946dpj}. 
As expected we find that the dominant component is $m=0$.  We also find evidence for an  $m=1$ mode and a $m=2$ component, all other harmonic components have a significance of less than $2\sigma$. \par

\begin{figure*}
\begin{center}
\includegraphics[scale=0.40]{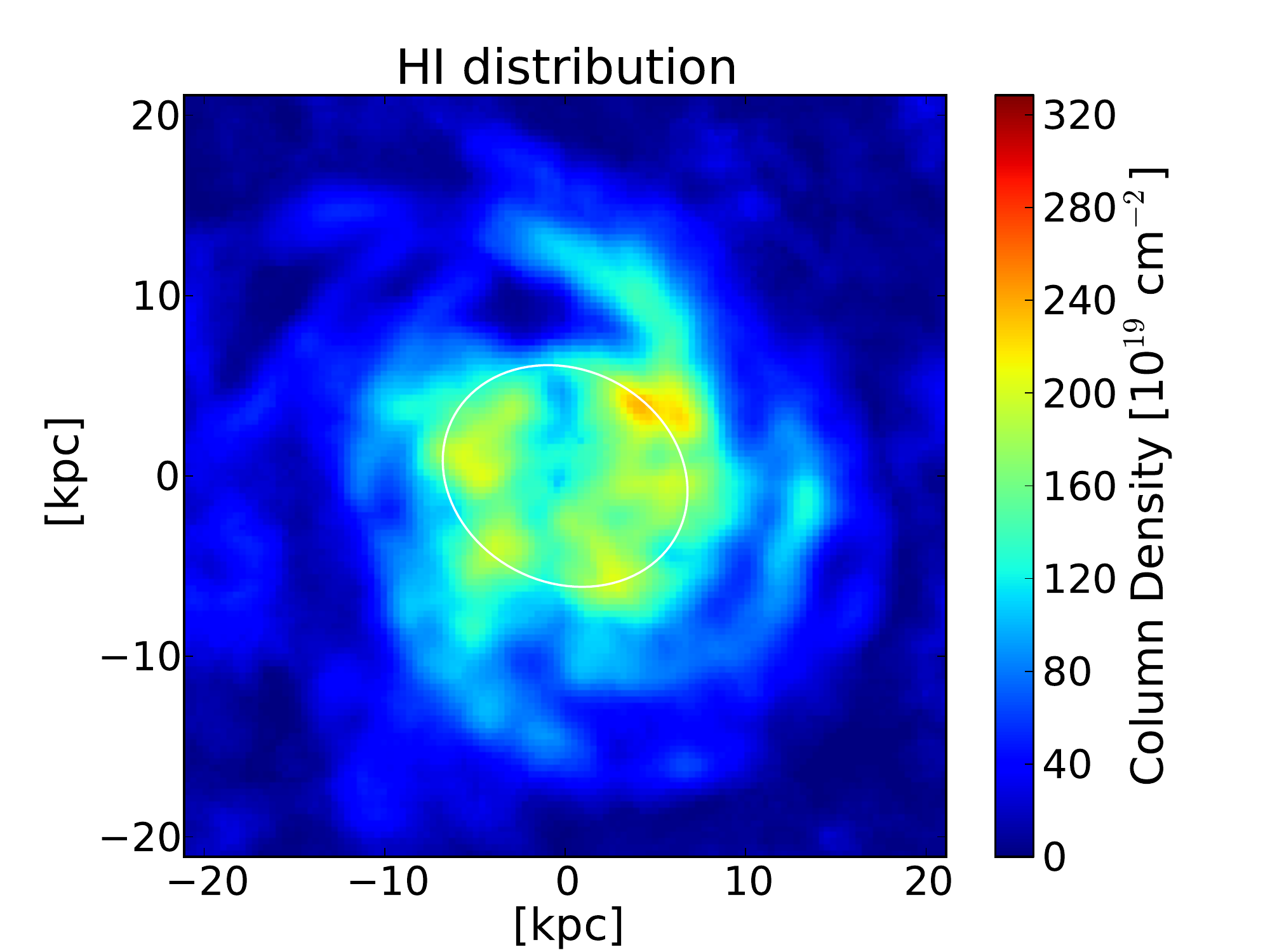}
\includegraphics[scale=0.40]{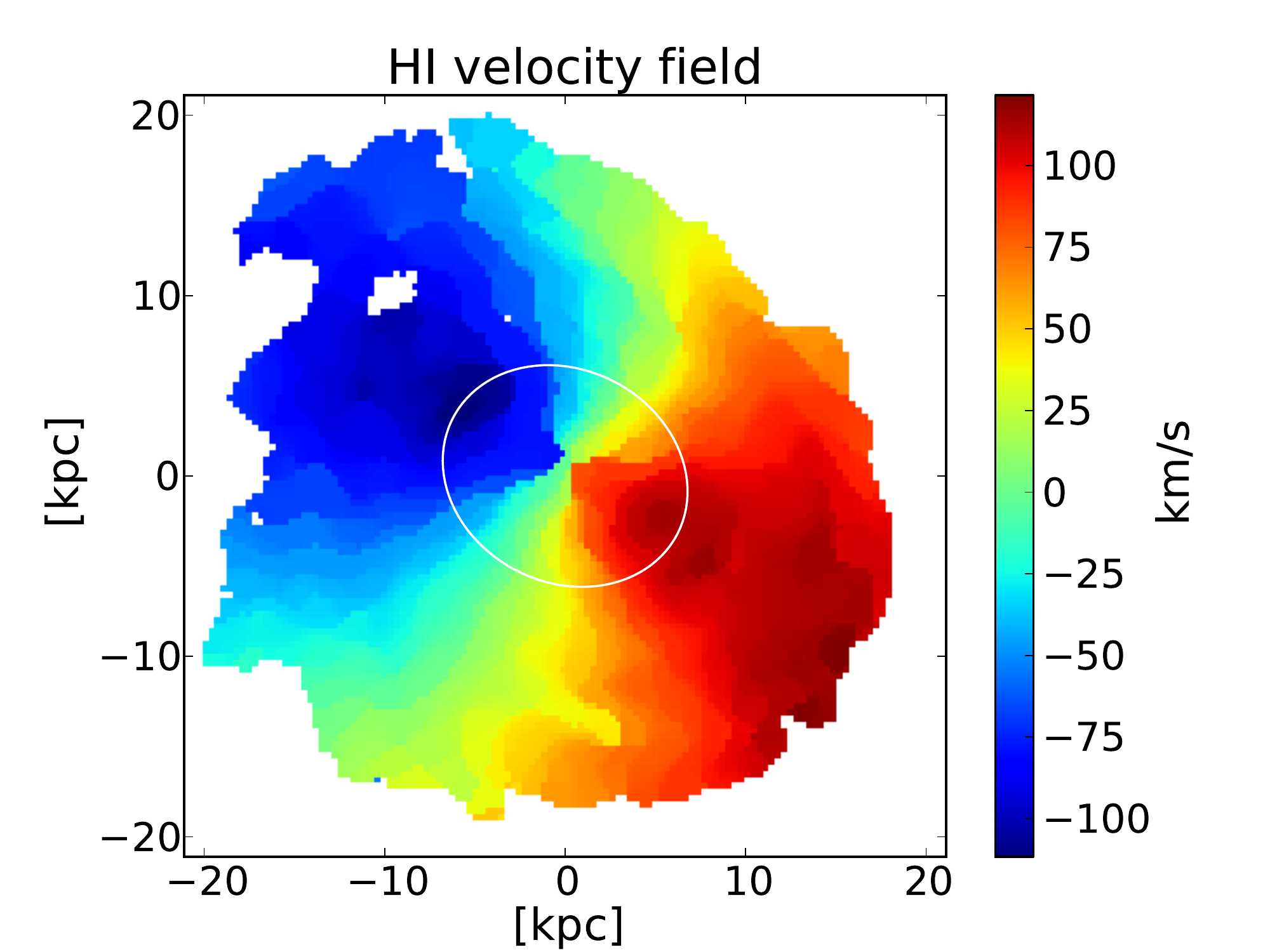}
\caption{NGC~6946: high sensitivity maps (beam 60 arcsec, $\sim$1.7kpc) in the observed orientation. Left: The atomic gas distribution (in column densities, corrected for inclination). Right: HI velocity field. The white ellipse indicates the region within 7~kpc radius that is excluded from the fit.} 
\label{fig:NGC6946nat}
\end{center}
\end{figure*}

\begin{figure*}
\begin{center}
\includegraphics[scale=0.4]{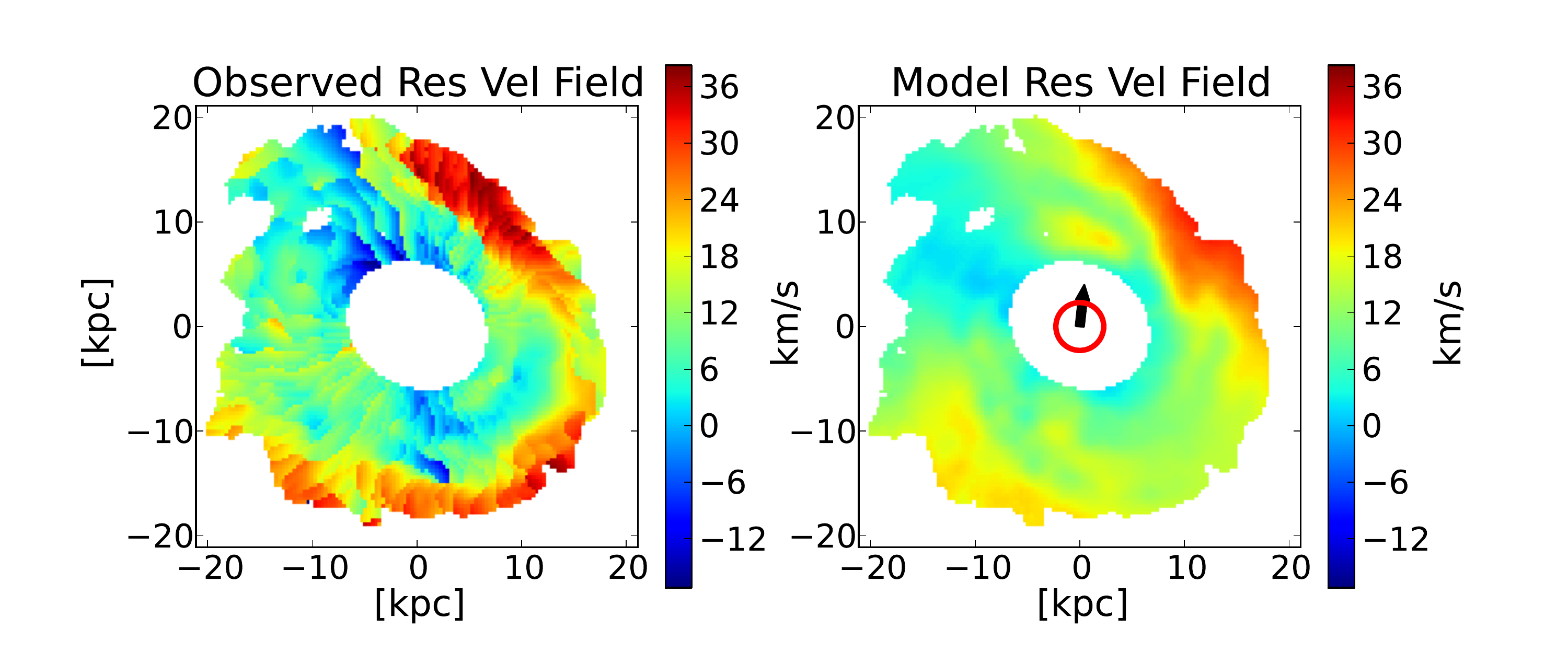}
\includegraphics[scale=0.3]{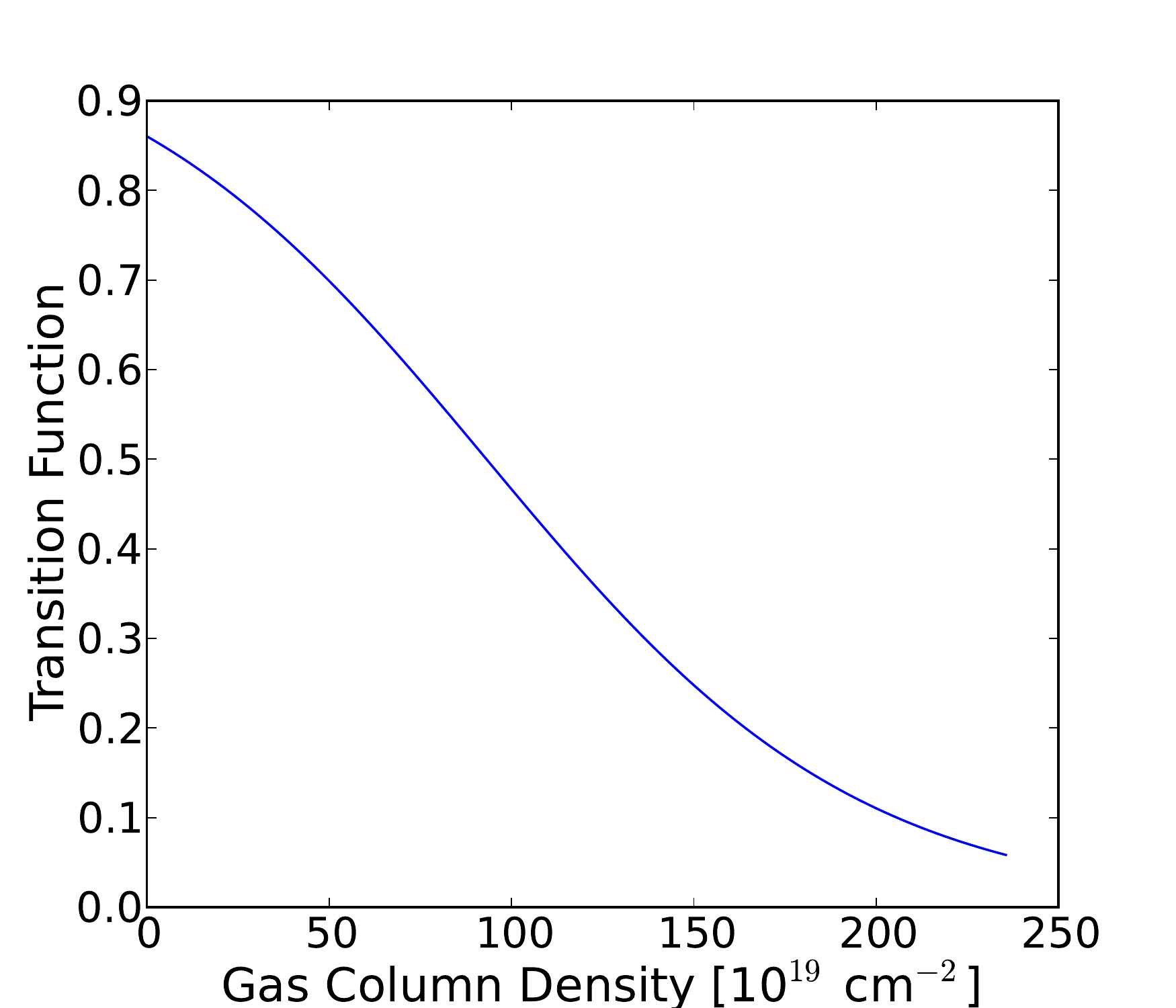}
\caption{NGC~6946: Comparison between observed (left) and model residual velocity field (middle). The residual fields are defined as difference between circular rotation and observed velocity field or model velocity field, respectively. The black arrow in the center of the model shows the direction of the ram wind in the sky plane (XY), while the radius of the red ring indicates the relative velocity contribution from the line-of-sight direction (Z). The ram wind  is in receding direction (red ring, into the page, approaching direction would be indicated as blue ring). Note that the substructure in the model is due to the fact that the ram pressure is also a function of column density (see text). Right panel: The scaling function $\eta(\rho_{ISM})$ (see Eq.~\ref{eq:sfunct}) characterizing the effectiveness of the ram pressure on the HI kinematics as function of HI column density due to the transition from clumpy to diffuse gas towards the outer disk.} 
\label{fig:NGC6946_fig5}
\end{center}
\end{figure*}

In a second step we apply our fitting procedure to decompose the residual velocity field into ram pressure terms as well as contributions of a systematically warped disk (see Eq.~\ref{eq:ramwarp_res}). For this fitting we have first convolved the data cube to a larger beamsize (60 arcsec FWHM). The resulting moment-0 and moment-1 maps are shown in Fig~\ref{fig:NGC6946nat}. Moreover, we have carried out all subsequent analysis in the resulting native image frame, i.e. excluding any further smoothing, rotation or deprojection, to eliminate errors due to pixel interpolation. As noted above, we have excluded the kinematic data within a radius of 7~kpc. We subtracted the nominal circular velocity model with constant $PA$ and inclination from the observed velocity field. The residual velocity field shows a strong $m=0$ component increasing toward the outer disk. The residual velocity field has been fit by our ram and warp model as expressed in Eq.~\ref{eq:ramwarp_res}. The derived model is presented in Fig~\ref{fig:NGC6946_fig5}, which shows a very good agreement with the observed residual velocity field. An overview of the model parameters is given in Tab.~\ref{tab:obs_par} and Tab.~\ref{tab:obs_res}. \par

\begin{figure*}
\begin{center}
\includegraphics[scale=0.45]{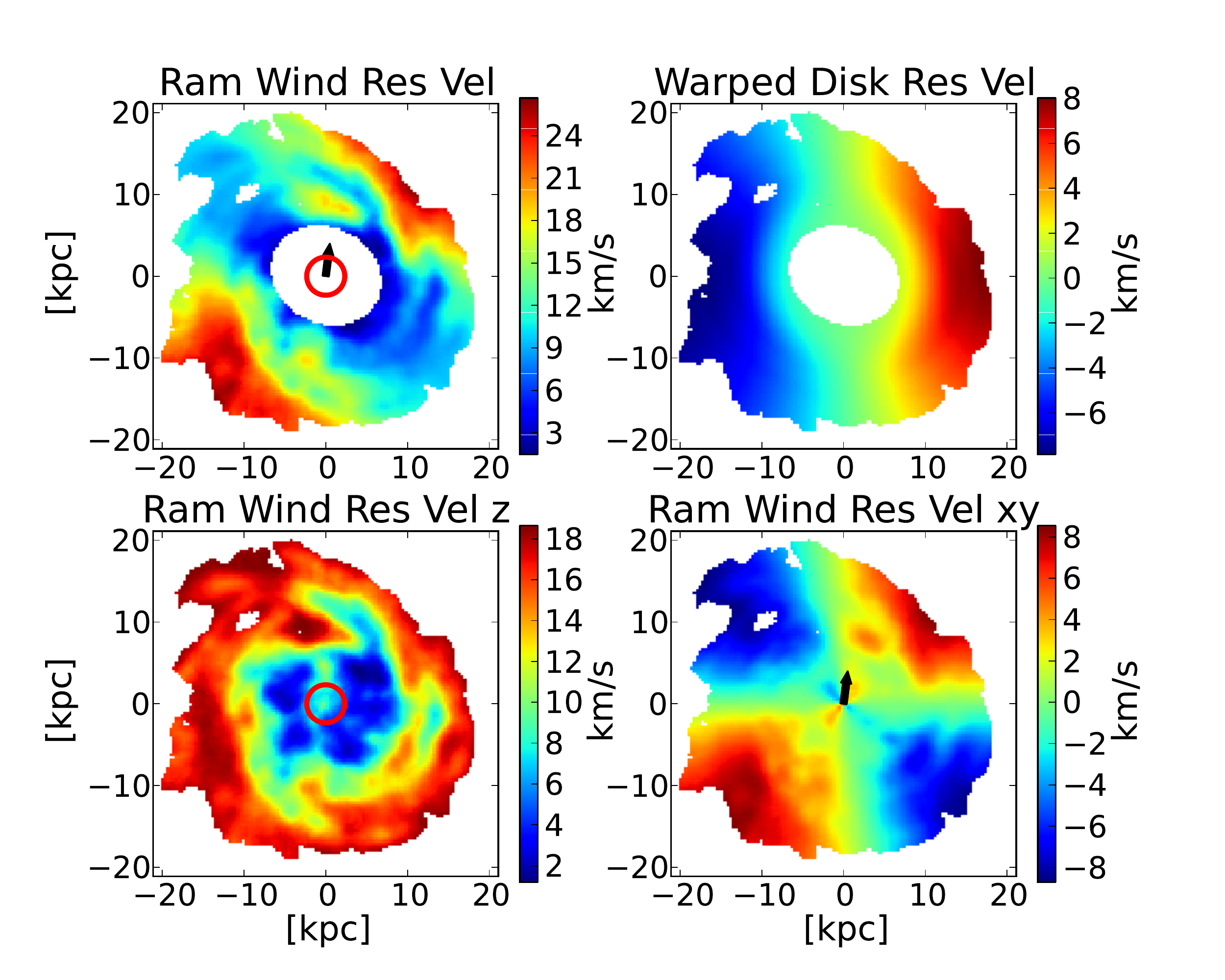}
\caption{NGC~6946: Top: Decomposition of model velocity field into velocities due to ram pressure  (left) and velocity contributions due to systematic deviations such as change in position angle and inclination of disk (right). Bottom: Decomposition of ram pressure velocity field into velocities due to ram wind perpendicular ($z$-axis, bottom left)  and parallel ($xy$-plane, bottom right) to the disk. The arrow and ring are defined as in Fig.~\ref{fig:NGC6946_fig5}} 
\label{fig:NGC6946_fig6}
\end{center}
\end{figure*}

As shown in Fig.~\ref{fig:NGC6946_fig6} our results demonstrate that the dominant residual velocity in NGC~6946 is consistent with a ram pressure wind inducing an effective velocity change of $\mid \Delta v\mid=32\pm3$~km~s$^{-1}$. We find only a small contribution from an $m=1$ mode of $\sim5$~km~s$^{-1}$ which is likely due to uncertainty in the inclination (or rotation velocity) since there is no sign for a significant change in position angle ($<$3$\deg$). The direction of the ram wind is given by the angles $\theta_{Ram}=118\deg$ and $\gamma_{Ram}=56\deg$ with an effective ram wind component perpendicular and parallel to the disk of $\Delta v\bot\approx26$~km~s$^{-1}$ and $\Delta v\|\approx18$~km~s$^{-1}$ which results in a line-of sight velocity change of $vc_0\approx22$~km~s$^{-1}$ and $vc_2\approx10$~km~s$^{-1}$, respectively. The best-fit scaling function of the effective ram pressure as function of gas density is quite smooth rather than abrupt as shown in Fig~\ref{fig:NGC6946_fig5} with a transition near $\rho_{crit}=\sim95\times10^{19}$~cm$^{-2}$.

%\begin{figure}
%\begin{center}
%\includegraphics[scale=0.45]{NGC6946_harm_fig6.pdf}
%\caption{NGC~6946. Decomposition of residual velocity field into its main harmonic components. Top left: sum of all components, top right: vc$_0$, bottom left: vc$_1$, and bottom right: vc$_2$  .} 
%\label{fig_NGC6946_harm}
%\end{center}
%\end{figure}

\clearpage

\twocolumn

\subsection{NGC~3621: An example of ram pressure interaction in a warped gas disk.}
\label{subsec:ngc3621} 
\begin{figure*}
\begin{center}
\includegraphics[scale=0.40]{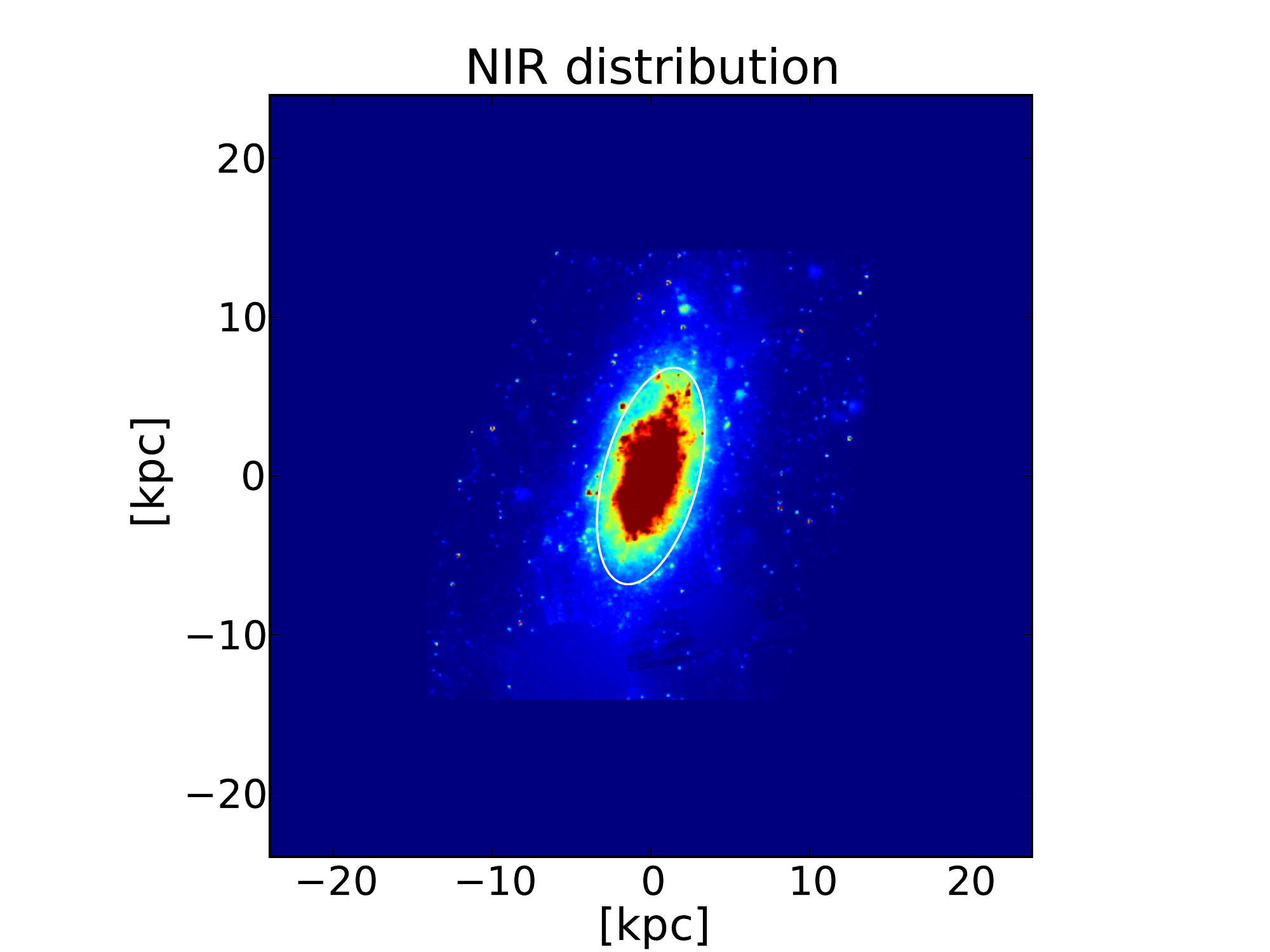}
\includegraphics[scale=0.40]{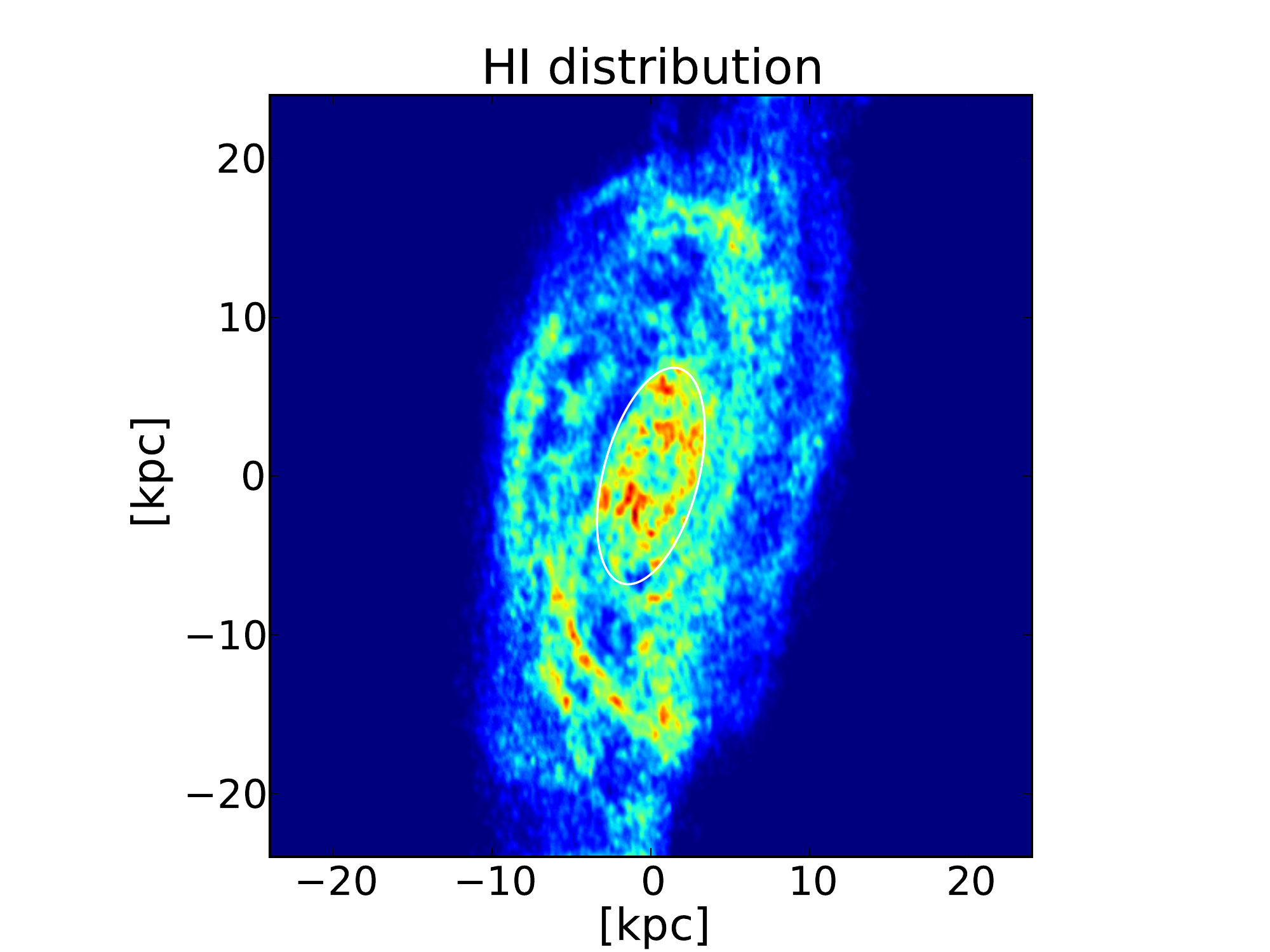}
\includegraphics[scale=0.40]{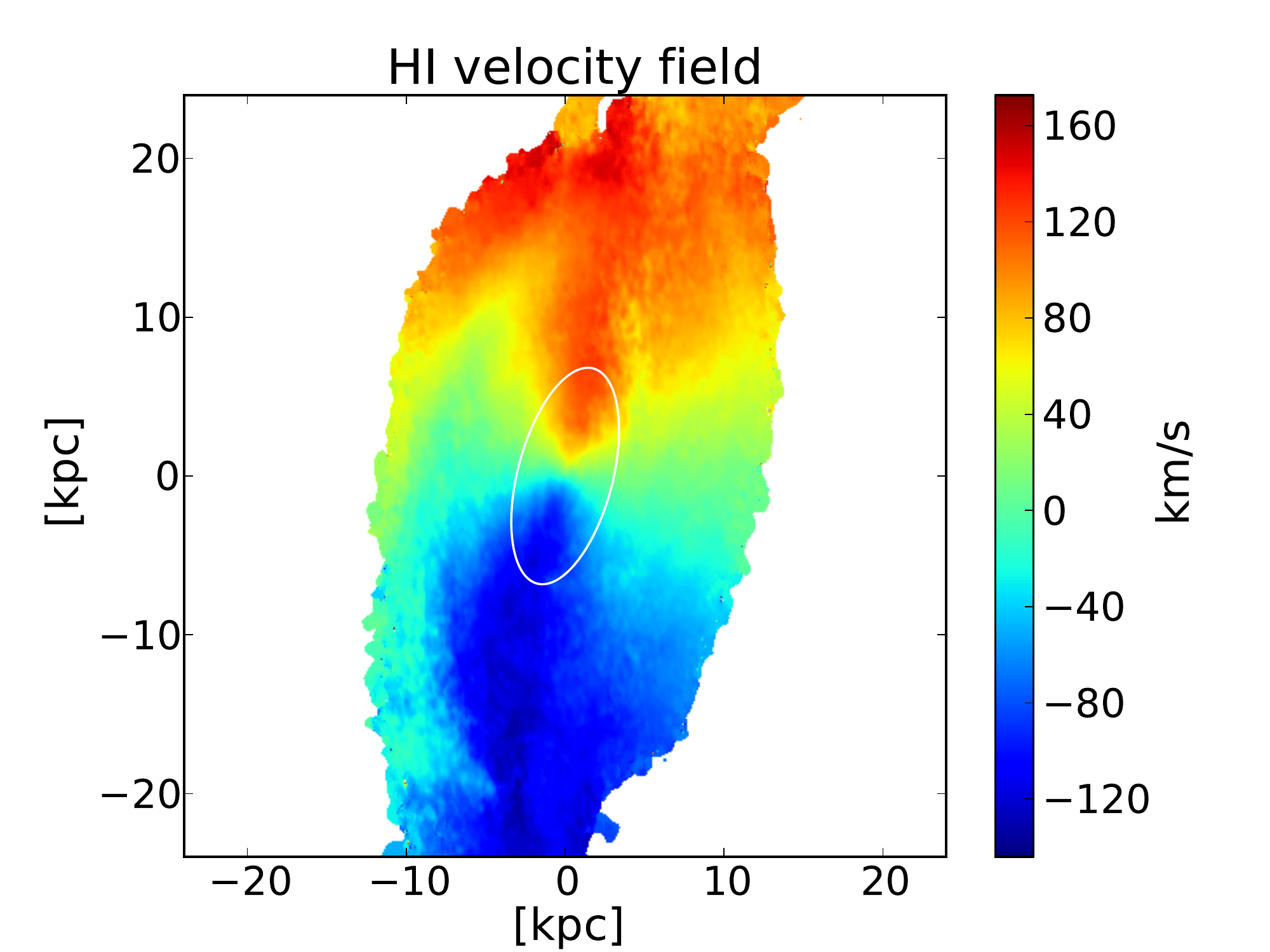}
\includegraphics[scale=0.40]{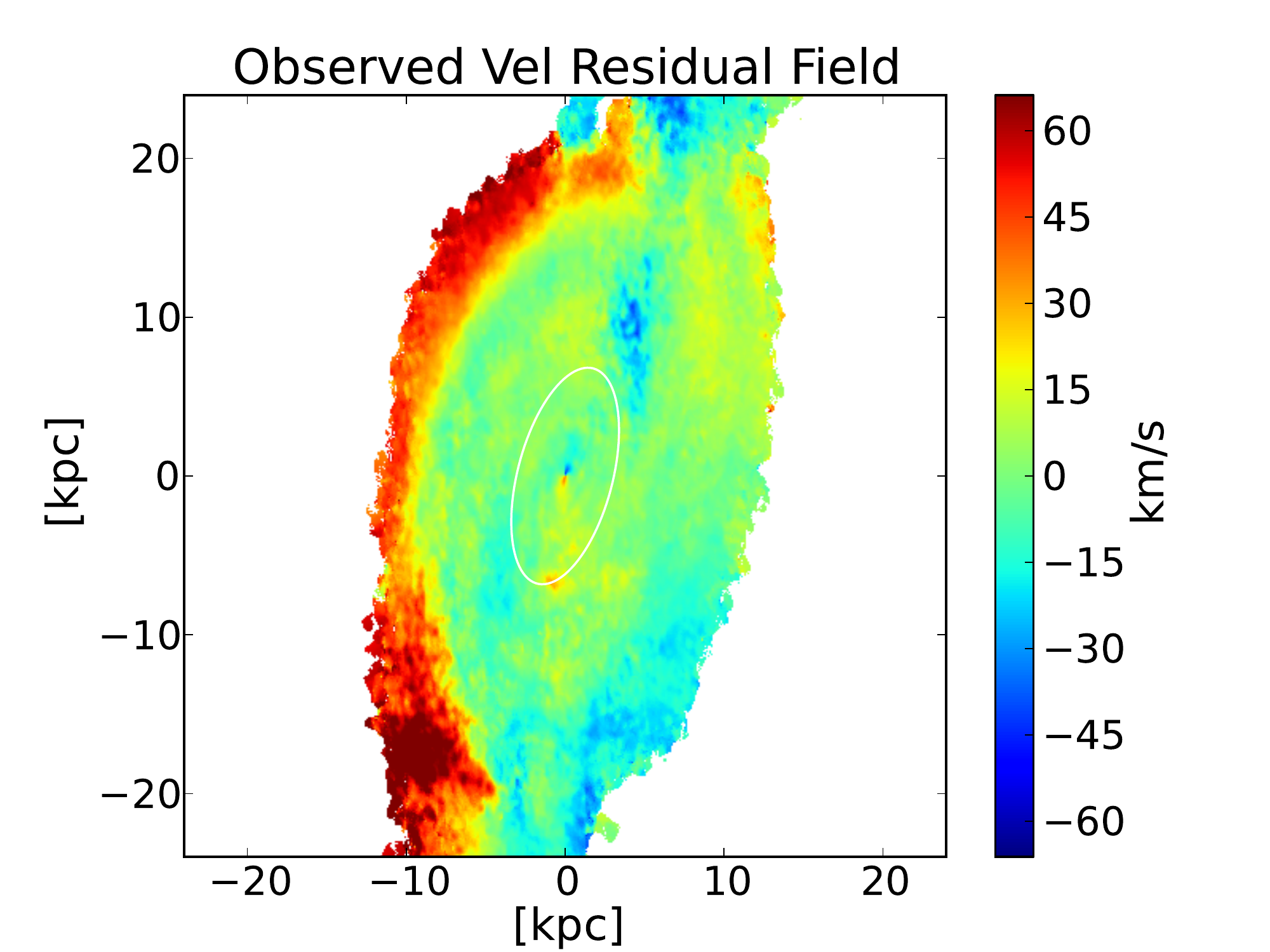}
\caption{NGC~3621: high spatial resolution maps. Top left: Stellar distribution as imaged in the near-IR by Spitzer IRAC at 3.6$\mu$m. Top right: atomic gas distribution (HI mom0 map). Bottom left: HI velocity field (mom1). Bottom right: HI residual velocity field created from a circular rotation model with constant P.A. and inclination of 345$\deg$ and 65$\deg$, respectively. The white circle indicates the region within 7~kpc radius that has been excluded from the fit.} 
\label{fig:NGC3621dpj}
\end{center}
\end{figure*}

\begin{figure*}
\begin{center}
\includegraphics[scale=0.40]{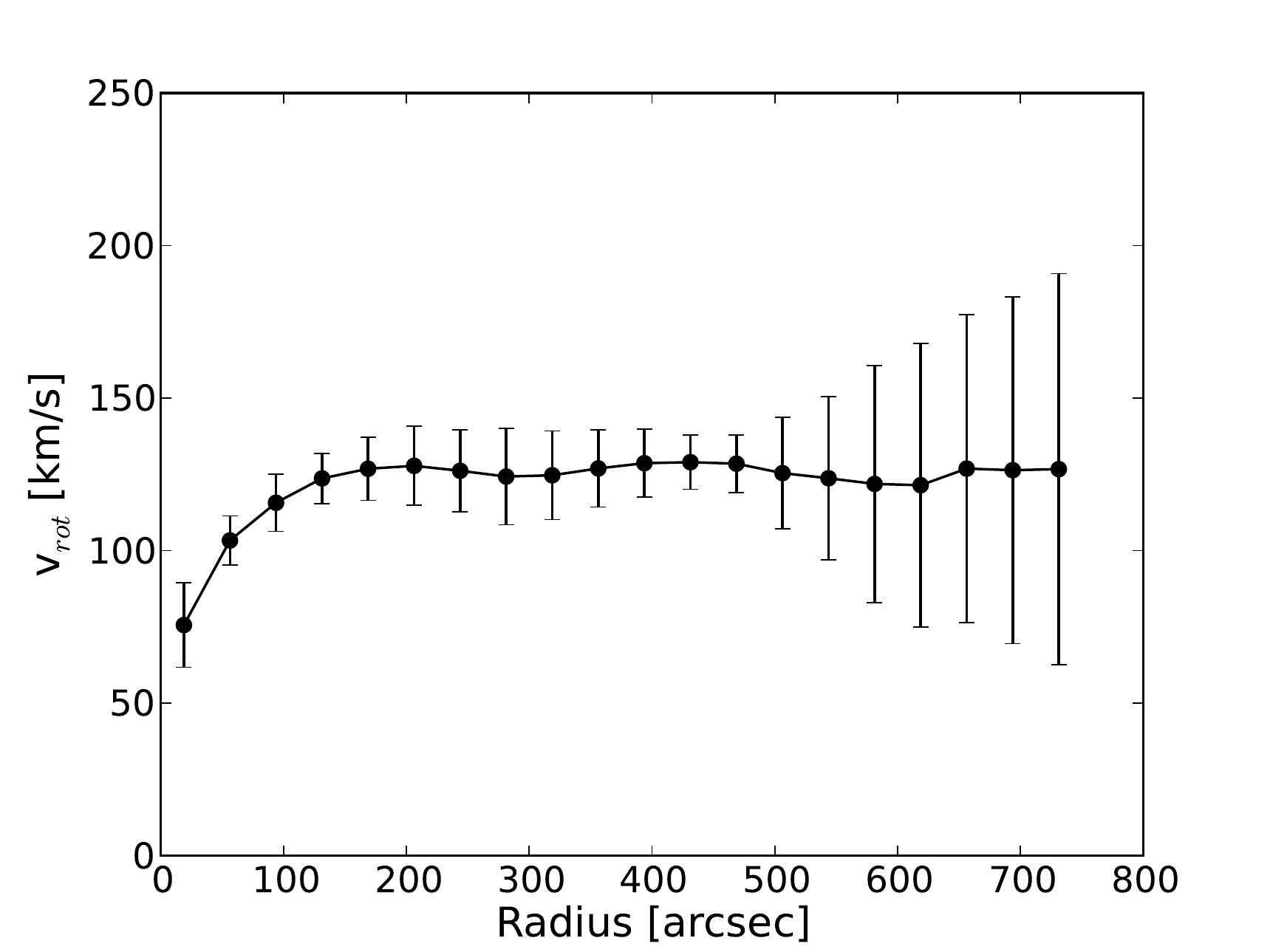}
\caption{NGC~3621. Rotation curve derived from velocity field with constant inclination and position angle. The errorbars represent the standard deviation of the velocity within each radial bin.} 
\label{fig:NGC3621rotc}
\end{center}
\end{figure*}

\begin{figure*}
\begin{center}
\includegraphics[scale=0.40]{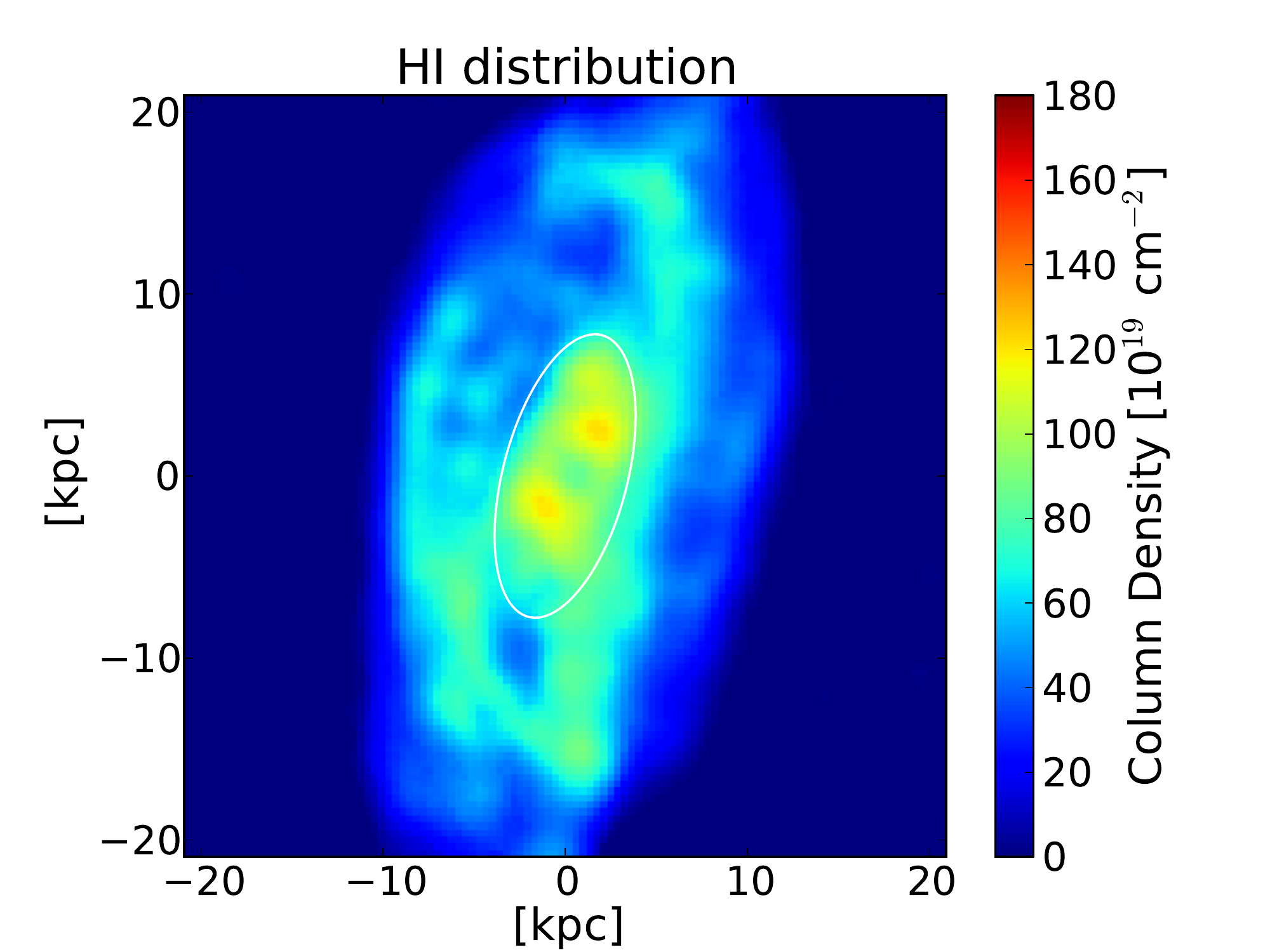}
\includegraphics[scale=0.40]{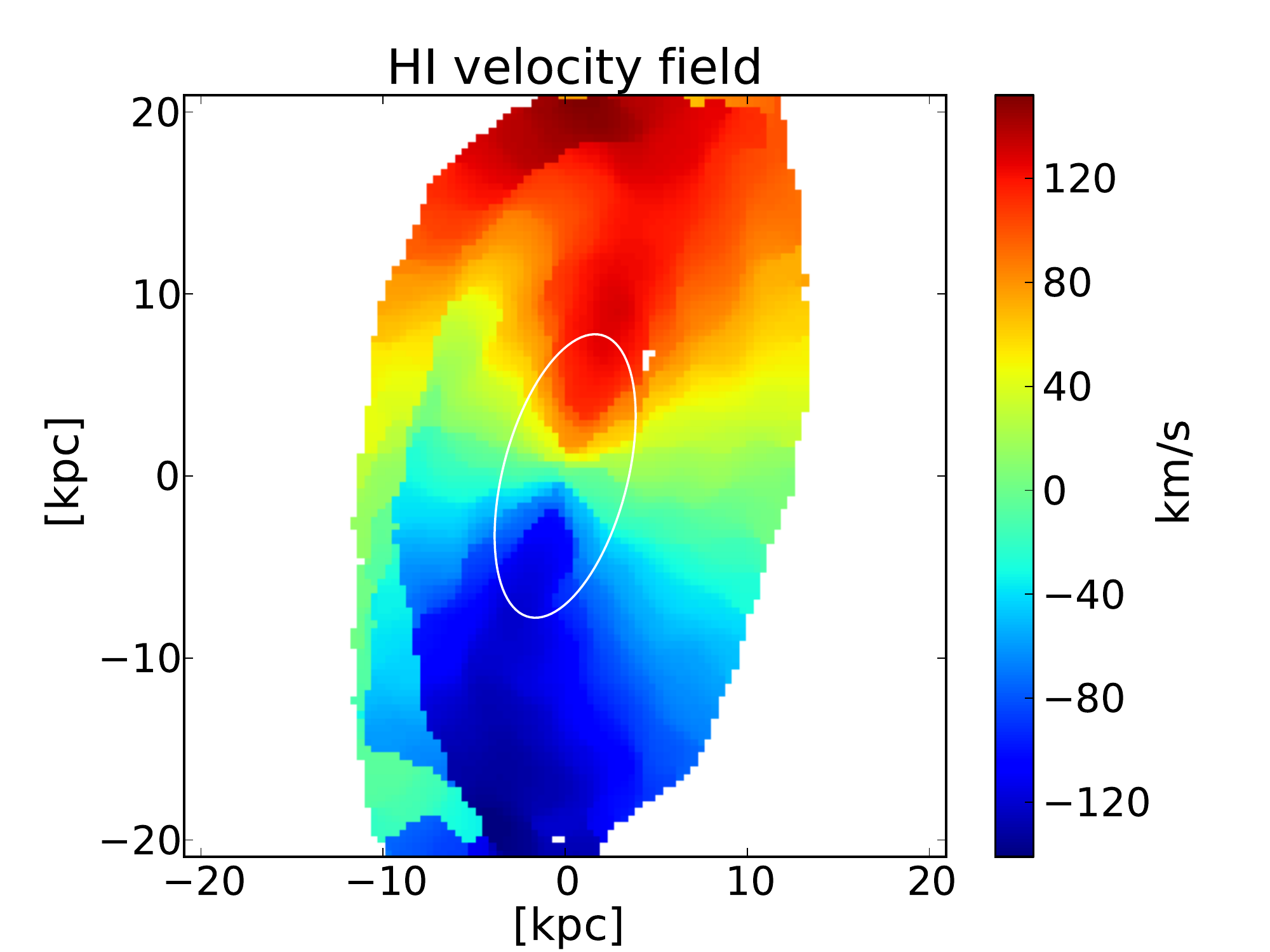}
\caption{NGC~3621: high sensitivity maps (beam 60 arcsec, $\sim$1.9~kpc) in the observed orientation. Left: The atomic gas distribution (in column densities, corrected for inclination). Right: HI velocity field. The white ellipse indicates the region within 7~kpc radius that is excluded from the fit.} 
\label{fig:NGC3621nat}
\end{center}
\end{figure*}

NGC~3621 exhibits a similar regular gas and stellar distribution as NGC~6946 but with larger kinematic irregularities. Deep observations with \textit{GALEX} in the UV-light show evidence for a strong warp in the outer disk \citep[][]{Gil07}. However, NGC~3621 has no nearby companion and shows no sign of tidal interaction. In particular at both ends of the major axis the gas extends beyond the regularly rotating disk with strong irregular velocities that are not necessarily associated with the main rotation of the disk \citep[for a large-scale image see][]{deB08}. Previous studies have suggested a warped disk which might explain some of the irregularities \citep{Wal08, deB08}.  Another possibility might be that parts of the gas are in the early process of being stripped from the main gas disk. Therefore we focus on the regular rotating disk as shown in Fig.~\ref{fig:NGC3621dpj}. The rotation curve and residual velocity field are derived in the same way as described in the previous section for NGC~6946. Fig.~\ref{fig:NGC3621rotc} shows the circular rotation velocity which rises to a maximum of $\sim$135~km~s$^{-1}$ and remains flat with increasing radius for a constant inclination and position angle. The residual velocity field exhibits a strong velocity offset on one half of the disk perpendicular to the major axis at a (deprojected) radial scale of $\sim$15--25~kpc while the opposite side shows only a small deviation from circular rotation (roughly $\pm(10-20)$~km~s$^{-1}$, see left panel of Fig.~\ref{fig:NGC3621nat}). This pattern can neither be explained by a warped disk nor by uniform kinematic ram pressure perturbations on its own, which suggests the possibility of a combination of both. Indeed, such a combined scenario is a natural outcome of a long-duration ram pressure interaction (as discussed below). \par

\begin{figure*}
\begin{center}
\includegraphics[scale=0.4]{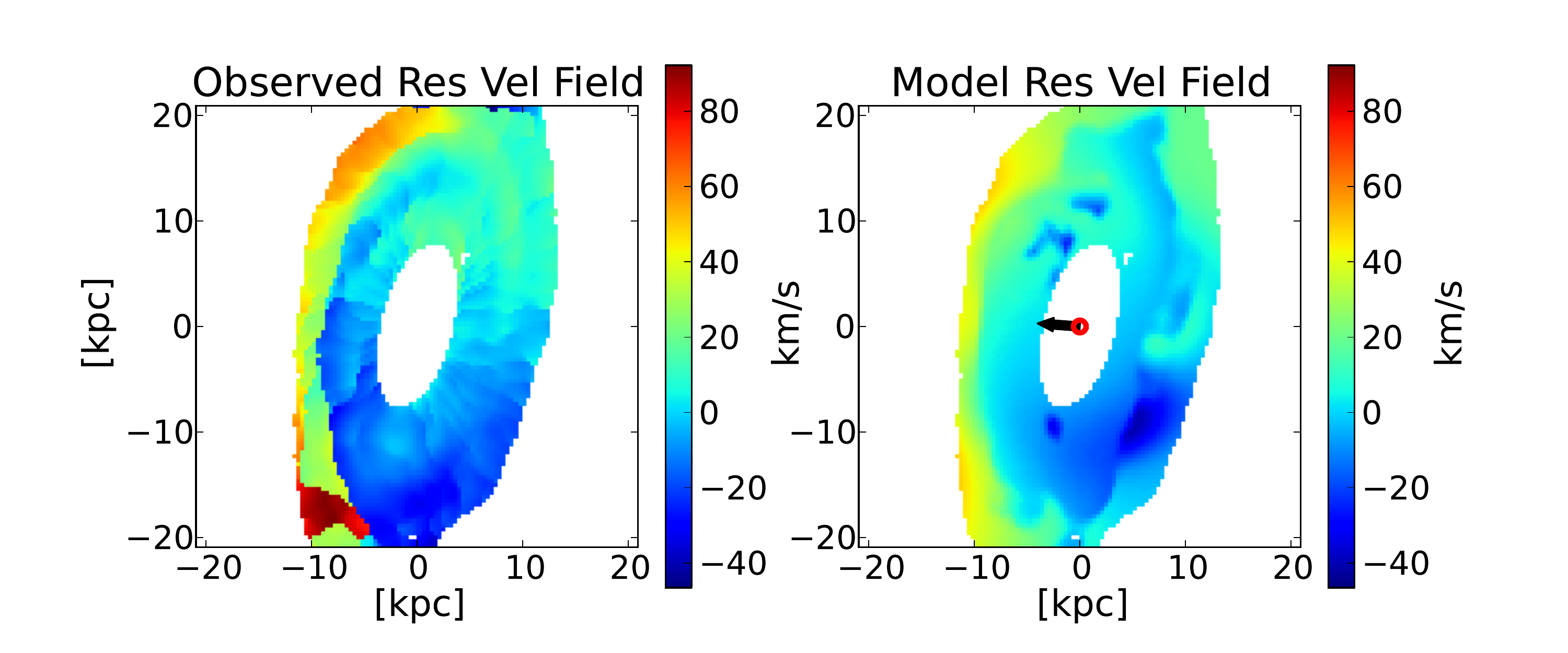}
\includegraphics[scale=0.3]{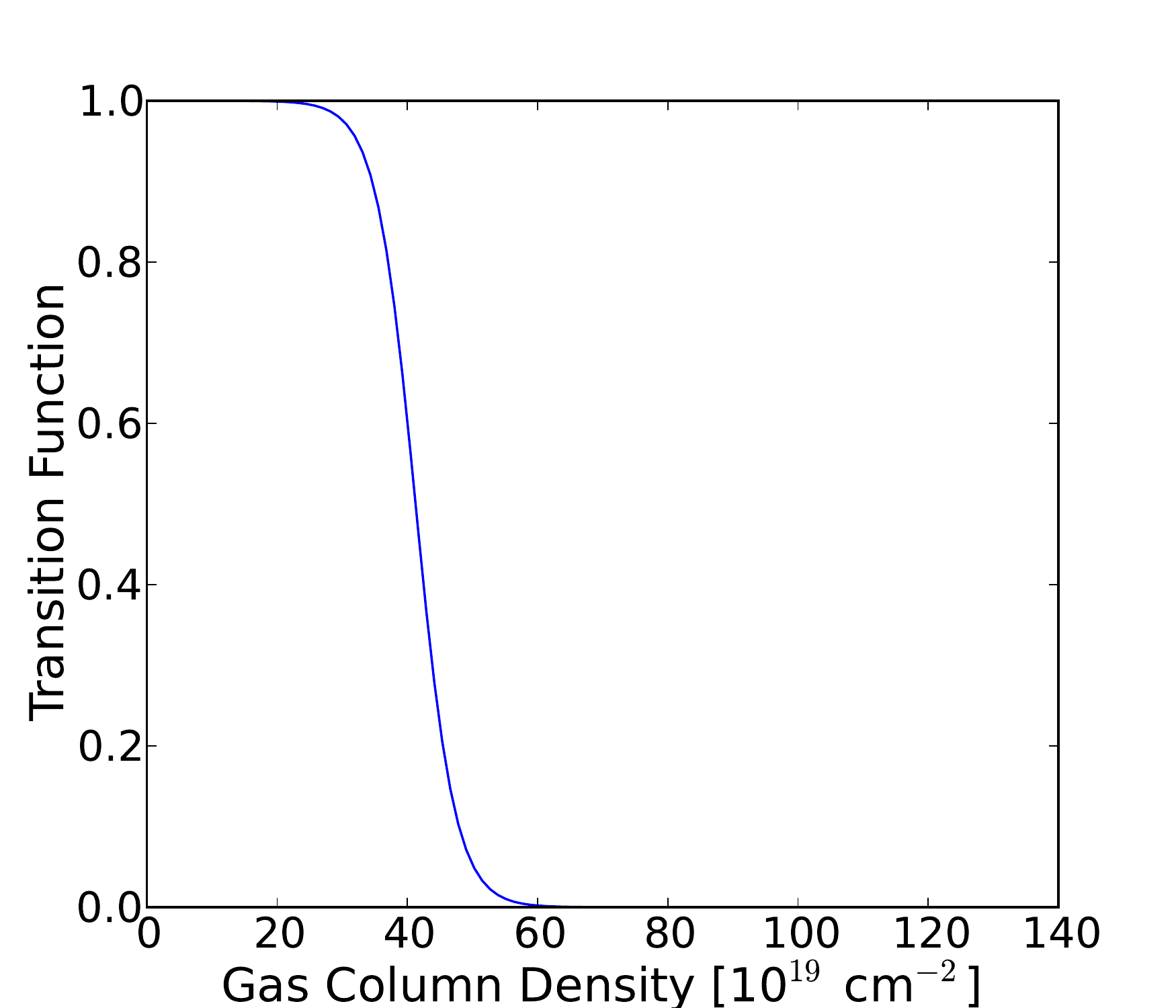}
\caption{NGC~3621, Comparison between observed (left) and model residual velocity field (middle). The residual fields are defined as difference between circular rotation and observed velocity field or model velocity field, respectively. The arrow and ring are defined as in Fig.~\ref{fig:NGC6946_fig5}. Right panel: The scaling function $\eta(\rho_{ISM})$ (see Eq.~\ref{eq:sfunct}) characterizing the effectiveness of the ram pressure on the HI kinematics as function of HI column density.} 
\label{fig:NGC3621_fig5}
\end{center}
\end{figure*}

\begin{figure*}
\begin{center}
\includegraphics[scale=0.4]{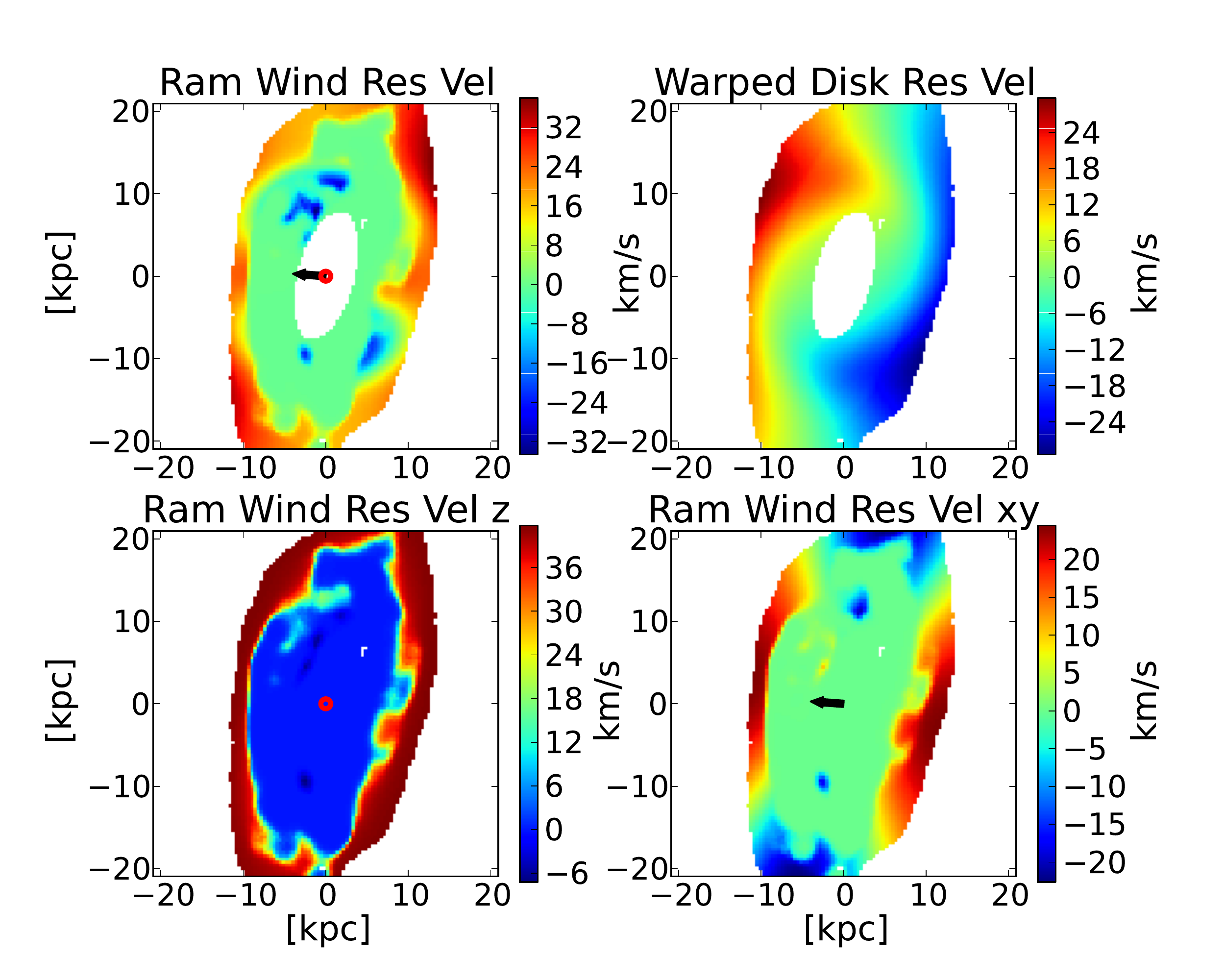}
\caption{NGC~3621: Top: Decomposition of model velocity field into velocities due to ram pressure (left) and velocity contributions due to systematic deviations such as change in position angle and inclination of disk (right). Bottom: Decomposition of ram pressure velocity field into velocities due to ram wind perpendicular ($z$-axis, bottom left)  and parallel ($xy$-plane, bottom right) to the disk. The arrow and ring are defined as in Fig.~\ref{fig:NGC6946_fig5}.} 
\label{fig:NGC3621_fig6}
\end{center}
\end{figure*}

To test this scenario we have fit our combined ram pressure and warped disk model to the gas kinematics shown in the 60$\arcsec$ beamsize image of Fig.~\ref{fig:NGC3621nat}. The region within 7~kpc radius is excluded from the fit. We have used the same fitting procedure and model as described above for NGC~6946. A comparison between the observed and model residual velocity field is shown in Fig.~\ref{fig:NGC3621_fig5}. The decomposition of the residual velocity field reveals (see Fig.~\ref{fig:NGC3621_fig6}) a combination of strong m=0 and m=1 modes with $vc_0$=33~km~s$^{-1}$ and $vc_1$=42~km~s$^{-1}$. These modes correspond to the perpendicular component of the ram pressure and to the residual velocity component of a warped disk, respectively. We also find a significant contribution of an m=2 mode with $vc_2$=24~km~s$^{-1}$. The warp is characterized by a significant change in the position angle of $ \Delta \phi =20\deg$ and inclination of $ \Delta i=47\deg$.  The ram wind direction is given by the angles $\theta_{Ram}=89\deg$ and $\gamma_{Ram}=-78\deg$ with an effective ram wind component perpendicular and parallel to the disk of $\Delta v\bot\approx -114$~km~s$^{-1}$ and $\Delta v\|\approx 27$~km~s$^{-1}$, respectively.  The transition of the effective ram pressure as function of HI gas density is relatively sharp near $N \sim41\times10^{19}$~cm$^{-2}$ as shown in Fig~\ref{fig:NGC3621_fig5}. Our results demonstrate that the kinematics of the outer disk of NGC~3621 can be well-characterised by ram pressure acting on a moderately warped gas disk.\par

This result is not surprising, since ram pressure can produce a warped gas disk over a time-scale of several orbits \citep[][]{Haa13}. The measured amplitude, P.A., and sign of the warp, are in good agreement with the formation of a warped disk under ram pressure which depends on the ram wind angle and the galaxy sense of rotation (here clockwise). However, we note that the current ram pressure environment can significantly differ, in particular in ram pressure strength, from the environment when the warp has formed.  While this scenario might be a possible explanation for the peculiar velocity field of NGC~3621, we cannot rule out other possibilities such as an unwarped galaxy experiencing a one-sided disturbance. However, the cause for such a one-sided disturbance is not obvious since NGC~3621 has no nearby companion and its stellar distribution does not show any significant interaction features. 

\clearpage

\twocolumn

\subsection{NGC~628: An example of a strongly warped disk.}

\begin{figure*}
\begin{center}
\includegraphics[scale=0.40]{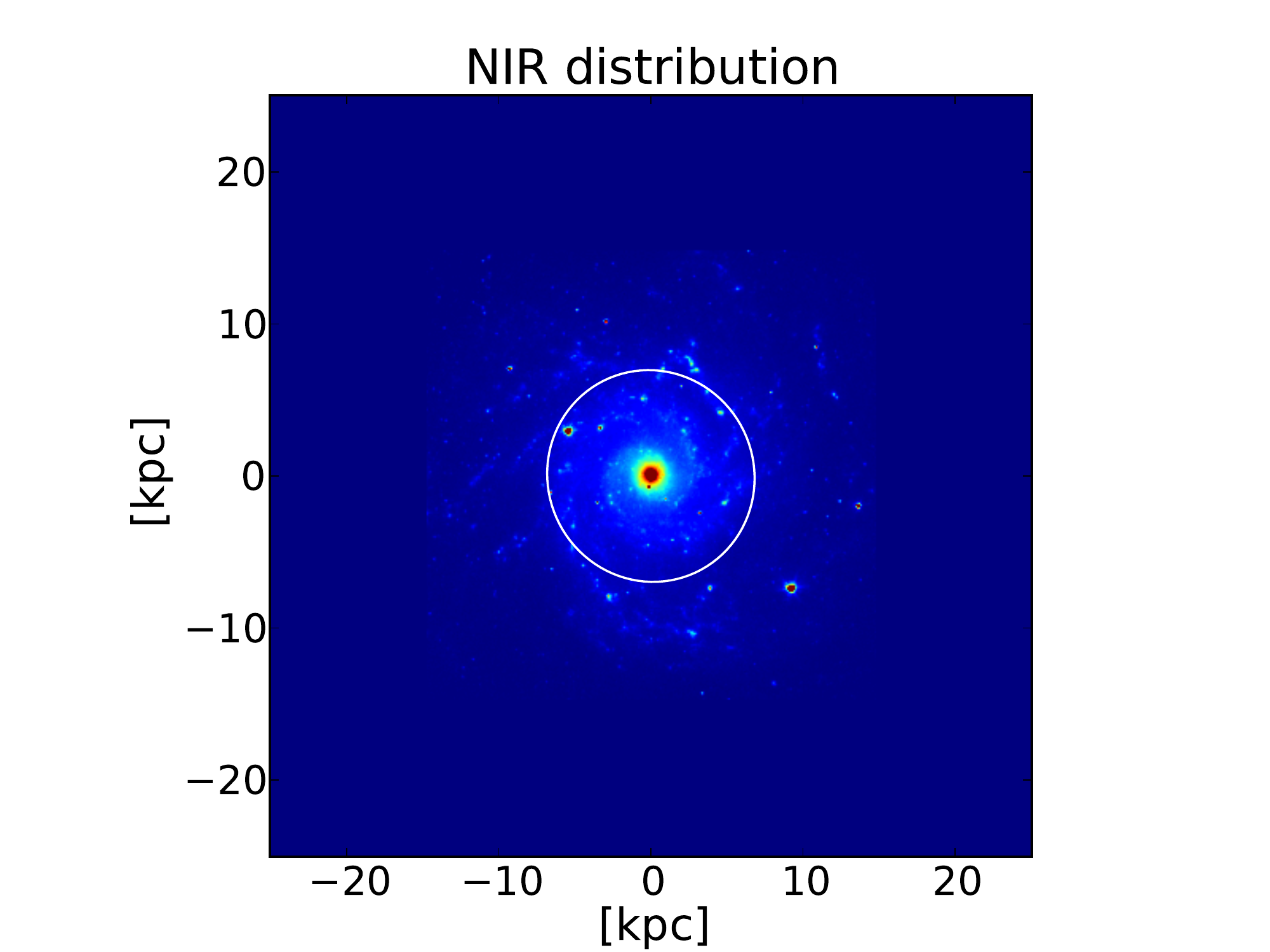}
\includegraphics[scale=0.40]{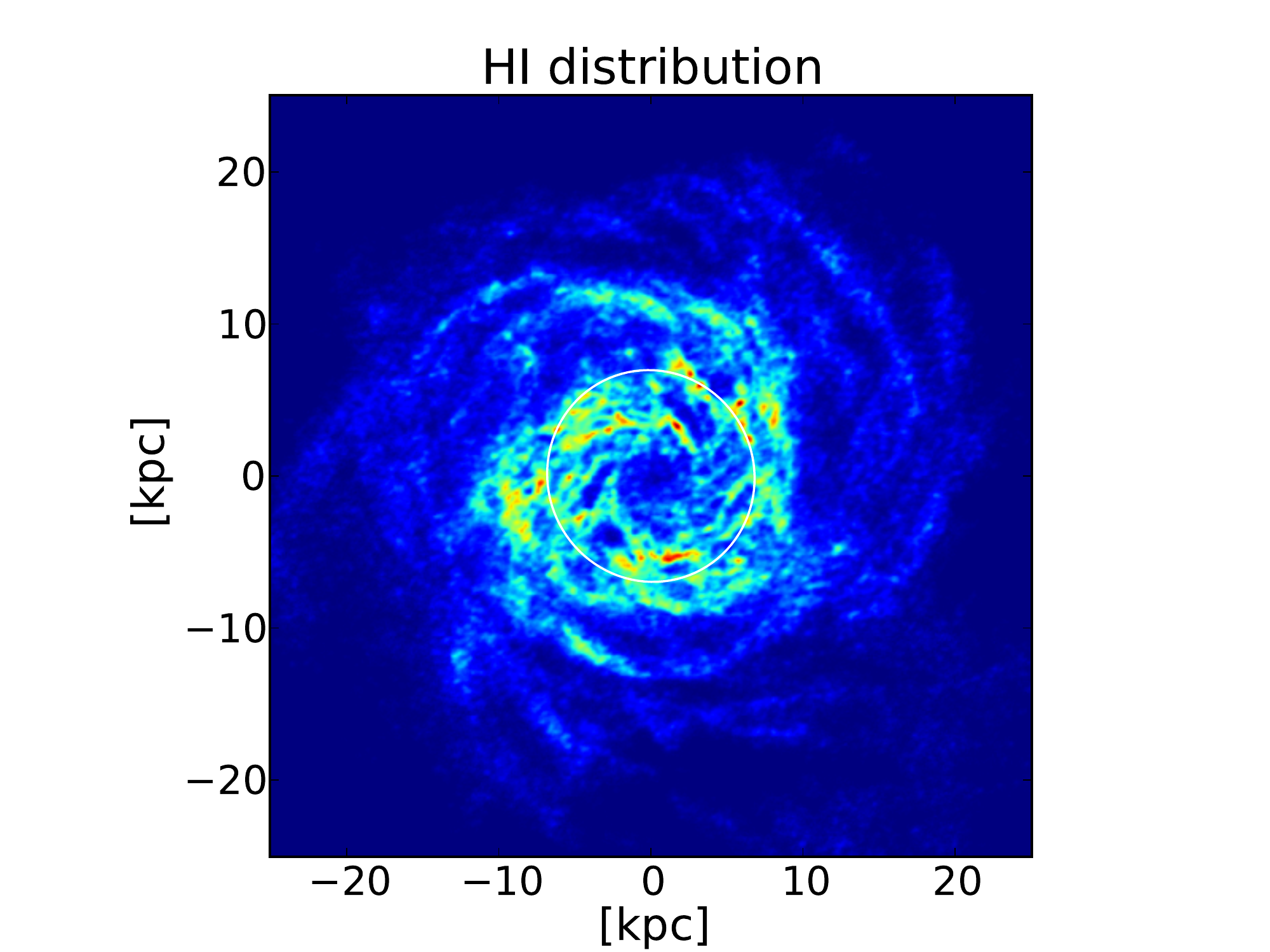}
\includegraphics[scale=0.40]{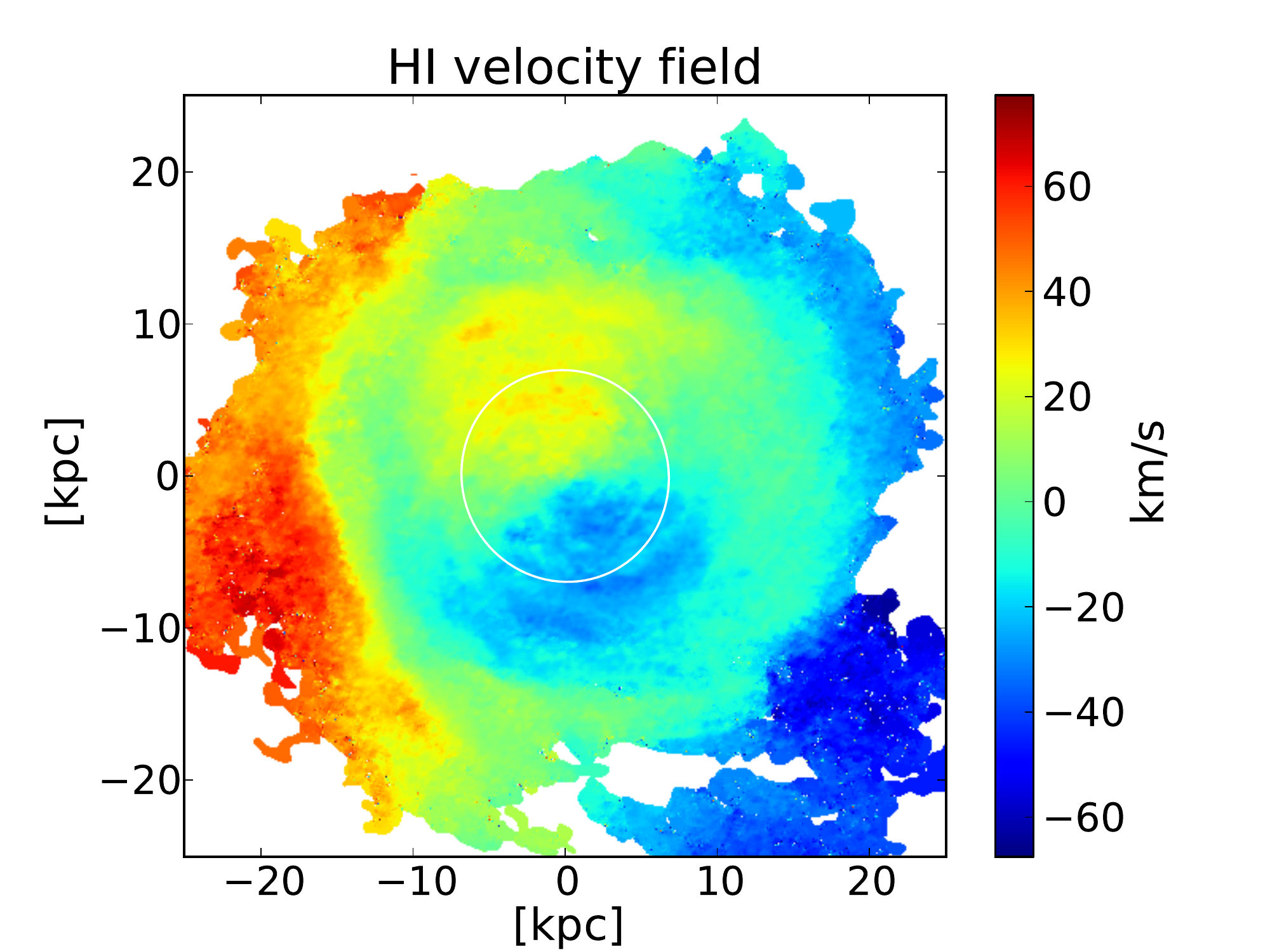}
\includegraphics[scale=0.40]{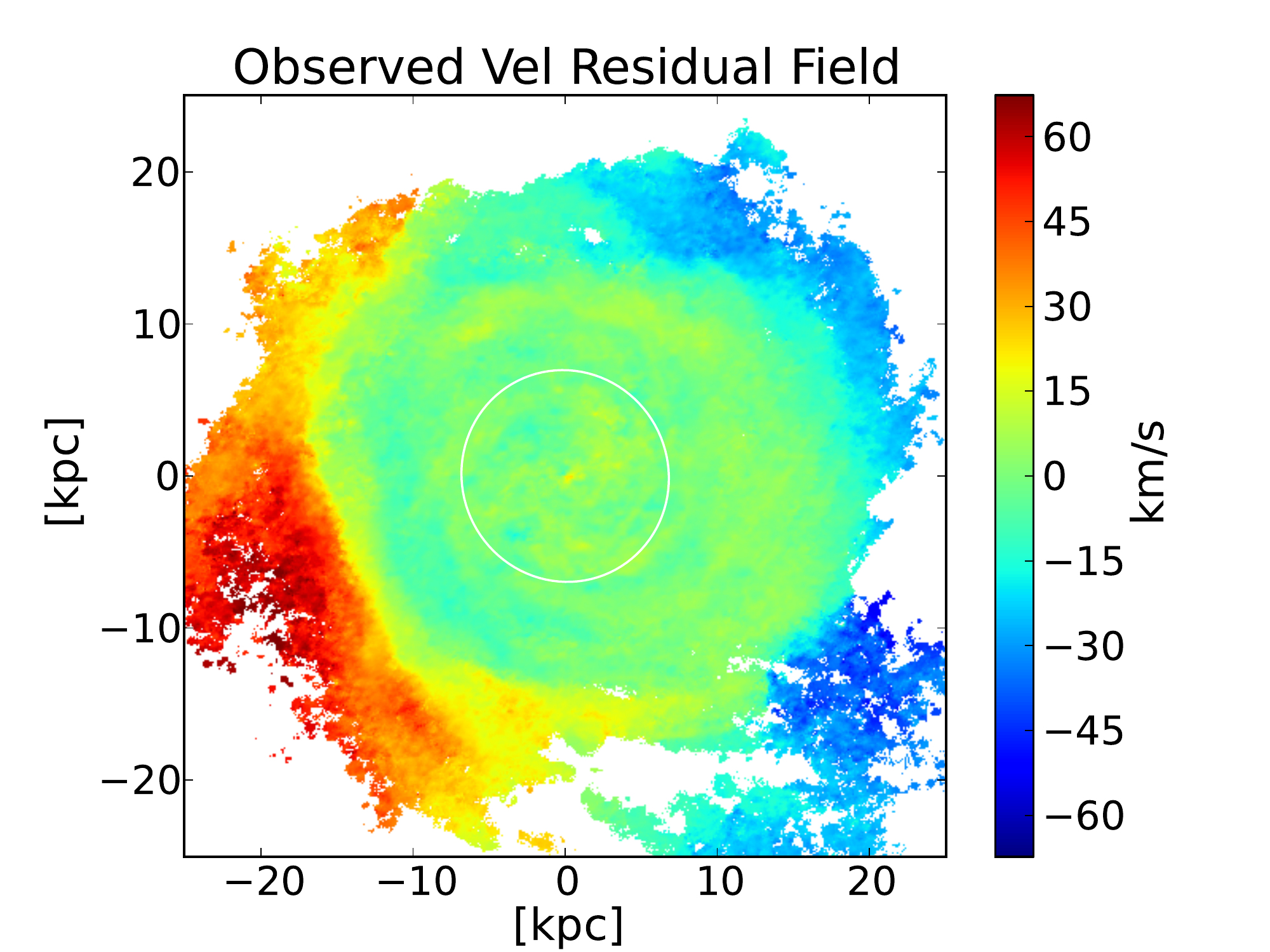}
\caption{NGC~628: high spatial resolution maps. Top left: Stellar distribution as imaged in the near-IR by Spitzer IRAC at 3.6$\mu$m. Top right: atomic gas distribution (HI mom0 map). Bottom left: HI velocity field (mom1). Bottom right: HI residual velocity field created from a circular rotation model with constant P.A and inclination of 23$\deg$ and 14$\deg$, respectively.  The white circle indicates the region within 8~kpc radius that has been excluded from the fit.} 
\label{fig:NGC628dpj}
\end{center}
\end{figure*}

\begin{figure*}
\begin{center}
\includegraphics[scale=0.40]{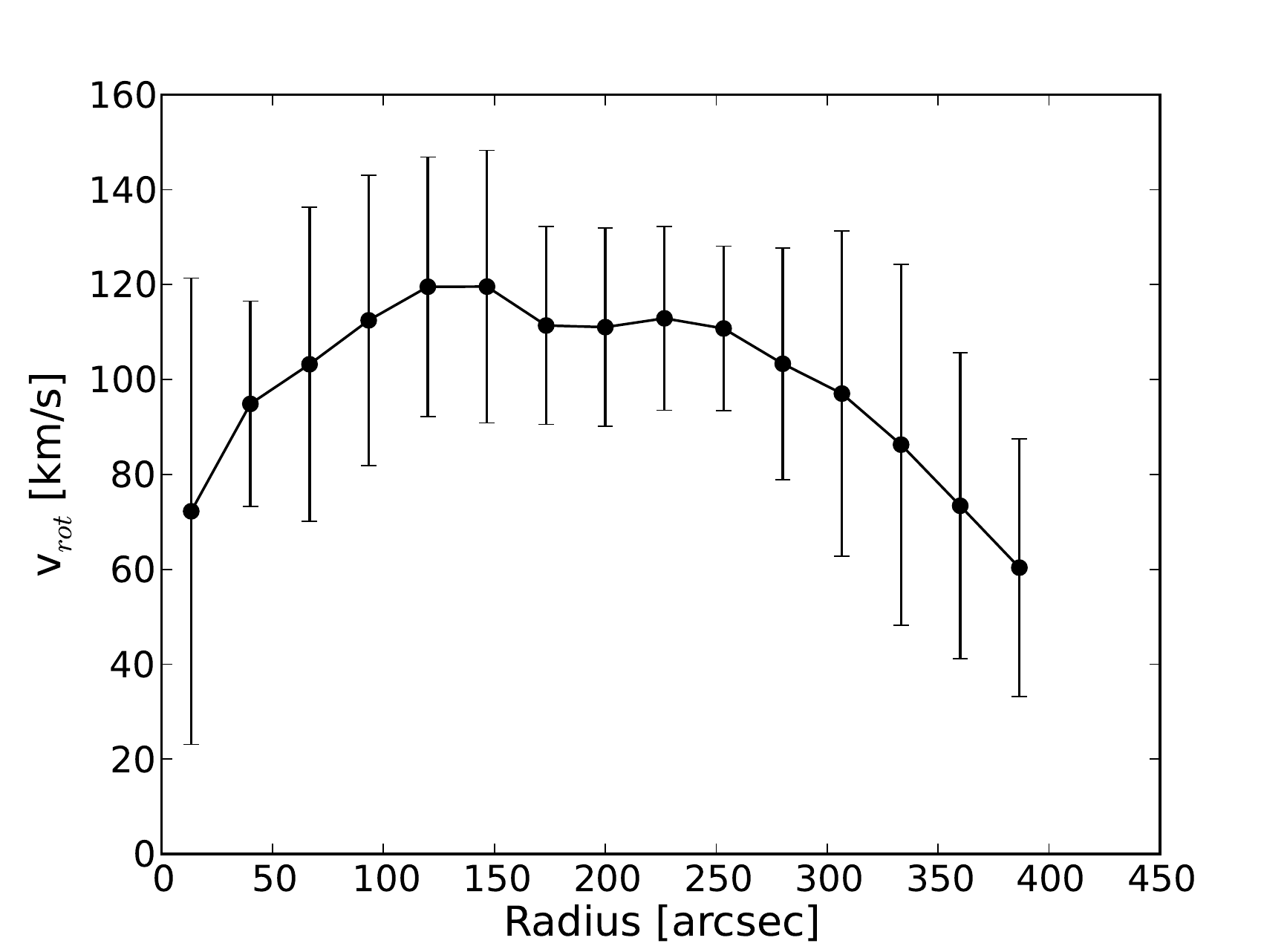}
\caption{NGC~628. Rotation curve derived from velocity field with constant inclination and position angle. The errorbars represent the standard deviation of the velocity within each radial bin. Right: The circular nominal velocity model in the observed projected orientation.} 
\label{fig:NGC628rotc}
\end{center}
\end{figure*}

NGC~628 is a nearly face-on spiral galaxy (inclination of the inner disk is $\sim14\deg$) and exhibits a very regular gas and stellar distribution as shown in Fig.~\ref{fig:NGC628dpj}. However, the major axis of the velocity field shows a strong variation as function of radial distance from the center which is caused by a strongly warped disk \citep{Sho84, Kam92}. The residual velocity pattern is consistent with a strong warp which causes the apparent drop in rotation velocity towards larger radii if one would assume a constant P.A. and inclination over the entire disk as plotted in Fig.~\ref{fig:NGC628rotc}. Therefore, this galaxy represents a good test of whether our fit method can accurately describe strong warped disks as well.\par

\begin{figure*}
\begin{center}
\includegraphics[scale=0.40]{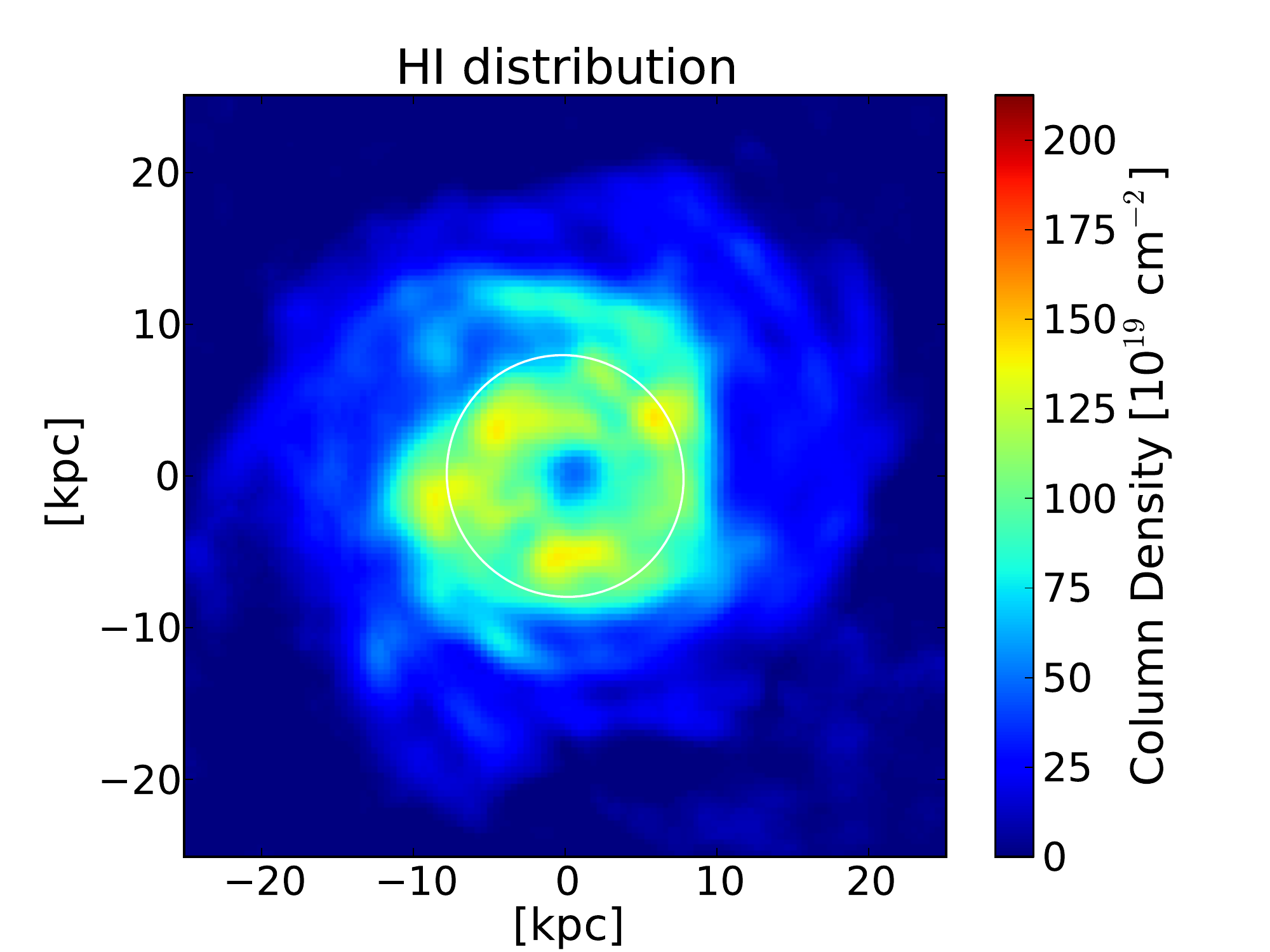}
\includegraphics[scale=0.40]{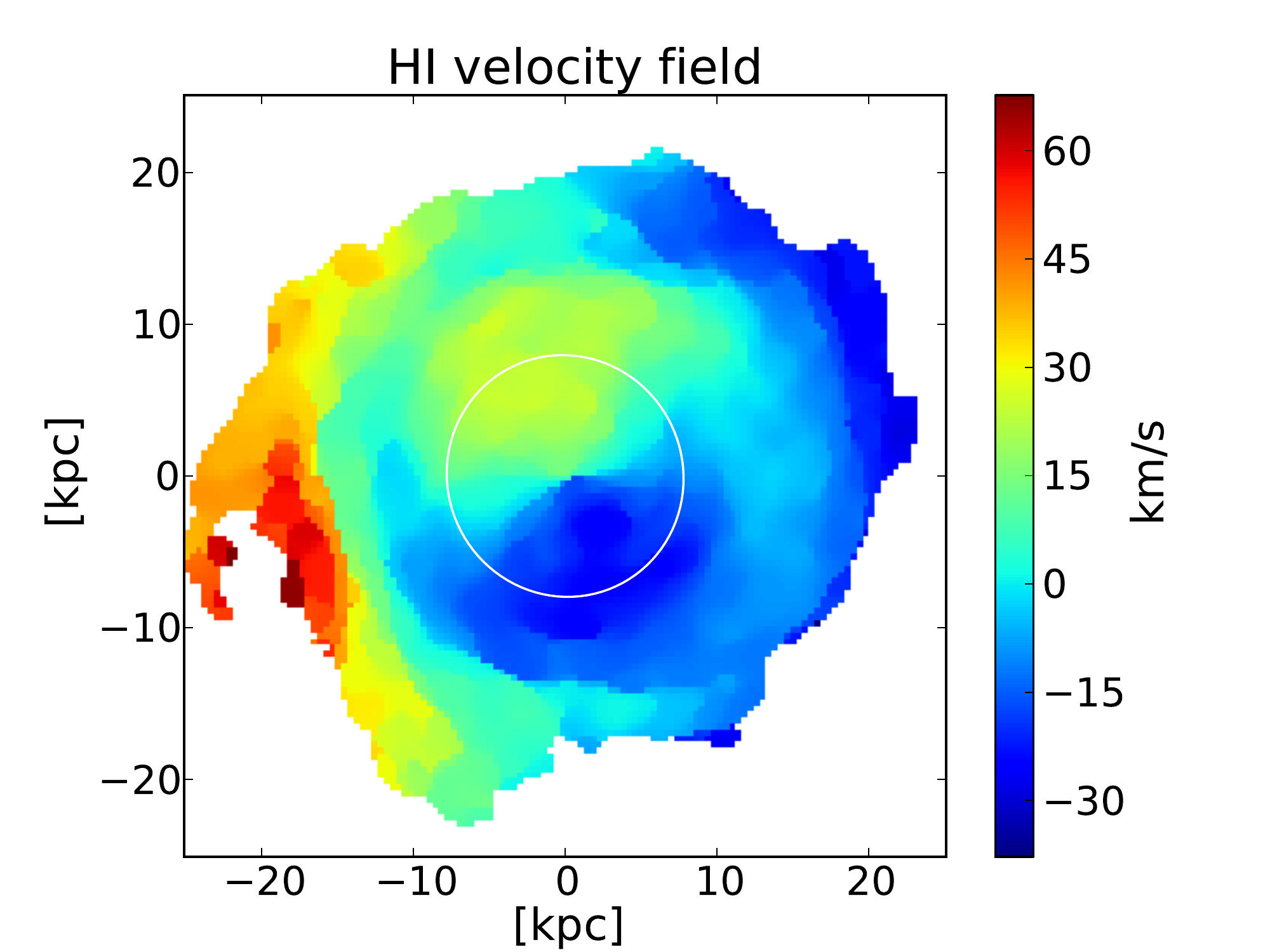}
\caption{NGC~628: high sensitivity maps (beam 60 arcsec, $\sim$2~kpc) in the observed orientation. Left: The atomic gas distribution. Right: HI velocity field. The white ellipse indicates the region within 8~kpc radius that is excluded from the fit.} 
\label{fig:NGC628nat}
\end{center}
\end{figure*}

\begin{figure*}
\begin{center}
\includegraphics[scale=0.35]{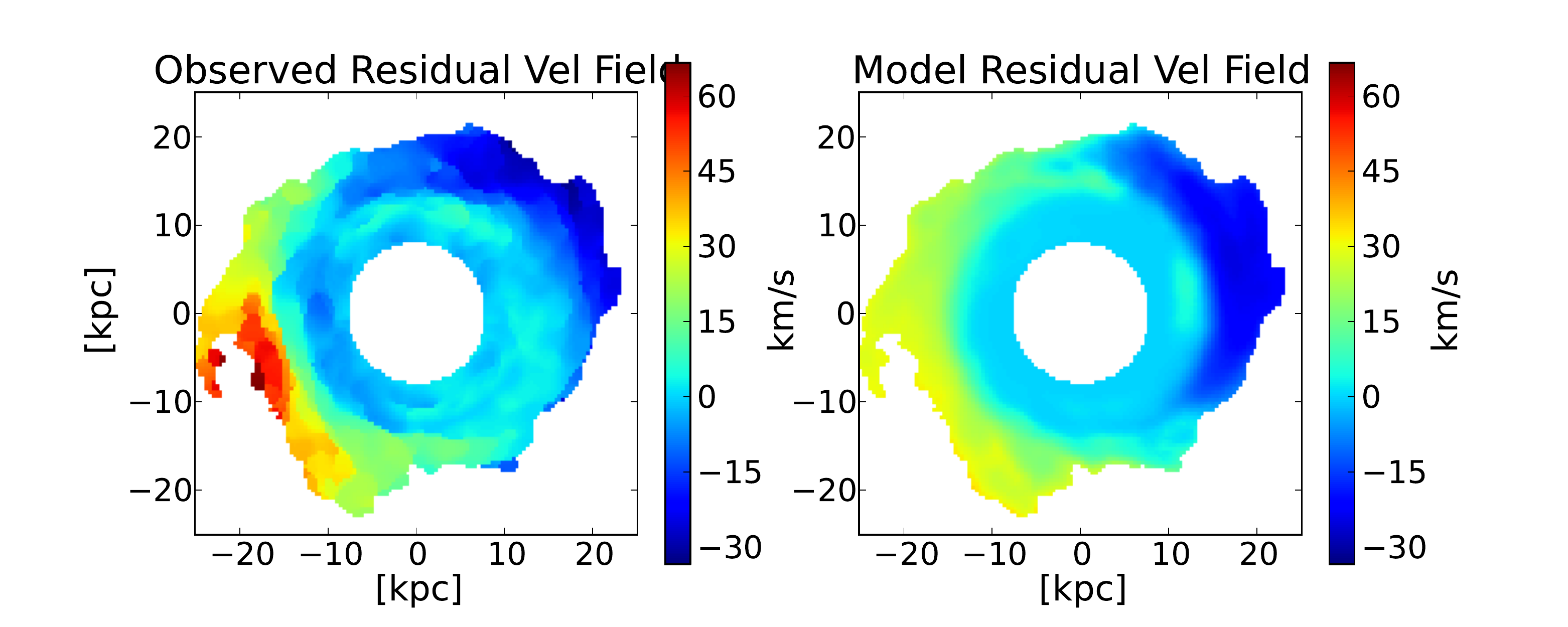}
\caption{NGC~628: Comparison between observed (left panel) and model residual velocity field (right panel). The residual fields are defined as difference between circular rotation and observed velocity field or model velocity field, respectively.} 
\label{fig:NGC628_fig5}
\end{center}
\end{figure*}

\begin{figure*}
\begin{center}
\includegraphics[scale=0.40]{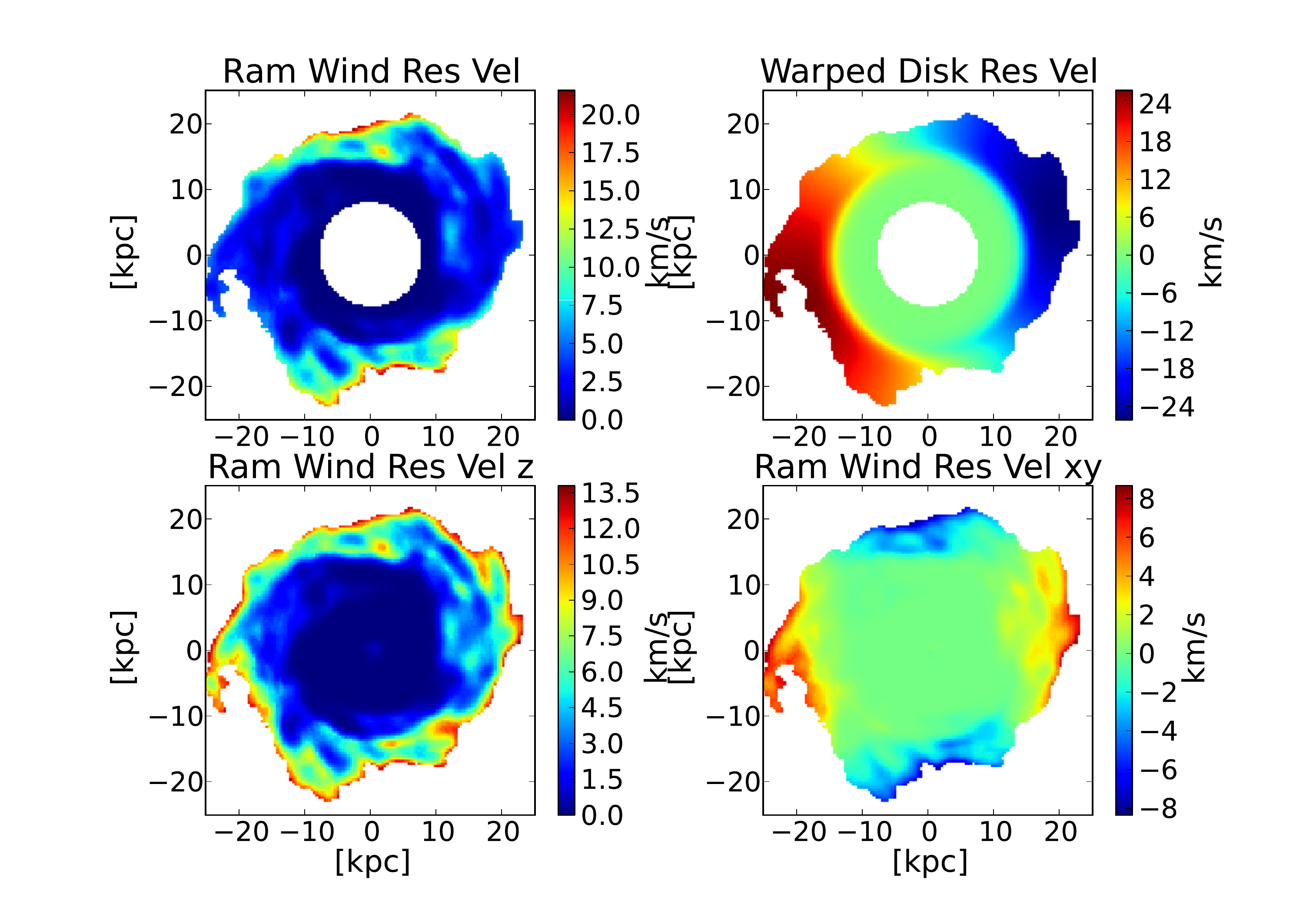}
\caption{NGC~628: Top: Decomposition of model velocity field into velocities due to ram pressure  (left) and velocity contributions due to systematic deviations such as change in position angle and inclination of disk (right). Bottom: Decomposition of ram pressure velocity field into velocities due to ram wind perpendicular ($z$-axis, bottom left)  and parallel ($xy$-plane, bottom right) to the disk.} 
\label{fig:NGC628_fig6}
\end{center}
\end{figure*}

As described above, the residual velocity field (see Fig.~\ref{fig:NGC628nat}) is fit with the combined ram pressure and warped disk model. We exclude the inner region of 8~kpc from the fit. Our results show an excellent agreement with a strongly warped disk given a dominant m=1 mode with $vc_1\approx36$~km~s$^{-1}$ as presented in Fig.~\ref{fig:NGC628_fig5} and Fig.~\ref{fig:NGC628_fig6}. The warped disk is characterized by a change in inclination of $\Delta i=14\deg$ and position angle of $\Delta \phi=73\deg$ which occurs at a radius of $\sim$16~kpc . This is in good agreement with the warp geometry described in \cite{Kam92} who found a change in inclination of $\sim11\deg$ and PA of $\sim75\deg$ at a transition radius of $\sim$7--11$\arcmin$ $(\simeq$14--22~kpc). While the fit also indiates small contributions from m=0 and m=2 modes, their significance is too low to make any conclusions about ram pressure interaction in NGC~628.

\clearpage

\twocolumn
\section{Discussion}
\label{sec:dis}

\subsection{Long term Implications of Ram Pressure on Disk Geometry and Kinematics}
\label{subsec:sim}

In this study we focus on the systematic velocity perturbation from pure circular rotation that is a consequence of a ram pressure interaction. The momentum of the intercepted IGM is absorbed by the diffuse, dissipative ISM in a small fraction of the rotational period (50 to few hundred Myrs), but constantly evolves due to the ongoing ram pressure interaction. This interaction eventually yields a change of the orbital paths of the entire gas disk, which has two long-term implications. The in-plane component of the interaction will tend to make the disk orbits more elliptical by compressing them toward the wind and elongating them away from the wind. The out-of-plane component of the interaction tends to push the orbits away from the mid-plane. Both these tendencies are moderated by the effects of differential rotation within a dissipative medium and the gravitational restoring forces of the potential, although ultimately there will be significant changes in the orbital paths of the gaseous component. The long-term consequences on the geometry of the gas disk are investigated in detail in another study \citep[][]{Haa13} and are summarised only briefly here. In most circumstances, namely if the ram wind is moderately inclined to the disk, a constant ram pressure over long time intervals induces an ``S-shaped'' warped disk (m=1 mode) with an orientation that depends on the direction of the galaxy's motion through the IGM and whether the galaxy is rotating in a clockwise or counter-clockwise sense. However, the orbital paths of the gas clouds change continuously, increasing the inclination of the warped disk with time.\par

\begin{figure}
\begin{center}
\includegraphics[scale=0.42]{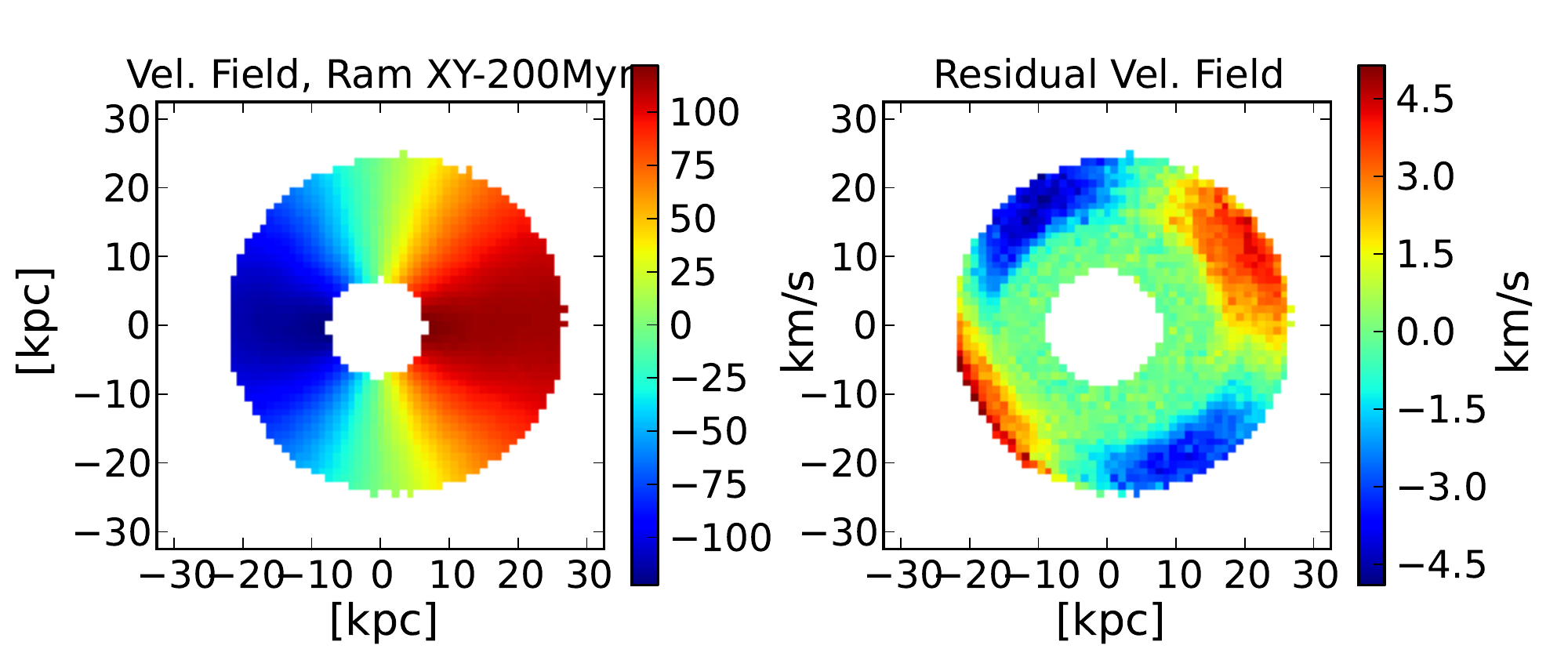}
\includegraphics[scale=0.42]{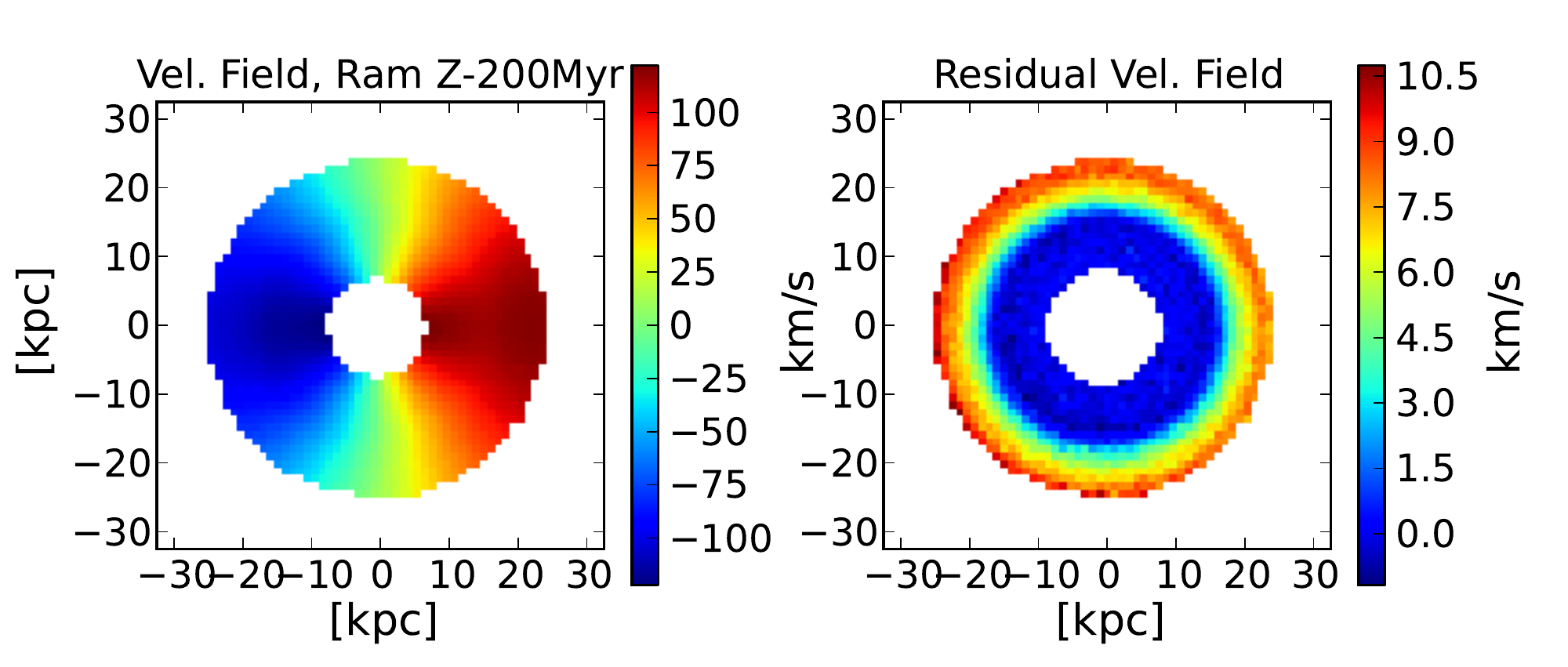}
\includegraphics[scale=0.42]{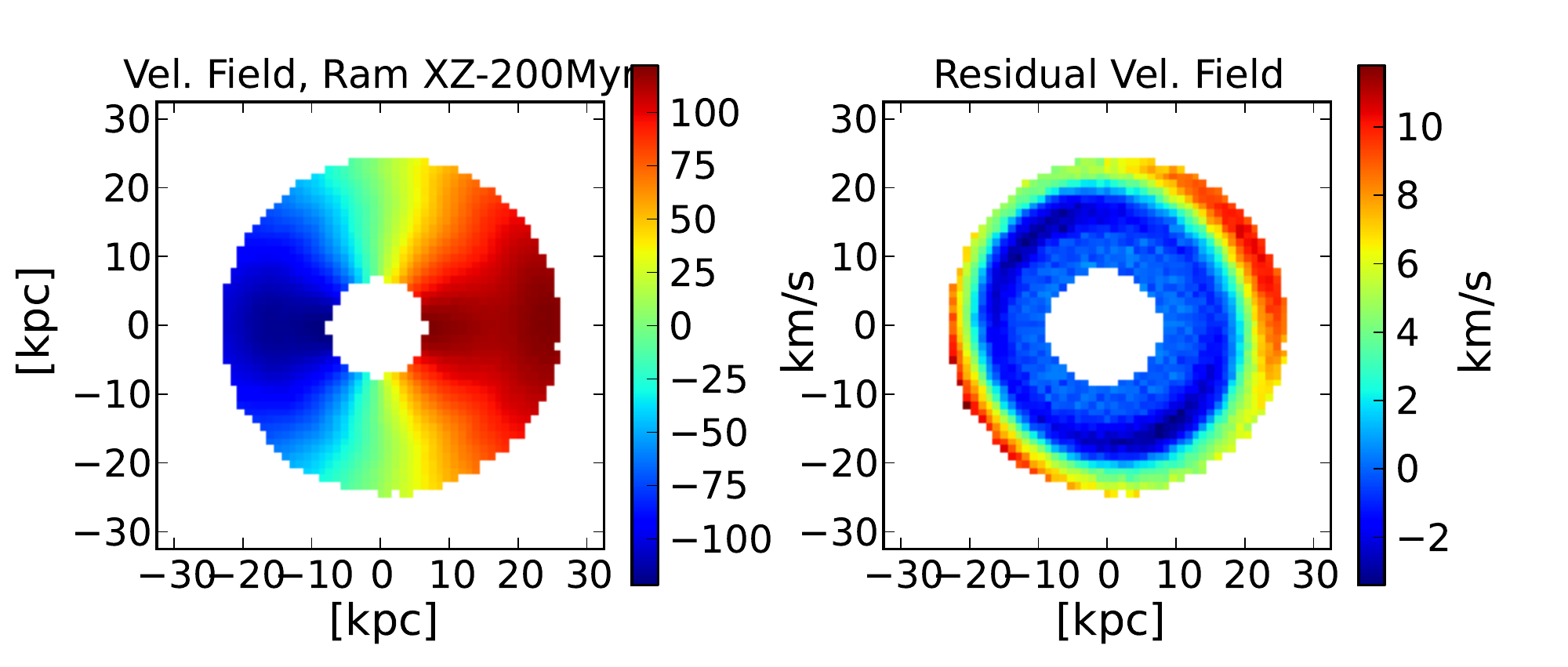}
\includegraphics[scale=0.42]{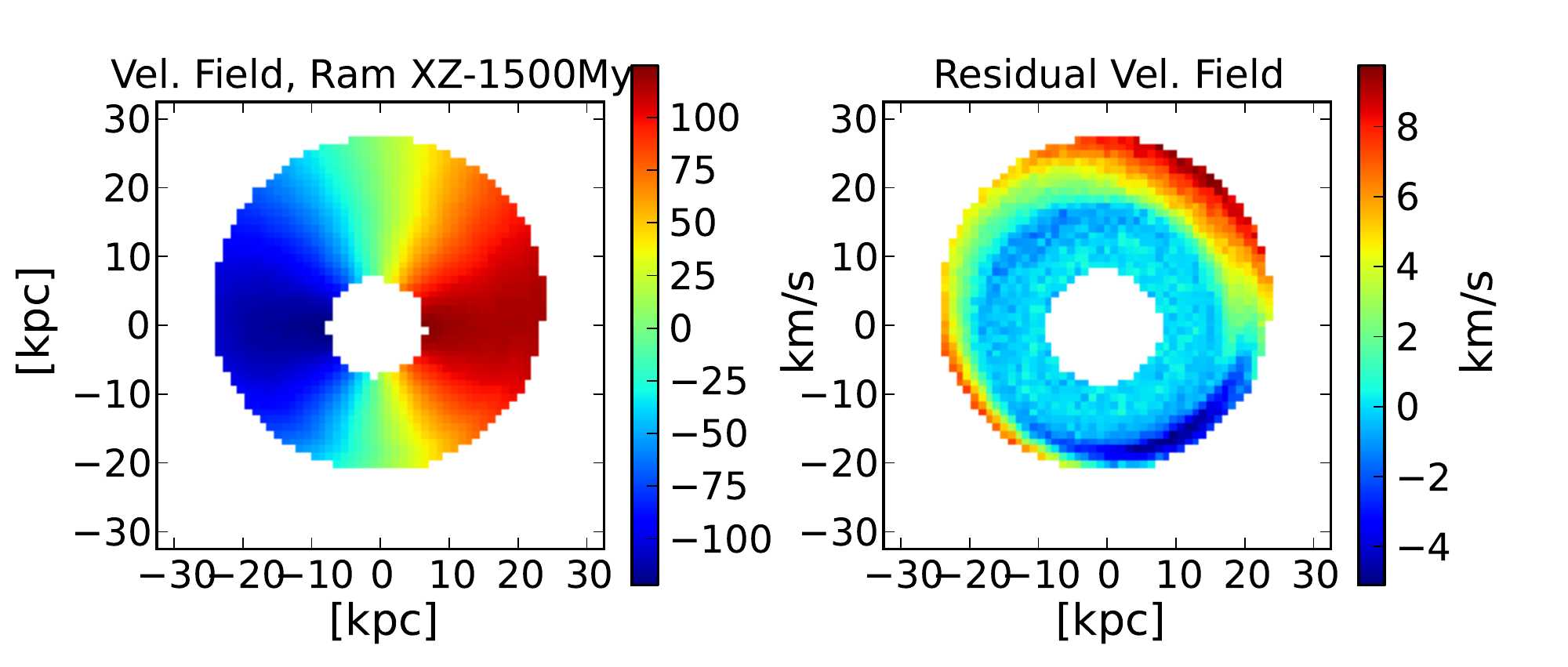}
\caption{The simulated velocity field (left) and residual velocity field (right) of particles in a static gravitational potential after 200~Myrs of ram pressure from three different directions: from top to bottom: a) parallel to plane of the disk, b) perpendicular to disk, c) 45$\deg$ inclined to disk, and d) simulation after 2 rotation periods, which shows a strong asymmetry in the velocity field due to the combination of warped morphology and kinematic perturbation.} 
\label{fig:sim_short}
\end{center}
\end{figure}

\begin{figure}
\begin{center}
\includegraphics[scale=0.45]{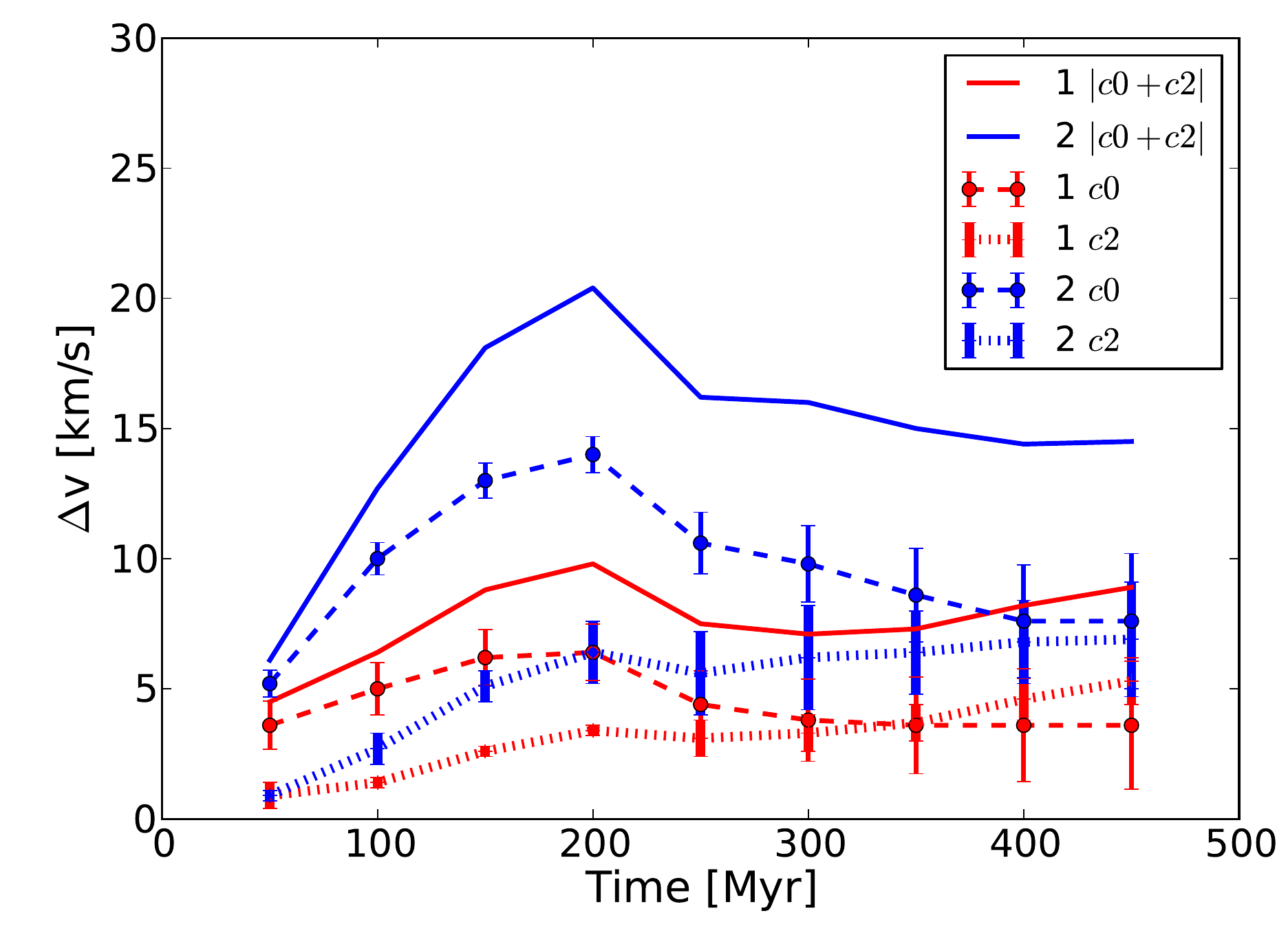}
\caption{Amplitudes of the kinematic modes as a function of time derived from simulated velocity fields of gas clouds in a static galactic gravitational potential with a ram pressure wind that is 45$\deg$ inclined to the disk. The ratio of ram pressure to gravitational force at  R=20~kpc is $F_{ram}/F_{grav}$=1/24 (red lines) and $F_{ram}/F_{grav}$=1/12 (blue lines).} 
\label{fig:sim_time}
\end{center}
\end{figure}

While the ``cumulative'' ram pressure induced warped disk is formed over several rotation periods (a few Gyrs), the ``instantaneous'' ram pressure induced velocity perturbation is established within a fraction of the rotation period and remains relatively constant. To assess the acceleration timescale $t_{acc}$ that corresponds to this velocity perturbation, we have derived model velocity fields from simulating the gas trajectories in a static galactic potential as a function of the ram wind vector and time. The galactic potential and gas trajectories are described in detail in \citep{Haa13}. 
%Velocity fields are derived for a ram wind inclination angle, $\gamma_{ram}=$45$\deg$. 
We have produced residual velocity fields by subtracting the velocity field of the undisturbed disk which is rotating counter-clockwise at a speed of $\sim$160~km~s$^{-1}$ at 20~kpc radius.  Figure \ref{fig:sim_short} depicts the velocity field of the  gas after 200~Myrs of ram pressure interaction for three different ram wind angles. We find that the m=0 mode and m=2 mode are associated with the ram pressure components perpendicular and parallel to the disk, respectively, in excellent agreement with our simple geometric models. However, deviations from circular motion due to elliptical orbits produce an additional m=0 and m=1 mode whose amplitudes are dependent on the strength of elongation and the direction of the elongation with respect to the position angle. For instance, an orbital elongation of 10\% can contribute up to 10 (4)~km~s$^{-1}$ to the m=0 (1) mode, which might be a limitation of the current simulation since it does not take into account the dissipative nature of the gas. Figure \ref{fig:sim_time} shows the derived Fourier components (velocity modes) as a function of time and strength of ram pressure. We find that the kinematic m=0 and m=2 modes are established rapidly; within one quarter of the rotation period (or $\sim$150--200~Myrs). Within this short timescale, the $c_0$ term is enhanced relative to the $c_2$ by up to a factor of two. For longer timescales ($>$200~Myrs),  both the $c_0$ and $c_2$ terms become roughly equal as expected geometrically from a $\gamma_{ram}=$45$\deg$ wind orientation.
We find that the amplitude of the velocity perturbation is proportional to the strength of the ram pressure, $\mid \Delta v \mid=t_c \times a_{ram}$, with a coefficient $t_{c2}=90 \pm 15$~Myr and $t_{c0}=90 \pm 35$~Myr based on the  velocity amplitude of the $c_2$ and $c_0$ mode, respectively.  While the m=2 mode  remains very constant over time (from 200~Myrs to Gyrs), the amplitude of the m=0 mode can fluctuate about 50\% over one rotation period, which is accompanied with the build-up of an  m=1 mode if there is a significant ram wind component perpendicular to the disk. 
%The calibration coefficient appropriate for the the m=0 mode, $t_{c0}=135 \pm 45$~Myr, has a significantly larger uncertainty due to the anticipated fluctuations. %These coefficients can be applied to calculate the relative ram velocity components, which results in an accuracy of the 3d vector measurement of the ram wind of $\Delta\theta_{ram}\sim \pm10\deg$ and $\Delta\gamma_{ram}\sim \pm20\deg$, which does not include fit uncertainties.
\par

Only if the ram wind direction is exactly perpendicular to the disk will the ram pressure cause a displacement in the vertical direction, inducing an additional ``flaring'' and a minor ``U-shaped'' warp \citep[][]{Haa13}, which is balanced by the gravitational restoring force. Averaged over the velocity field of a rotating disk, this results in an m=0 mode, which can not be distinguished from the kinematic perturbation. If the ram wind angle is not exactly perpendicular to the disk, as is most often the case, the gas orbits constantly change with time and the m=0 and m=2 kinematic modes instead trace the``instantaneous'' ram pressure induced perturbations from those orbits in the out-of-plane and in-plane directions, as quantified in this paper. \par

The combination of a warped disk and a kinematic velocity perturbation due to ram wind can cause an asymmetric velocity field such as e.g. shown in the bottom right panel of  Fig. ~\ref{fig:sim_time}, which might be a possible explanation for the peculiar velocity field of, e.g., NGC~3621 (see \S~\ref{subsec:ngc3621}). We note that the ram pressure environment that is responsible for the warped disk can significantly differ, in particular in its strength, from the current ram pressure environment due to the long timescales associated with warp formation ($>$1~Gyr).

\clearpage

\subsection{Application to extragalactic 3D velocity measurements and IGM studies}

One of the main applications of kinematic ram pressure detection in galaxies is the measurement of a galaxies' movement with respect to the IGM rest-frame in all three dimensions as shown for example in Fig.~\ref{fig:NGC6946_fig5}. 
Moreover, the effective velocity change in the gas disk is approximately equal to the product of relative IGM:ISM velocity with column density contrast.
If either of these quantities can be independently constrained, the other follows. 
The relation for a dissipative collision can be approximated as (see \S \ref{subsec:math}) 
\begin{equation}
\mid v_{Ram} \mid \approx \mid \Delta v \mid \eta(N_{ISM}) \frac{N_{ISM}}{N_{IGM}}.
\end{equation}
where  $\mid \Delta v \mid$, $\eta(N_{ISM})$ and $N_{ISM}$ are given by the HI observations and fit results. The relevant column density of IGM is given by $N_{IGM} = n_{IGM}v_{Ram}t_{c}$, in terms of a critical acceleration timescale, $t_{c}$. This is equivalent to an effective ram pressure acceleration  of  
\begin{equation}
a_{Ram} \approx v_{ram}^2 \ \frac{n_{IGM}}{\eta(N_{ISM}) N_{ISM}},
\end{equation}
which generates a velocity change in the ISM given by
\begin{equation}
\mid \Delta v \mid = t_c \; v_{ram}^2 \frac{n_{IGM}}{\eta(N_{ISM}) N_{ISM}}. 
\end{equation}
While the critical timescale, $t_{c0}$, for the wind component perpendicular to the disk has a relatively large uncertainty, we can  estimate $t_c$ based on the m=2 kinematic mode derived from our simulation of gas trajectories (see \S~\ref{subsec:sim}),  with $t_{c2}$=90$ \pm 15$~Myr given by the linear coefficient between ram pressure acceleration and ram pressure induced velocity amplitude.  Note that this coefficient is relatively constant over time and independent of the strength of the gravitational potential, at least for the m=2 mode.\par
 
%In this case the ram pressure acceleration $a_{ram}$ can be estimated by the offset in z-direction $\delta z$ given by the measured velocity offset of the warp. 

\begin{figure}
\begin{center}
\includegraphics[scale=0.45]{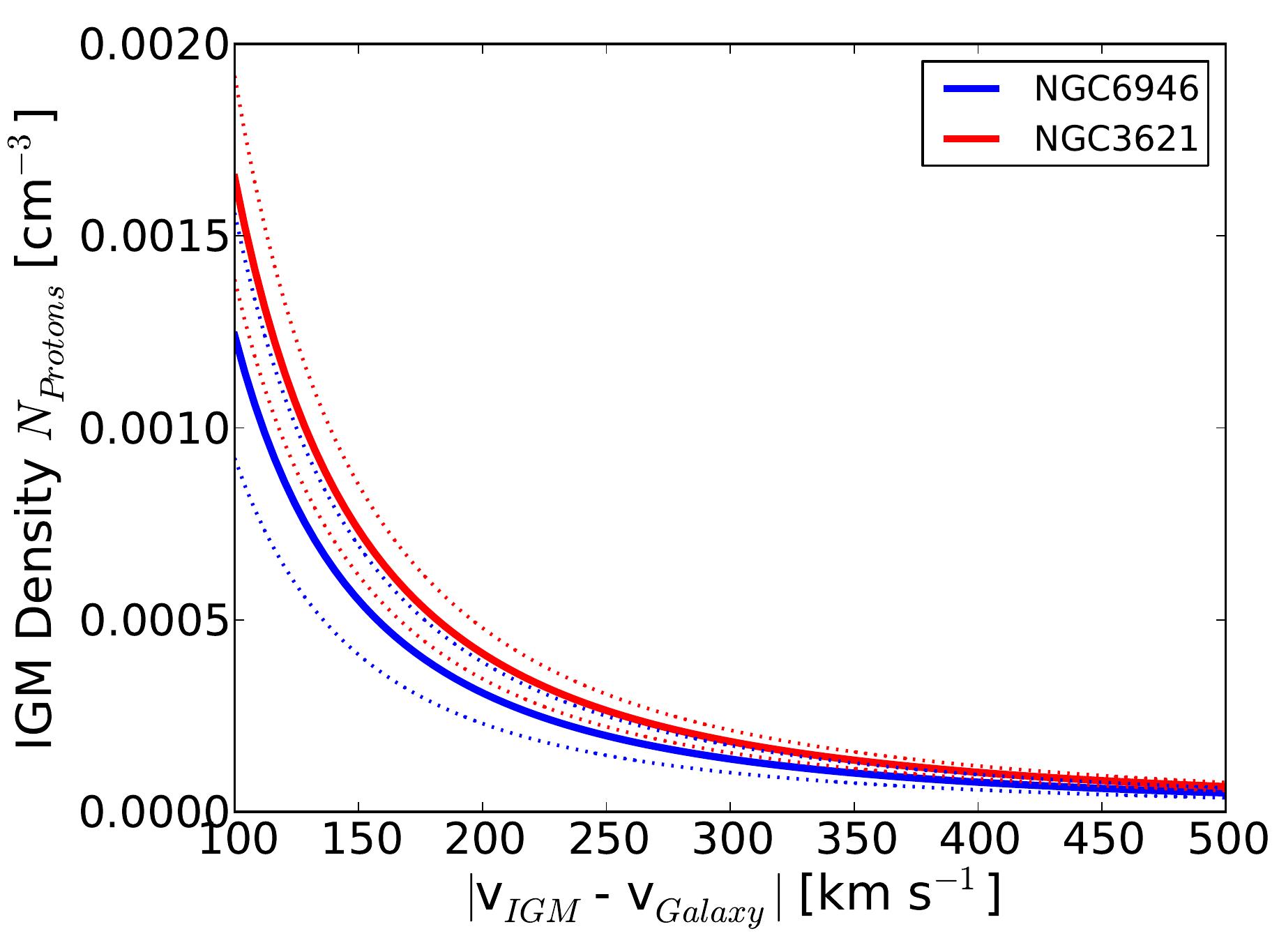}
\caption{The IGM density-velocity phase space diagram derived from our ram pressure fit models for NGC~6946 (blue) and NCG~3621 (red). The velocity is defined as the relative velocity between the galaxy and the IGM. The dotted lines indicate the upper and lower limits of our ram pressure fit results.} 
\label{fig:discussion}
\end{center}
\end{figure}

%Given the rotation velocity, $v_{rot}$, at radius, $R$, and the measured  velocity component change parallel to the disk, $\Delta v_\| $, the relation between density of the IGM and ram wind velocity can be estimated as,
%\begin{equation}
%n_{IGM} \simeq \frac{2\;N_{ISM}\;\eta(N_{ISM})}{\pi\;R} \frac{v_{rot}(R)\;\Delta v_\| }{v_{ram}^2}.
%\end{equation}
%\mid v_{Ram} \mid \approx \frac{1}{\cos{\gamma_{ram}}} [ \frac{2}{\pi} \frac{\Delta v_\| \eta(\rho_{ISM} v_{rot(R)}}{R} \frac{N_{ISM}}{\rho_{IGM}}]^{-0.5}.
This relation allows us to determine not only the ram wind vector $\mathbf{\vec{v}_{ram}}/\mid v_{ram} \mid$, but also to calculate the density - velocity phase space given by the velocity amplitude of the ram wind and the density of the IGM. 
In Fig.~\ref{fig:discussion} we show the calculated density - velocity diagram for NGC~6946 and NGC~3621.
We assume here that the ISM in the outer part of the disk ($>$10~kpc) is dominated by atomic hydrogen and have not applied a correction factor to either the ISM or IGM density for He abundance which can range from $\sim$(1.1 -- 1.3) $\times$ M$_{HI}$. The contribution of molecular gas can also be neglected at these radii, given the sharp decline in molecular gas surface density at about $\sim$7~kpc, where atomic gas starts to dominate the total gas mass \citep[see. e.g.][]{Sch11, Big12}. 
A similar sharp decline is observed for the atomic gas surface density, but at radii of about $\sim$16~kpc, more than twice the scale length of molecular gas. 
For instance, high resolution observations of  NGC~6946 with the IRAM 30-m telescope as part of the  HERA CO-Line Extragalactic Survey \citep[HERACLES,][]{Ler09} show no CO detections at radii $>250\arcsec = 7.2$~kpc.\par 

An interesting outcome of our study is that the estimated densities of the IGM for a given velocity are very similar for both galaxies, which might indicate a very similar environment.
The estimated density of the IGM for our two galaxies, e.g., at a typical velocity of $\sim$300~km~s$^{-1}$ is $(15 \pm 5) \times 10^{-5}$~cm$^{-3}$. This is consistent with the median properties expected for galaxies in group environments derived in \citet{Haa13}.
% of $<\sigma_{3D}> =$ 260~km~s$^{-1}$ and $n_H(<R>) = 10 \times 10^{-5}$~cm$^{-3}$.
These IGM densities are one to two orders of magnitude lower than in the center of galaxy clusters but converge towards similar densities at the outskirts of clusters at radii of $>$1~Mpc based on radial extrapolation of X-ray measurements \citep[see e.g.][]{Vik06}. However, it is not the goal of this study to derive a complete dynamical picture of these particular galaxies. Our examples are intended as a proof-of-concept of the kinematic ram pressure measurement approach and should merely serve as an outlook of possible future applications.\par

%The m=1 mode is established at a similar ram pressure force and the warp direction depends on the direction of the galaxy’s motion through the IGM and whether the galaxy is rotating in a clockwise or counter-clockwise direction 

%motion can be approximated by a constant momentum transfer (m=0 and m=2 kinematic mode) integrated over a timescale of less than one orbit plus an ``S-shaped'' warped disk (m=1) that evolves over longer timescales.

Another result of our kinematic ram pressure measurements is the increasing importance of the ram pressure interaction below HI column densities of N$_{HI}$=(5--12) $\times 10^{20}$~cm$^{-2}$, which roughly corresponds to the radial scale where a steep decline of the HI column density occurs as a function of radius. This turnover radius of $\sim(15\pm2)$~kpc for the two galaxies likely marks the transition from a clumpy, two phase neutral ISM \citep{Wol03}, to a diffuse single phase atomic gas, where the impact of the ram pressure wind on the ISM is significantly enhanced \citep[see e.g.][]{Vol01}.\par

\subsection{Contrasting the weak and strong ram pressure regimes}

In the strong ram pressure regime such as encountered in the Virgo Cluster, significant stripping of even dense ISM gas is observed, together with highly asymmetric morphologies of the stripped gas \citep[see e.g.][]{Vol99, Vol01, Ken04, Chu09, Vol11, Vol12}. Such asymmetries have been interpreted as providing evidence for incomplete penetration of the ram wind through the galaxy disk. In those cases where gas stripping has significantly thickened the distribution or enhanced confinement pressure has greatly increased the surface covering factor of dense gas clouds, it is possible that the penetration timescale becomes comparable to the rotation period, as outlined in \S~\ref{subsec:math}. Under these circumstances, the ``cloud-shadowing'' phenomenon may be of relevance. 

In contrast, we consider in our study the most commonly occurring circumstances which apply to galaxies, namely isolated galaxies or membership in galaxy groups of intermediate mass, where IGM densities and relative velocities (with the ram force scaling as the square of the velocity) result in ram pressure forces that are one to two orders of magnitude lower than those in clusters. Neither significant disk thickening by stripping nor enhanced cloud covering fraction are observed in such galaxies. In this case, cloud-shadowing is likely negligible for ram-disk interaction angles of $\gamma_{Ram}\ge6\deg$ due to the small volume filling factor of dense gas, as discussed in \S~\ref{subsec:math}.\par

Other effects such as magnetic fields, an inhomogeneous ISM, Rayleigh-Taylor and Kelvin-Helmholtz instabilities may complicate the interaction process and influence the stability of the gas disk \citep[see e.g.][]{Mor00, Cro05, Roe08, Clo13}.  Future studies might include additional terms in the kinematic model to account for ram-wind gradients and the evolution of the ram-wind angle, but these effects would go far beyond the aim of this study, which is simply to demonstrate the viability of kinematic ram pressure measurements.

%than cluster galaxies: a) ram forces are at least one order of magnitude larger in galaxy clusters where ram pressure can penetrate the dense gas which is much more prone to self-shielding, and b) the gradient of the intracluster gas, and thus the possibility of uneven ram pressure, is much larger in galaxy clusters than in galaxy groups. While this can lead to ram pressure asymmetries in cluster galaxies, in particular for infalling galaxies at the boundary of a cluster, such conditions are much less likely for isolated galaxies or group galaxies. 

%\clearpage

\section{Summary}
\label{sec:sum}
In this study we develop a novel method to measure the ram pressure properties of the surrounding intergalactic material (IGM) for relatively isolated galaxies based on a kinematic decomposition of the outer HI velocity field. In principle this allows us to detect ram pressure interaction at IGM densities that are an order of magnitude lower than in galaxies that show ram pressure stripping as observed in galaxy clusters.
Our results demonstrate that kinematic ram pressure perturbations from circular rotation are characterized by m=0 and m=2 modes in the residual velocity field (defined as residual between observed velocity and circular rotation), corresponding to a ram wind perpendicular and parallel to the gas disk, respectively. These velocity perturbations probe the current ram pressure environment at timescales of a fraction of the rotation period of a galaxy. Over longer timescales, ram pressure will change the orbital paths of gas clouds which can lead to the formation of warped disks after a few rotational periods. However, warps and uncertainties in the disk geometry typically generate m=1 and m=3 modes, which are clearly distinguishable from current ram pressure perturbations. We derive a mathematical description of ram pressure induced kinematic perturbations of HI velocity fields as a function of HI column density and ram wind angle for galaxies with constant inclination as well as those with a warped disk geometry. These results are in agreement with velocity fields derived from simulated gas clouds orbiting in a static gravitational potential within a ram pressure environment. \par

We have tested our models for three nearby galaxies (NGC~6946, NGC~3621, and NGC~628) observed in the HI 21~cm line emission as case studies of very different dynamical states. The impact of ram pressure on the disk kinematics, as measured by effective velocity change, increases towards lower gas column densities due to the transition from clumpy to diffuse neutral atomic gas. Utilizing Markov Chain Monte Carlo fitting of the 2-dimensional velocity field, we find that the main residual velocity in the outer disk ($\gtrsim$10~kpc radius) is primarily due to ram pressure in NGC~6946 and NGC~3621 while NGC~628 is dominated by a warped disk. Our fit models reveal the three-dimensional vector of the galaxies' movement through the IGM and provide constraints on the amplitude of the velocity and the density of the surrounding IGM.  We find evidence for an increasing impact of ram pressure at HI column densities below (4--10)$\times10^{20}$~cm$^{-2}$ which corresponds to radii greater than $\sim$15~kpc in our targets. Moreover, we find very similar densities of the intergalactic material for both galaxies, NGC~6946 and NGC~3621, at a nominal ram wind velocity (e.g. at 300~km~s$^{-1}$: $\rho_{IGM}\simeq15\times10^{-5}$ cm$^{-3}$ assuming a fully dissipative interaction).
This work demonstrates the feasibility of kinematic ram pressure measurements, even under the condition of warped disks, opening a new possibility for extragalactic velocity measurements and IGM studies.

\section*{Acknowledgements}

The authors wish to thank Jeff Kenney for a careful review of the manuscript and many valuable suggestions to improve this paper. The authors are very grateful to Peter Kamphuis for helpful discussions on warps in atomic gas disks and for valued comments on the manuscript. 
This research has made use of THINGS, ``The HI Nearby Galaxy Survey'' \citep{Wal08}. The model fitting have utilized the open source toolkit \textit{emcee}, a Markov Chain Monte Carlo ensemble sampler. This research has also made use of the NASA/IPAC Extragalactic Database (NED) which is operated by the Jet Propulsion Laboratory, California Institute of Technology, under contract with the National Aeronautics and Space Administration. 

%\clearpage

%%%%%%%%%%%%%%%%%%%%%%%%%
%%%%%%%  TABLES  %%%%%%%%%%%%
%%%%%%%%%%%%%%%%%%%%%%%%%

\onecolumn

\begin{table}
\caption{Sample Overview}
\begin{small}
\begin{tabular}{@{}lrrrrrr@{}}
\hline
Source Name & RA & DEC & Dist & $v_{sys}$ & P.A. & Inclination \\
                     &[deg]&[deg] & [Mpc]& [km~s$^{-1}$]&[deg]&[deg]\\
\hline 
NGC~6946&20h34m52.3s&+60d09m14s&5.9&42&243&33\\
NGC~3621&11h18m16.5s&-32d48m51s&6.6&728.5&345&65\\
NGC~628&01h36m41.7s&+15d47m01s&6.9&657&23&14\\
\hline
\end{tabular}
\label{tab:obs_ov}
\\
\small{Column (1): Source Name, Column (2): right ascension (J2000), Column (3): declination (J2000) , Column (4): Distance, Column (5): The systemic velocity, Column (6): The position angle of the disk (see text), Column (7): The inclination of the disk.}
\end{small}
\end{table}

\begin{table}
\caption{Fit Parameter Results}
\begin{small}
\begin{tabular}{@{}lrrrrrrrrr@{}}
\hline
Source Name & $\rho_{crit}$&$f_{smooth}$&$v_{ram}$ X&$v_{ram}$ Y&$v_{ram}$ Z& $\Delta \phi$&$\Delta i$&$r_w$&$s_w$ \\
                     &[$19^{19}$~cm$^{-2}$]&[$19^{19}$~cm$^{-2}$]&  [km~s$^{-1}$]& [km~s$^-1$]& [km~s$^{-1}$&[deg]&[deg]&[kpc]&[kpc]\\
\hline 
NGC~6946&94.9$\pm$1.8 & 98.6$\pm$13.2 & 3.6$\pm$0.8 & 27.1$\pm$1.1 & 16.0$\pm$3.2 &2.6$\pm$0.1&2.9$\pm$0.0&10.5$\pm$0.2&4.6$\pm$0.1\\
NGC~3621&41.1$\pm$0.4&6.1$\pm$0.7 & -116.4$\pm$3.8 & 8.5$\pm$0.5 & 17.0$\pm$1.4 & 19.6$\pm$0.6 & 47.4$\pm 0.1$& 11.5$\pm$0.3 & 5.8$\pm$0.1\\
NGC~628&34.0$\pm$1.1 &16.9$\pm$10.8 & 2.5$\pm$1.6 & -6.9$\pm$1.0 & 8.1$\pm$6.5 & 73.0$\pm$4.9 & 14.1$\pm$1.8 & 15.8$\pm$0.7 & 0.9$\pm$0.9\\
\hline
\end{tabular}
\label{tab:obs_par}
\\
\small{Column (1): Source Name, Column(2): The critical column density treshold, Column (3): the smooth transition range, Column (4): Effective ram component in X-direction, Column (5): Effective ram component in Y-direction (6):  Effective ram component in Z-direction (7): The parameter $\Delta \phi$ of a warped disk (see text), Column (8): The parameter $\Delta i$ of a warped disk, Column (9): The radius where the warp occurs, Column (10): the radial transition scale of the warp.}
\end{small}
\end{table}

\begin{table}
\caption{Ram Pressure Properties}
\begin{small}
\begin{tabular}{@{}lrrrrrrrrrr@{}}
\hline
Source Name & $\mid \Delta v_{ram}\mid$&$\gamma_{ram}$&$\theta_{ram}$&$vc_0$&$vc_1$&$vc_2$&$\Delta v\bot$&$\Delta v\|$ \\
                     & [km~s$^{-1}$]&[deg]&[deg]& [km~s$^{-1}$]& [km~s$^{-1}$]& [km~s$^{-1}$]& [km~s$^{-1}$]& [km~s$^{-1}$]\\
\hline 
NGC~6946&31.7$\pm$2.6 & 56.2$\pm$2.8 &117.7$\pm$0.2 & 21.6$\pm$2.4 & 5.2$\pm$0.2 & 10.1$\pm$0.1 & 26.2$\pm$3.0 & 17.8$\pm$0.3\\
% &2.6$\pm$0.1&2.9$\pm$0.0&10.5$\pm$0.2&4.6$\pm$0.1\\
NGC~3621&118.0$\pm$3.9 & -77.6$\pm$0.4 & 89.0$\pm$0.1 & 32.7$\pm$1.7 & 41.9$\pm$1.4 & 24.3$\pm$0.1 & -113.7$\pm$4.0 & 27.4$\pm$0.2\\
NGC~628&10.9$\pm$5.7 & 61.5$\pm$19.2 & 110.1$\pm$12.1 & 9.3$\pm$6.2 & 36.4$\pm$15.0 & 1.3$\pm$0.4 & 9.7$\pm$6.3 & 4.9$\pm$0.7\\
\hline
\end{tabular}
\\
\label{tab:obs_res}
\small{Column (1): Source Name, Column (2): The amplitude of the total change in velocity due to ram pressure, Column (3): The ram wind angle between the disk and the ramwind vector (4):  the azimuthal angle of the ram wind vector projected onto the disk (5): the effective change of the line-of-sight velocity component due to m=0 mode, Column (6): effective change due to m=1 mode, Column (7): effective change due to m=2 mode, Column (8): the effective ram wind velocity component perpendicular to the disk, Column (9): the effective ram wind velocity component parallel to the disk.}
\end{small}
\end{table}

\clearpage

\twocolumn

\onecolumn

\appendix
\section{Model Velocity Field and Decomposition for a Warped Disk}
\label{app:a}

\begin{figure}
\begin{center}
\includegraphics[scale=0.6]{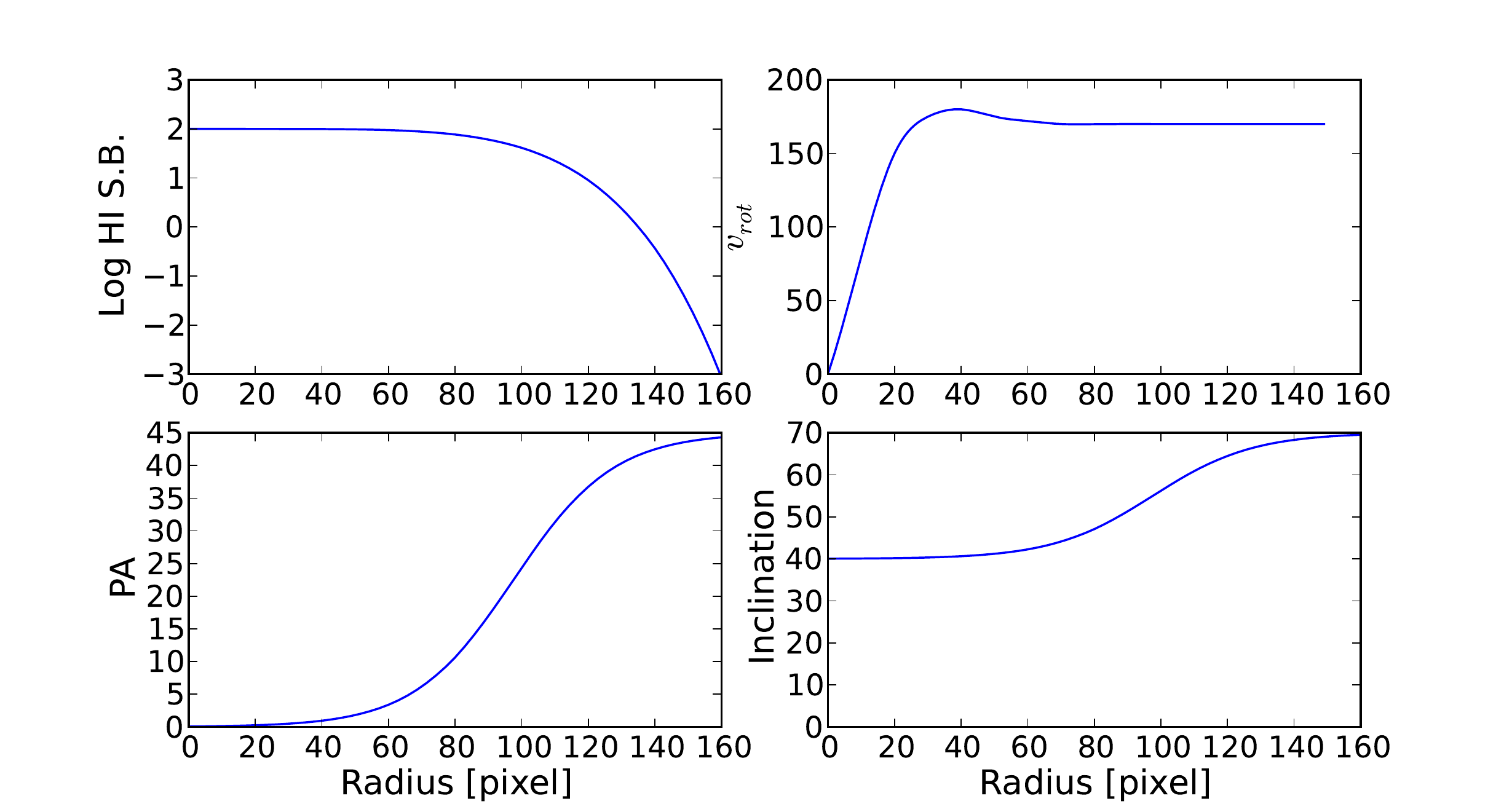}
\caption{Radial profile of the  disk surface brightness (top left), circular rotation velocity (top right), position angle (bottom left), and inclination (bottom right) for a warped disk}. 
\label{fig:warpmodel_radial}
\end{center}
\end{figure}

We have tested our fit routines and analysis of the velocity field pattern with a simple warped disk, which is described in the following. First we have created a disk with circular rotation velocity and a disk surface brightness as function of radius as described in \S~\ref{subsec:model}. The change in inclination and position angle is described as a smooth transition between the inner disk and an outer disk with a change in inclination and PA given by the transition functions,
\begin{equation}
i(r)=i_0 + \Delta i[0.5+0.5\tanh((r -r_W)/s_W)],
\end{equation}
and 
\begin{equation}
\phi(r)=\phi_0 + \Delta\phi[0.5+0.5\tanh((r -r_W)/s_W)],
\end{equation}
in terms of the amplitude change, $\Delta i$ in inclination and $\Delta\phi$ in PA, and the transition radius $r_W$ and scale of transition $s_W$. The azimuthal angle in the plane of the galaxy is given by,
\begin{equation}
\cos[\theta_W(r)]=\frac{-X\sin[\phi(r)] + Y\cos[\phi(r)]}{r}, \\
\sin[\theta_W(r)]=\frac{-X\cos[\phi(r)] - Y\sin[\phi(r)]}{r cos[i(r)]}. 
\end{equation}
The radial profiles of the model are shown in Fig.~\ref{fig:warpmodel_radial} given a change of inclination of $\Delta i=30\deg$ (in addition to $i_0=40\deg$) and a position angle shift of the warp $\Delta\phi=45\deg$ with $r_W=2/3\times$disk length and $s_W=0.2\times$disk length. The velocity field of a warped disk is then described by, 
\begin{equation}
v_W = v_C \cos(\theta_W)\sin[i(r)],
\end{equation}
and the residual field (for which the warp is simply neglected) by,
\begin{equation}
v_{Res}= v_W - v_C \cos(\theta_0)\sin(i_0),
\end{equation}
which is plotted in Fig.~\ref{modelwarp}.
 %created a simple warped disk model with a  transition in inclination and position angle which is described in detail in App.~\ref{app:a}. The results as shown in Fig.~\ref{fig_warpmodel} reveals a velocity field typical for observed warped disk.  The residual velocity field results in a $m=1$ component (bimodal) as function of radial distance which clearly distinguishes from the observed ram interaction pattern ($m=0$ and $m=2$ component).
 
\section{Deriving the Velocity Components for a Ram Wind with a Warped Disk}
\label{app:b}

%The inclination $i(r)$ and $\theta(r)$ of a warped disk as a function of radius can be described as a transition function at the transition radius $r_W$ and with the smoothness parameter $s_W$ characterising the length of the transition, as given as
%\begin{equation}
%i(r)=i_0+i_W(0.5+0.5\tanh[(r-r_W)/s_W]),
%\end{equation}
%and
%\begin{equation}
%\theta(r)=\theta_0+\theta_W(0.5+0.5\tanh[(r-r_W)/s_W]),
%\end{equation}
%where $\theta_0$ and $i_0$ are the PA and inclination of the inner disk and $i_0+i_W$ and $\theta_0+\theta_W$ the PA %and inclination of the outer warped disk.\par

The line-of-sight velocity component for a warped disk with ram pressure terms is given by, 
\begin{equation}
\label{eq:ramwarpvelocity1}
v_{LoS}=v_{Sys}+[v_{Rot}(r)+v_{RamRot}]\cos[\theta_W(r)]\sin[i(r)] + v_{RamExp}\sin[\theta_W(r)]\sin[i(r)] + v_{Ram\bot} \cos[i(r)].
\end{equation}
with the ram velocity terms as derived in \S~\ref{subsec:velfield}:
\begin{eqnarray}
v_{Ram\bot}&=&\mid \Delta v\mid \eta(\rho_{ISM})\sin[\gamma_{Ram}(r)], \\
v_{RamRot}&=&\mid \Delta v\mid \eta(\rho_{ISM})\sin[\theta_W(r)-\theta_{Ram}(r)]\cos[\gamma_{Ram}(r)],\\
v_{RamExp}&=&\mid \Delta v\mid \eta(\rho_{ISM})\cos[\theta_W(r)-\theta_{Ram}(r)]\cos[\gamma_{Ram}(r)]. 
\end{eqnarray}
While the ram wind vector field is fixed with respect to the sky plane (X,Y,Z),  the angles $\theta_{Ram}$ and $\gamma_{Ram}$ depend on the relative alignment of the plane of the galaxy and the sky plane as described in App.~\ref{app:c}. 
The change in inclination and PA are given by,
\begin{eqnarray}
i(r)&=&i_0 + i_W(r),\\
\phi(r)&=&\phi_0 + \phi_W(r).
\end{eqnarray}
where $i_W(r)$ and $\phi_W(r)$ can be described, e.g., by a transition function between inner and outer disk given as described in  App.~\ref{app:a}.
The azimuthal angle $\theta$ measured in the plane of a warped disk is then defined as
\begin{eqnarray}
\cos[\theta_W(r)]&=&\frac{-X\sin[\phi(r)] + Y\cos[\phi(r)]}{r} \\
\sin[\theta_W(r)]&=&\frac{-X\cos[\phi(r)] - Y\sin[\phi(r)]}{r\cos[i(r)]}
\label{eq:thetawarp}
\end{eqnarray}
Subtracting the ``nominal'' velocity field of an unperturbed, fixed orientation disk, 
\begin{equation}
\label{eq:circvelocity}
v_{Nom}=v_{Sys}+v_{Rot}(r)\cos(\theta)\sin(i_0),
\end{equation}
with 
\begin{equation}
\cos(\theta)=\frac{-X\sin[\phi_0] + Y\cos[\phi_0]}{r}
\end{equation}
yields the equation for the residual velocity field,
\begin{eqnarray}
\label{eq:ramwarpres1}
v_{Res}&=&v_{Rot}(r)\sin[i(r)]\{\cos[\theta_W(r)]-\cos(\theta)\} \\
& &+v_{RamRot}\cos[\theta_W(r)]\sin[i(r)] + v_{RamExp}\sin[\theta_W(r)]\sin[i(r)] + v_{Ram\bot} \cos[i(r)].
\end{eqnarray}
The first part of the equation describes the residual term due to the warped disk with a changing rather than constant inclination $i(r)$ and PA $\phi(r)$. The second and third terms characterize the ram interaction terms of the warped disk. The first term can be simplified as,
\begin{eqnarray}
\label{eq:ramwarpres}
v_{ResW}&=&v_{Rot}(r)\{\sin[i_W(r)]\cos[\theta_W(r)]-\sin[i_0]\cos(\theta)\} \\
&=&v_{Rot}(r)\bigg\{\sin[i_W(r)]\frac{-X\sin[\phi_0+\phi_W(r)] + Y\cos[\phi_0+\phi_W(r)]}{r}   - \sin[i_0]\frac{-X\sin[\phi_0] + Y\cos[\phi_0]}{r}\bigg\}
\end{eqnarray}
Here we make use of the trigonometric rules for linear combinations (phasor addition):
\begin{eqnarray}
a\sin(x) + b\sin(x+\beta)&=&c\sin(x+\delta)\\
a\cos(x) + b\cos(x+\beta)&=&c\cos(x+\delta)
\end{eqnarray}
with 
\begin{equation}
c=\sqrt{a^2+b^2+2ab\cos(\beta)}
\end{equation}
and
\begin{equation}
\delta=\arctan(\frac{b\sin(\beta)}{a+b\cos(\beta)}) + 
\begin{cases}
0 , \; a+b\cos[\beta] \geq0 ,\\
\pi , \;  a+b\cos[\beta]<0.
\end{cases}
\end{equation}
If we substitute $a$ with $-\sin(i_0)$,  $b$ with $\sin[i_W(r)]$, $x$ with $\phi_0$, and $\beta$ with $\phi_W(r)$, we obtain
\begin{eqnarray}
v_{Res}&=&v_{Rot}(r)\;c\frac{-X\sin[\phi_0+\delta]+Y\cos[\phi_0+\delta]}{r}\\
&=&v_{Rot}(r) \; c\; \cos[\theta_{W*}(r)].
\end{eqnarray}
which has the form of a first order Fourier component in terms of the modified azimuthal angle, $\theta_{W*}$, that scales as a function of radius with
\begin{eqnarray}
c(r)&=&\sqrt{\sin^2[i_0] + \sin^2[i_W(r)] - 2\sin[i_0]\sin[i_W(r)]\cos[\phi_W(r)]}\\
\theta_{W*}(r)&=&\frac{-X\sin[\phi_0+\delta]+Y\cos[\phi_0+\delta]}{r}\\
\delta &=&\arctan\{\frac{\sin[i_W(r)]\sin[\phi_W(r)]}{-\sin[i_0]+\sin[i_W(r)]\cos[\phi_W(r)]}\} +
\begin{cases}
0 , \; -\sin[i_0]+\sin[i_W(r)]\cos[\phi_W(r)] \geq0 ,\\
\pi , \;  -\sin[i_0]+\sin[i_W(r)]\cos[\phi_W(r)]<0.
\end{cases}
\end{eqnarray}
%where we can substitute $a$ with $\theta_0-\phi_W(r)$ and $b$ with $\theta_0$ to show that,
%\begin{equation}
%\cos[\theta_0-\phi_W(r)]-\cos(\theta_0) = 2\sin[\phi_W(r)/2]\sin[\theta_0-\phi_W(r)/2].
%\end{equation}
The entire residual velocity field is then given as,
\begin{eqnarray}
\label{eq:ramwarpres2}
v_{Res}&=&v_{Rot}(r) c(r) \cos[\theta_{W*}(r)] \\
& &+\mid \Delta v\mid \eta(\rho_{ISM}) \{ \cos[\gamma_{Ram}(r)]\sin[i(r)]\sin[2\theta_W(r)-\theta_{Ram}(r)] +  \cos[i(r)]\sin[\gamma_{Ram}(r)] \}
\end{eqnarray}
where the trigonometric identity,
\begin{equation}
\sin(2a-b)=\sin(a-b)\cos(a)+\cos(a-b)\sin(a),
\end{equation}
has been used for the second term. The first term represents a first order Fourier component ($\cos[\theta_{W*}(r)]$) and describes the main residual of a warped disk in comparison to a circular rotating disk with constant inclination and position angle, while the second part of the equation shows the second order Fourier component ($\sin[2\theta_W(r)-\theta_{Ram}(r)]$) and the zeroth (independent of $\theta$) order, which characterize the ram interaction terms of a warped disk parallel and perpendicular to the disk, respectively. 

\section{Coordinate System Relations}
\label{app:c}

\begin{figure}
\begin{center}
\includegraphics[scale=0.6]{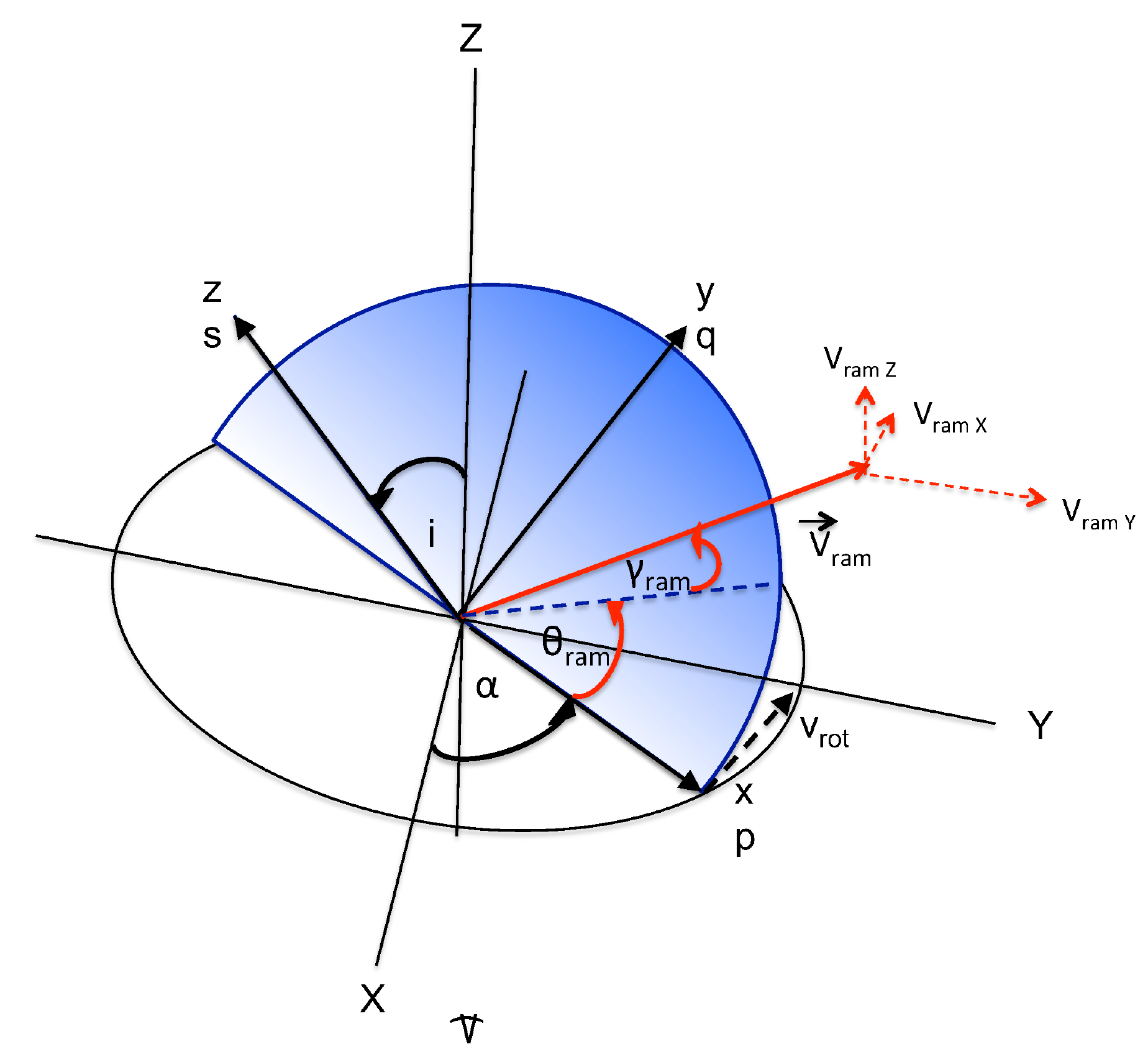}
\caption{Illustration of the alignment of the ram wind vector and the plane of the galaxy (x,y,z) with respect to the sky plane (X,Y,Z) given by the inclination $i$ and the position angle $\alpha = \pi/2 - \phi$.  The projection of the ram wind vector $v_{Ram}$(X,Y,Z) onto the plane of the galaxy is described by $\gamma_{Ram}$ and the azimuthal angle $\theta_{Ram}$ (see text). The line-of-sight is along the $Z$-axis.}
\label{geometry2}
\end{center}
\end{figure}

The coordinate system of the ram wind vector field can be described either with respect to the plane of the galaxy (x,y,z) or the plane of the sky (X,Y,Z) as illustrated in Fig.~\ref{geometry2}. The transformation between the two coordinate systems is described in detail in Vogelaar 2006\footnote{\href{http://www.astro.rug.nl/~gipsy/inspector/}{http://www.astro.rug.nl/~gipsy/inspector/}}.  While we assume that the ram wind vector is constant in the sky plane, the position angle $\alpha = \pi/2 - \phi$ and inclination $i$ of the plane of the galaxy can change (tilted ring model). Thus the relative contribution of the ram wind to the perpendicular, rotational, and radial velocity of the tilted ring will change as function of $\phi$ and $i$ as well. In the following we will provide a mathematical transformation of coordinates between the plane of the sky and the plane of a tilted ring.\par

The components of the rotation axis $\overrightarrow{s}$ of the tilted ring are:
\begin{eqnarray}
s_X&=&\sin(\alpha)\sin(i)\\
s_Y&=&-\cos(\alpha)\sin(i)\\
s_Z&=&\cos(i).
\end{eqnarray}
The components of the unit vector $\overrightarrow{p}$ along the receding part of the major axis of the tilted ring are:
\begin{eqnarray}
p_X&=&\cos(\alpha)\\
p_Y&=&\sin(\alpha)\\
p_Z&=&0.
\end{eqnarray}
The components of the unit vector perpendicular to $\overrightarrow{s}$ and $\overrightarrow{p}$, $\overrightarrow{q}=\overrightarrow{s} \times \overrightarrow{p}$, are
\begin{eqnarray}
q_X&=&-\sin(\alpha)\cos(i)\\
q_Y&=&\cos(\alpha)\cos(i)\\
q_Z&=&\sin(i).
\end{eqnarray} 
Now we define the transformation between the sky coordinate system and the plane of the tilted ring system which is given by the coordinates $z\| \overrightarrow{s}$, $x\| \overrightarrow{p}$, and $y\|\overrightarrow{q}$. If we use the unit vectors $\overrightarrow{e}$ then $\overrightarrow{e_x}=\overrightarrow{p}$, $\overrightarrow{e_y}=\overrightarrow{q}$, and $\overrightarrow{e_z}=\overrightarrow{s}$. The relation between the tilted ring system and the system of the sky plane is
\begin{eqnarray}
\overrightarrow{e_x}&=&\cos(\alpha)\overrightarrow{e_X} + \sin(\alpha)\overrightarrow{e_Y}\\
\overrightarrow{e_y}&=&-\sin(\alpha)\cos(i)\overrightarrow{e_X} + \cos(\alpha)\cos(i)\overrightarrow{e_Y}+ \sin(i)\overrightarrow{e_Z}\\
\overrightarrow{e_z}&=&\sin(\alpha)\sin(i)\overrightarrow{e_X} - \cos(\alpha)\sin(i)\overrightarrow{e_Y}+ \cos(i)\overrightarrow{e_Z},
\end{eqnarray} 
and the inverse:
\begin{eqnarray}
\overrightarrow{e_X}&=&\cos(\alpha)\overrightarrow{e_x} - \sin(\alpha)\cos(i)\overrightarrow{e_y}+\sin(\alpha)\sin(i)\overrightarrow{e_z}\\
\overrightarrow{e_Y}&=&\sin(\alpha)\overrightarrow{e_x} + \cos(\alpha)\cos(i)\overrightarrow{e_y}- \cos(\alpha)\sin(i)\overrightarrow{e_z}\\
\overrightarrow{e_Z}&=&\sin(i)\overrightarrow{e_y}+ \cos(i)\overrightarrow{e_z}.
\end{eqnarray} 
Hence the transformation of the ram vector $\overrightarrow{\theta_{Ram}}$ between the coordinate system is given by:
\begin{eqnarray}
v_{Ram\;x}&=&v_{RamX} \cos(\alpha)+ v_{RamY}\sin(\alpha)\\
v_{Ram\;y}&=&-v_{RamX}\sin(\alpha)\cos(i) + v_{RamY}\cos(\alpha)\cos(i)+ v_{RamZ}\sin(i)\\
v_{Ram\;z}&=&v_{RamX}\sin(\alpha)\sin(i)- v_{RamY}\cos(\alpha)\sin(i)+ v_{RamZ}\cos(i).
\end{eqnarray} 
The sum of the components of the ram wind is,
\begin{equation}
|\overrightarrow{v_{Ram}}|=\sqrt{v_{RamX}^2+v_{RamY}^2+v_{RamZ}^2}
\end{equation}
The angle $\gamma_{Ram}$ is then given  as
\begin{eqnarray}
\sin(\gamma_{Ram})&=&\frac{v_{Ram\;z}}{\mid \overrightarrow{v_{Ram}} \mid}\\
&=&\frac{v_{RamX}\sin(\alpha)\sin(i)- v_{RamY}\cos(\alpha)\sin(i)+ v_{RamZ}\cos(i)}{\mid \overrightarrow{v_{Ram}} \mid},
\end{eqnarray} 
and $\theta_{Ram}$ as
\begin{eqnarray}
\cos(\theta_{Ram})&=&\frac{v_{Ram\;x}}{\mid \overrightarrow{v_{Ram}} \mid \cos(\gamma_{Ram})}\\
&=&\frac{v_{RamX} \cos(\alpha)+ v_{RamY}\sin(\alpha)}{\mid \overrightarrow{v_{Ram}} \mid \cos(\gamma_{Ram})}.
\end{eqnarray} 
\

\clearpage

\label{lastpage} 

\begin{thebibliography}{}
% see papers in proposal
 \bibitem[Abramson et al.(2011)]{Abr11} Abramson, A., Kenney, J.~D.~P., Crowl, H.~H., et al.\ 2011, \aj, 141, 164 
\bibitem[Arrigoni Battaia et al.(2012)]{Arr12} Arrigoni Battaia, F., Gavazzi, G., Fumagalli, M., et al.\ 2012, \aap, 543, A112 
\bibitem[Bigiel \& Blitz(2012)]{Big12} Bigiel, F., \& Blitz, L.\ 2012, \apj, 756, 183 
\bibitem[Braun(1991)]{Bra91} Braun, R.\ 1991, \apj, 372, 54 
\bibitem[Briggs(1990)]{Bri90} Briggs, F.~H.\ 1990, \apj, 352, 15 
 \bibitem[Cayatte et al.(1990)]{Cay90} Cayatte, V., van Gorkom, J.~H., Balkowski, C., \& Kotanyi, C.\ 1990, \aj, 100, 604 
\bibitem[Chung et al.(2009)]{Chu09} Chung, A., van Gorkom,  J.~H., Kenney, J.~D.~P., Crowl, H., \& Vollmer, B.\ 2009, \aj, 138, 1741 
\bibitem[Close et al.(2013)]{Clo13} Close, J.L.,  Pittard, J. M., Hartquist, T. W. , Falle, S. A. E. G.\ 2013 \mnras, in press
\bibitem[Cort{\'e}s et al.(2006)]{Cor06} Cort{\'e}s, J.~R., Kenney, J.~D.~P., \& Hardy, E.\ 2006, \aj, 131, 747 
\bibitem[Crowl et al.(2005)]{Cro05} Crowl, H.~H., Kenney, J.~D.~P., van Gorkom, J.~H., \& Vollmer, B.\ 2005, \aj, 130, 65
\bibitem[Crowl \& Kenney(2008)]{Cro08} Crowl, H.~H., \& Kenney, J.~D.~P.\ 2008, \aj, 136, 1623 
\bibitem[Cui et al.(2010)]{Cui10} Cui, H.-J., Xu, H.-G., Gu, J.-H., et al.\ 2010, Research in Astronomy and Astrophysics, 10, 301
\bibitem[de Blok et al.(2008)]{deB08} de Blok, W.~J.~G., Walter, F., Brinks, E., et al.\ 2008, \aj, 136, 2648 
\bibitem[Domainko et al.(2006)]{Dom06} Domainko, W., Mair, M., Kapferer, W., et al.\ 2006, \aap, 452, 795
\bibitem[Foreman-Mackey et al.(2012)]{For12} Foreman-Mackey, D., Hogg,D.W., Lang, D., Goodman, J. \ 2012, arXiv:1202.3665
\bibitem[Gil de Paz et al.(2007)]{Gil07} Gil de Paz, A., Boissier, S., Madore, B.~F., et al.\ 2007, \apjs, 173, 185 
\bibitem[Gunn \& Gott(1972)]{Gun72} Gunn, J.~E., \& Gott, J.~R., III 1972, \apj, 176, 1 
\bibitem[Haan et al.(2009)]{Haa09} Haan, S., Schinnerer, E., Emsellem, E., et al.\ 2009, \apj, 692, 1623 
\bibitem[Haan \& Braun(2014)]{Haa13} Haan, S. , \& Braun, R.,\ 2014, \mnras, 440, L21
\bibitem[Haynes \& Giovanelli(1986)]{Hay86} Haynes, M.~P., \& Giovanelli, R.\ 1986, \apj, 306, 466 
\bibitem[Helou et al.(1981)]{Hel81} Helou, G., Salpeter, E.~E., Giovanardi, C., \& Krumm, N.\ 1981, \apjs, 46, 267 
\bibitem[Hidaka \& Sofue(2002)]{Hid02} Hidaka, M., \& Sofue, Y.\ 2002, \pasj, 54, 33 
\bibitem[Hoffman et al.(1988)]{Hof88} Hoffman, G.~L., Helou, G., \& Salpeter, E.~E.\ 1988, \apj, 324, 75 
\bibitem[Kamphuis \& Briggs(1992)]{Kam92} Kamphuis, J., \& Briggs, F.\ 1992, \aap, 253, 335 
%\bibitem[Kamphuis(2008)]{Kam08} Kamphuis, P.\ 2008, Ph.D.~Thesis
\bibitem[Kamphuis et al.(2013)]{Kam13} Kamphuis, P., Rand, R.~J., J{\'o}zsa, G.~I.~G., et al.\ 2013, \mnras, 1790 
\bibitem[Kenney et al.(2004)]{Ken04} Kenney, J.~D.~P., van Gorkom, J.~H., \& Vollmer, B.\ 2004, \aj, 127, 3361 
\bibitem[Leroy et al.(2009)]{Ler09} Leroy, A.~K., Walter, F., Bigiel, F., et al.\ 2009, \aj, 137, 4670 
\bibitem[Lucero et al.(2005)]{Luc05} Lucero, D.~M., Young, L.~M., \& van Gorkom, J.~H.\ 2005, \aj, 129, 647 
\bibitem[Machacek et al.(2006)]{Mac06} Machacek, M., Jones, C., Forman, W.~R., \& Nulsen, P.\ 2006, \apj, 644, 155 
\bibitem[Mori \& Burkert(2000)]{Mor00} Mori, M., \& Burkert, A.\ 2000, \apj, 538, 559
\bibitem[Nakanishi et al.(2006)]{Nak06} Nakanishi, H., Kuno, N., Sofue, Y., et al.\ 2006, \apj, 651, 804
\bibitem[Portas et al.(2009)]{Por09} Portas, A., Brinks, E., Usero, A., et al.\ 2009, IAU Symposium, 254, 52P 
\bibitem[Rangarajan et al.(1995)]{Ran95} Rangarajan, F.~V.~N., White, D.~A., Ebeling, H., \& Fabian, A.~C.\ 1995, \mnras, 277, 1047  
\bibitem[Roediger \& Hensler(2008)]{Roe08} Roediger, E., \& Hensler, G.\ 2008, \aap, 483, 121
\bibitem[J{\'o}zsa(2007)]{Joz07} J{\'o}zsa, G.~I.~G.\ 2007, \aap, 468, 903 
\bibitem[Schindler et al.(2005)]{Sch05} Schindler, S., Kapferer, W., Domainko, W., et al.\ 2005, \aap, 435, L25
\bibitem[Schruba et al.(2011)]{Sch11} Schruba, A., Leroy, A.~K., Walter, F., et al.\ 2011, \aj, 142, 37 
\bibitem[Shostak \& van der Kruit(1984)]{Sho84} Shostak, G.~S., \& van der Kruit, P.~C.\ 1984, \aap, 132, 20 
%\bibitem[Shull \& Draine(1987)]{Shu87} Shull, J.~M., \& Draine, B.~T.\ 1987, Interstellar Processes, 134, 283 
\bibitem[van der Kruit \& Allen(1978)]{Kru78} van der Kruit, P.~C., \& Allen, R.~J.\ 1978, \araa, 16, 103 
\bibitem[van der Kruit(2007)]{Kru07} van der Kruit, P.~C.\ 2007, \aap, 466, 883
\bibitem[Veilleux et al.(1999)]{Vei99} Veilleux, S., Bland-Hawthorn, J., Cecil, G., Tully, R.~B., 
\& Miller, S.~T.\ 1999, \apj, 520, 111 
\bibitem[Vikhlinin et al.(2006)]{Vik06} Vikhlinin, A., Kravtsov, A., Forman, W., et al.\ 2006, \apj, 640, 691 
\bibitem[Vollmer et al.(1999)]{Vol99} Vollmer, B., Cayatte, V., Boselli, A., Balkowski, C., \& Duschl, W.~J.\ 1999, \aap, 349, 411 
\bibitem[Vollmer et al.(2001)]{Vol01} Vollmer, B., Cayatte, V., Balkowski, C., \& Duschl, W.~J.\ 2001, \apj, 561, 708 
\bibitem[Vollmer(2003)]{Vol03} Vollmer, B.\ 2003, \aap, 398, 525 
\bibitem[Vollmer et al.(2004)]{Vol04} Vollmer, B., Balkowski, C., Cayatte, V., van Driel, W., \& Huchtmeier, W.\ 2004, \aap, 419, 35 
\bibitem[Vollmer(2009)]{Vol09} Vollmer, B.\ 2009, \aap, 502, 427 
\bibitem[Vollmer \& Leroy(2011)]{Vol11} Vollmer, B., \& Leroy, A.~K.\ 2011, \aj, 141, 24 
\bibitem[Vollmer et al.(2012)]{Vol12} Vollmer, B., Wong, O.~I., Braine, J., Chung, A., \& Kenney, J.~D.~P.\ 2012, \aap, 543, A33 
\bibitem[Vollmer et al.(2013)]{Vol13} Vollmer, B., Soida, M., Beck, R., et al.\ 2013, \aap, 553, A116 
\bibitem[Walter et al.(2008)]{Wal08} Walter, F., Brinks, E., de Blok, W.~J.~G., et al.\ 2008, \aj, 136, 2563 
\bibitem[Westmeier et al.(2011)]{Wes11} Westmeier, T., Braun, R., \& Koribalski, B.~S.\ 2011, \mnras, 410, 2217 
\bibitem[Wolfire et al.(2003)]{Wol03} Wolfire, M.~G., McKee,  C.~F., Hollenbach, D., \& Tielens, A.~G.~G.~M.\ 2003, \apj, 587, 278 
\end{thebibliography}
\end{document}